\documentclass[aps,prb,twocolumn,preprintnumbers,superscriptaddress,amsmath]{revtex4}
\usepackage{graphicx}
\usepackage{graphicx}
\usepackage{bm}
\usepackage{hyperref}
 \hypersetup{
	colorlinks = true,
	allcolors= blue,
	hypertexnames=false,
}
\usepackage[mathlines]{lineno}
\usepackage[utf8]{inputenc}
\usepackage{setspace}
\usepackage[T1]{fontenc}
\usepackage{physics}
\usepackage{chemformula}
\usepackage{amsfonts}
\usepackage{amsmath}
\usepackage{xcolor}
\usepackage[shortlabels]{enumitem}
\usepackage{makecell}

\usepackage[scr = dutchcal]{mathalfa}

\newcommand{\mtrx}[1]{
   \overset{\raisebox{-0.8ex}{\text{\tiny{$\leftrightarrow$}}}}{#1}
}

\newcommand{\LamR}{\mtrx{\bm{\Lambda}}{}_{\mathbf{R}}}
\newcommand{\LamP}{\mtrx{\bm{\Lambda}}{}_{\bf P}}
\newcommand{\rhon}{\mtrx{\varrho}{}^\text{n}}
\newcommand{\rhoan}{\mtrx{\varrho}{}^\text{an}}
\newcommand{\wt}[1]{\widetilde{#1}}

\renewcommand{\a}{\alpha}
\renewcommand{\b}{\beta}

\newcommand{\w}{\omega}

\renewcommand{\k}{\kappa}

\newcommand{\herm}{^\dagger}

\renewcommand{\H}{\mathcal{ H}}
\newcommand{\V}{\mathcal{V}}

\newcommand{\R}{\mathcal{R}}

\newcommand{\pert}{^{\scriptscriptstyle{(1)}} }
\newcommand{\unpert}{^{\scriptscriptstyle{(0)}} }

\newcommand{\HA}{\text{HA}}
\newcommand{\BO}{\text{BO}}
\newcommand{\anh}{\text{anh}}

\DeclareMathAlphabet{\mathsfit}{\encodingdefault}{\sfdefault}{m}{sl}
\SetMathAlphabet{\mathsfit}{bold}{\encodingdefault}{\sfdefault}{bx}{sl}
\newcommand{\tens}[1]{\bm{\mathsfit{#1}}}
\newcommand{\tenscomp}[1]{\mathsfit{#1}}

\usepackage{textcomp}
\usepackage{graphicx}

\newcommand{\myoverset}[2]{
  \mathrel{\vbox{\offinterlineskip\ialign{
    \hfil##\hfil\cr
    $\scriptstyle {#1}$\cr
    \noalign{\kern-.1ex}
    ${#2}$\cr
}}}}

\newcommand{\kket}[1]{\big|#1\big\rangle}
\newcommand{\bbra}[1]{\big\langle #1 \big|}
\newcommand{\bbrakket}[2]{\big\langle #1 \big| #2 \big\rangle}

\newcommand{\ddyad}[2]{\big|#1\big\rangle\big\langle#2\big|}

\newcommand{\scf}{\text{scf}}
\newcommand{\tot}{\text{tot}}

\usepackage{upgreek}

\usepackage{mathrsfs}
\DeclareMathAlphabet{\mathscrbf}{OMS}{mdugm}{b}{n}

\setcitestyle{numbers,square}

\usepackage{booktabs}
\usepackage{multirow}

\begin{document}
\title{Effective single particle picture for anharmonic lattice dynamics: a Rosetta stone for electronic and ionic response}

\author{Giovanni Caldarelli}
\email{giovanni.caldarelli@epfl.ch}
\affiliation{Dipartimento di Fisica, Universit\`a di Roma La Sapienza, Piazzale Aldo Moro 5, I-00185 Roma, Italy}
\affiliation{Theory and Simulation of Materials (THEOS),
Ecole Polytechnique Federale de Lausanne, 1015 Lausanne, Switzerland}

\author{Francesco Mauri}
\affiliation{Dipartimento di Fisica, Universit\`a di Roma La Sapienza, Piazzale Aldo Moro 5, I-00185 Roma, Italy}

\begin{abstract}
We establish a theoretical framework for the dynamics of a lattice of ions in a mean-field approach, where anharmonicity is included via self-consistency. In this picture, the many-body dynamics of a system of $N$ atoms in three dimensions is mapped onto two kinds of $6N$-dimensional vectors: the phonon condensate, describing the evolution of the average atomic positions, and the phonon spinors, describing the evolution of the atomic elastic constants. The phonon spinors are classified by a quantum number that behaves as a spin: the phonon pseudospin.

The many-body Liouville equation is replaced by two wave equations equivalent to the time-dependent Schrodinger equation for the electronic wave function in density functional theory. Exploiting this parallelism, we formulate the response of the anharmonic lattice in one-to-one correspondence with time-dependent density functional theory for electrons. In complete analogy with the electronic case, we express the ionic response in terms of matrix elements of operators representing external fields and forces. We show how anharmonicity screens external perturbations through a phonon analogue of the Hartree-exchange-correlation kernel.

We provide expressions for the lattice optical and thermal conductivity, showing how thermal conductivity depends on the phonon pseudospin. By approximating the density matrix as a Gaussian, we recover the equations of the time-dependent self-consistent harmonic approximation. In this case, the linear-response equations are formulated in terms of an anharmonic kernel including three- and four-phonon scattering.

By translating anharmonic lattice dynamics into the language of density functional theory, this work shows how theoretical and computational advances in modeling the dynamical response of interacting electrons can be directly applied to interacting ions.
\end{abstract}

\maketitle

\section{Introduction}
Ionic responses are ubiquitous in condensed matter physics, theoretical chemistry, and material science. Studying how a lattice of atoms responds to external stimuli provides the pivotal characteristics that make a material promising for technological applications or to unveil unexplained physical phenomena.

For example, the behaviour of an ionic lattice in a temperature gradient or in a static electric field provides, respectively, lattice thermal conductivity and electric conductivity, two of the most scrutinized properties for energy applications \cite{snyder2008complex, bachman2016inorganic, bell2008cooling}. In the field of lattice thermal transport, major theoretical advancements have been made in uncovering heat transport through coherent wave-like tunneling of phonons \cite{allen1993thermal, simoncelli2019unified, simoncelli2022wigner, fiorentino2023from,dangic2025lattice, castellano2025mode}, providing new pathways for material modeling of ultra-efficient thermal insulators \cite{xia2023unified, li2025high, dilucente2023crossover,xia2020high}. However, many open questions remain e.g.\ on how to properly compute the conductivity of systems undergoing structural phase transitions \cite{aseginolaza2019phonon, dangic2021origin}, or featuring unexplained ultrahigh conductivity \cite{kang2018experimental, niyikiza2025thermal}.

The rising demand in solid-state batteries of the recent decades has stimulated intense research in the mechanisms of charge transport in solid-state electrolytes \cite{kamaya2011lithium, kato2016high, mizuno2005new}. First-principles calculations of electric transport by lithium ions are limited by the complexity of the crystalline matrix of highly-conductive materials, which renders a clear picture of lattice interactions that regulate lithium conductivity out of reach \cite{jun2024nonexistence, gigli2024mechanism, smith2020low}.

Many experimental techniques like inelastic neutron scattering \cite{brockhouse1995slow}, infrared \cite{kamba2021soft}, Raman \cite{ferrari2013raman} or inelastic X-ray spectroscopy \cite{burkel2000phonon} have investigated for years how a lattice of ions reacts to probes with certain frequencies and momenta.
Recent measurements of phenomena arising from the angular momentum carried by circularly polarized lattice vibrations in solids have inflamed the interest in phonon chirality \cite{nova2017effective, basini2024terahertz,juraschek2025chiral,zeng2025photo}. The breaking of time-reversal symmetry of a lattice of ions spawns a plethora of exotic phonon physics like the phonon Hall effect \cite{strohm2005phenomenological,sun2021phonon,chen2022large,li2020phonon} and the phonon Einstein-de-Haas \cite{zhang2014angular} among others.
First-principles theoretical models of phonon chirality are in their infancy \cite{juraschek2019orbital, bistoni2021intrinsic, zhang2026comprehensive,zhang2022lattice, ren2024adiabatic, stephens1985theory, scherrer2016nuclear, chen2025gauge}, and still lack first-principles inclusion of phonon anharmonicity.

The shortcomings of the state of the art of phonon physics call for refinements of theoretical approaches used to discuss the response of interacting lattices of ions.

On the other hand, models for the response of interacting electrons are far more refined. First-principles calculations based on density functional theory (DFT) have done wonders in the field of in silico materials research \cite{hafner2006toward,marzari2021electronic, saal2013materials, pizzi2016aiida}. The DFT extensions developed to study the response of electronic systems are the density functional perturbation theory (DFPT) \cite{baroni2001phonons} for static properties, the time-dependent density functional theory (TD-DFT) \cite{runge1984density, botti2007time,rocca2008turbo,casida1995time} for dynamical properties, the time-dependent Hartree-Fock \cite{larsen2000hartree,ikemachi2018time} or the hybrid functional approaches \cite{heyd2003hybrid, paier2006screened,skone2014self} to include the effects of the exchange interactions in the response. We will refer to all these theoretical approaches as generalized Kohn-Sham (gKS) approximations.
In gKS approaches, the response of an interacting system of electrons is obtained in terms of the fluctuations of the electronic density $n\pert$ induced by an external field. The electronic density is obtained from the KS states $\{\ket{\psi_{i}^\text{KS}}\}$, which satisfy a self-consistent time-dependent Schrodinger equation 
\begin{equation} \label{ks td}
    i\hbar \pdv{}{t} \ket{\psi_i^{\text{KS}}(t)} =   H^\text{KS}_\scf(t)  \ket{\psi_i^{\text{KS}}(t)}
\end{equation}
where $ H^\text{KS}_\scf(t)$ is the self-consistent field KS Hamiltonian that approximates electronic interactions in a mean-field fashion. 
gKS approaches  solve Eq.\ \eqref{ks td} to obtain the response in terms of external perturbations averaged on the evolved electronic density 
\begin{equation} \label{n(t) gks}
      n(t) = \sum_i f(\epsilon_i) \dyad{\psi_i^{\text{KS}}(t)}{\psi_i^{\text{KS}}(t)},
\end{equation}
which encodes fermionic statistics and finite temperature effects in the Fermi-Dirac distribution factors $f(\epsilon_i)$.
Ab initio numerical implementations of Eq.\ \eqref{ks td} in gKS approaches --- facilitated by massively exploiting symmetries operations \cite{giannozzi2009quantum}---, have successfully predicted and reproduced optical responses \cite{sharma2011bootstrap,benedict1998optical}, excitonic effects \cite{sottile2003parameter}, charge-density wave instabilities \cite{calandra2011charge, leroux2012anharmonic}, transport properties \cite{madsen2006boltztrap, ponce,macheda2018magneto}, and many more physical observables associated with electronic responses.
Expressing the response in terms of matrix elements of the external perturbation on the KS evolved basis allows for: 
(a) a detailed description of the states involved in the transitions, eventually prohibited by selection rules, and (b) a deep understanding of how electronic interactions screen the external fields. 

For phonons, many self-consistent approaches exist to account for strong phonon interaction in anharmonic materials \cite{monacelli2021stochastic,tadano2018first,Hellman2011,souvatzis2009self,monserrat2013anharmonic,dauxois1993dynamics}. However, in the existing methods, the specifics of the phonon interactions are often hidden in the mathematical or in the numerical details. This complicates the comparison of different theories of phonon anharmonicity \cite{mcgaughey2025phonon}, leaving uncertainties in whether anharmonicity is overcounted in self-consistent routines \cite{monacelli2025analyzing} or how to properly account for anharmonicity to compute responses as thermal conductivity \cite{li2023first}.

The goal of this work is to provide an effective single-particle picture for the dynamics of interacting ions in one-to-one correspondence with the established mean-field theories of interacting electrons.

As the gKS self-consistent Hamiltonian approximates the all-electron Hamiltonian in electronic dynamics, a self-consistent approximation of the Born-Oppenheimer (BO) Hamiltonian is needed for a mean-field approach of lattice dynamics. We adopt as our fundamental framework the time-dependent self-consistent harmonic approximation (TD-SCHA) \cite{monacelli2021stochastic,lihm2021gaussian,siciliano2023wigner}, as it provides a clear functional formulation of the self-consistent relations between the approximated state and Hamiltonian. Nonetheless, our formalism is valid for any mean-field approximation of BO dynamics.

In particular, we derive the phonon equivalent of the time-dependent gKS equations where the role of the KS states is played by the phonon spinors {$\ket{E_{\mu\sigma}(t)}$ and the phonon condensate $\ket{G(t)}$. Respectively, the dynamics of $\ket{E_{\mu\sigma}(t)}$ and $\ket{G(t)}$ can be associated with the dynamics of the interatomic force constants and of the average positions of the ions.}

Phonon spinors $\ket{E_{\mu\sigma}(t)}$ are obtained concatenating the polarization vectors of phonons with positive and negative phases, and exist in a vectorial space with twice the dimension of standard phonon eigenvectors. Doubling the dimension of the space allows the separation of the dynamics of phonons with positive and negative frequencies, in a procedure extremely similar to how electrons and holes are separated in Bogolubov-de Gennes approaches to superconductivity \cite{bogoliubov1958new,degennes1966superconductivity}. Phonon spinors with positive and negative frequencies are classified by a phonon pseudospin $\sigma {=} \pm$. The phonon pseudospin is a quantum number conserved by standard equilibrium dynamics, while anharmonicity and external fields provide couplings between $\sigma {=} \pm$ states.
In the absence of time-reversal symmetry breaking, the phonon spectrum is doubly degenerate in the phonon pseudospin $\sigma$. Nonetheless, we discuss in presenting the expression of optical and thermal conductivity within the harmonic approximation, how response properties such as thermal conductivity are pseudospin sensitive. 

The phonon condensate $\ket{G(t)}$ instead has no fermionic analogue, since it follows from  the nonconservation of the phonon number. It is obtained from the mapping of the ionic position/momentum operators in the single-particle augmented space, and it describes how the average position and momenta of the ions change due to external perturbation or anharmonic effects. The dynamics of $\ket{G(t)}$ is an additional degree of freedom of lattice dynamics; anharmonic couplings between $\ket{E_{\mu\sigma}(t)}$ and $\ket{G(t)}$ describe scattering mechanisms physically possible for lattice dynamics but completely absent in the standard electronic case.

In the augmented space, the evolution of $\ket{E_{\mu\sigma}(t)},\ket{G(t)}$ is regulated by single-particle Hamiltonian $\H(t)$ and forces $\ket{F(t)}$ that play the role of the time-dependent KS Hamiltonian for phonons.
The mathematical structure of the mean-field equations allows for a Dirac algebra where the vectors are represented by bras and kets and operators are represented by matrices, which immensely facilitates the analytical derivations and further tightens the comparison with gKS theories.
Formulating the response, we obtain the ionic susceptibility $\chi(t)$ as a functional of the single-particle phonon density matrix $\varrho(t)$, which generalizes the single-particle density matrix used in the recently introduced Wigner formulation of thermal transport \cite{simoncelli2019unified,simoncelli2022wigner}. The phonon single-particle density matrix $\varrho(t)$ has the same structure as in gKS [Eq.\ \eqref{n(t) gks}], featuring time-dependent projectors on the phonon spinors $\ket{E_{\mu\sigma}(t)}$ weighted by Bose-Einstein distribution factors.

In the linear regime, we formulate the response in terms of self-consistent potentials and forces, where anharmonicity acts as a phonon screening of the external fields, absent in the harmonic approximation. 
Phonon interactions are thus regulated by anharmonic kernels, just as Hatree-exchange correlation kernels encode electronic interactions in gKS approaches.

We formulate a closed self-consistent cycle of coupled equations for the calculation of the anharmonic response, in the spirit of Sternheimer formulations of gKS equations \cite{baroni2001phonons}. The necessary ingredients for the solution of the cycle are the description of the boundary conditions (phonon equilibrium) and an approximation of the anharmonic kernels. Different approximations of the many-body state of the lattice will result in different approximation for the anharmonic kernels, just as in the electronic case for the approximations of the Hartree-exchange-correlation functional. 
In the TD-SCHA, the anharmonic kernels feature three-phonon and four-phonon scattering vertices coupling phonon spinors and the phonon condensate. 

Besides refining the theory of anharmonicity, we believe that this work may serve as a Rosetta stone to bridge between electronic and ionic response theoretical approaches. Experts in gKS theories will find that the equations they are solving to obtain the electronic response have a precise phonon analogue, while the research in phonon anharmonicity can now benefit by the last decades of theoretical and numerical advancements done in the modeling of the electronic response.

This work is structured as follows. In Sec.\ \ref{sec: II}, we establish the 
single-particle mapping of the mean-field lattice dynamics. We first introduce the idea of the
single-particle mapping of lattice dynamics in the simplest nontrivial case: the harmonic approximation with external perturbing field. We introduce the single-particle mapping of a quadratic Hamiltonian including external field and forces. 
We introduce the generalized single-particle density matrix and the phonon condensate, and subsequently derive the effective single-particle Liouville equation that governs their pseudounitary time evolution. Then, the phonon spinors are introduced and the time-dependent Schrodinger equation for lattice dynamics is derived.
In Sec.\ \ref{sec: response}, we develop the response theory for the harmonic approximation. Here, we show how the ionic susceptibility can be expressed as a functional of the single-particle phonon density and phonon condensate. In the linear response regime, we derive the phonon response in terms of matrix elements of the bare external field and forces acting on the lattice. In Sec.\ \ref{sec: conductivities}, we demonstrate the capabilities of our method by deriving the equation for the lattice optical and thermal conductivity in the simple case of harmonic crystals. Here we show that even if the phonon spectrum is pseudospin degenerate, the response functions are in general pseudospin sensitive, as for thermal conductivity. 
In Sec.\ \ref{sec: espald}, we apply the single-particle mapping outlined previously to self-consistent approaches to anharmonicity,
developing an effective single-particle picture for anharmonic lattice dynamics. Adopting the TD-SCHA, we derive wave equations for the phonon spinors and condensate in one-to-one correspondence with gKS equation for electrons. Anharmonicity is included in the self-consistency of the single-particle Hamiltonian and forces. 
In Sec.\ \ref{sec: SC response}, we extend the response formalism to the self-consistent case.
In linear response, we derive an expression for the ionic susceptibility in terms of self-consistent potential and forces, that correspond to the mean-field potential and forces that the phonons feel, screened by anharmonicity. 
We derive a formal expression for the anharmonic interaction kernels, that in the TD-SCHA amounts to three and four phonon scattering vertices. We formulate the self-consistent cycle for the calculation of the linear response in the TD-SCHA, and show how our formulation recovers existing results for the TD-SCHA. Finally, in Sec.\ \ref{sec: conclusions}, we draw our conclusions and outline future directions.

\section{Effective single particle picture for lattice dynamics} \label{sec: II}
The paradigm of density functional theory is the simplification of the many-body electronic problem, described by an all-electron Hamiltonian {$  H_\text{el}(t) = \sum _i -\frac{\hbar^2}{2m}\nabla_i^2 + \sum_i   V_i^{\text{el-ion}} + \frac{1}{2} \sum_{ij}   V_{ij}^{\text{el-el}} + \sum_i   H_{\text{pert},i}(t) $}, where $i$ runs on all the electrons, to a single-particle problem.
In gKS approaches, the dynamics of interacting electrons is mapped into the dynamics of the density matrix $  n(t)$, regulated by a single-particle Hamiltonian $  H_\text{KS}^{[\psi(t)]}(t)$, where electronic interactions are included via self-consistency. 
In practice, the core of the single-particle mapping of gKS is that many-body operators {$\sum_{ij}   V_{ij}$} in the all-electron Hamiltonian are replaced by single-particle operators $  V(  r)$ acting on the density.
Most of the quantities we discuss in electronic structure theory, like Bloch wavefunctions or electronic band energies, are single-particle concepts. The description of the solid via its single-particle features is possible thanks to the single-particle mapping of gKS approaches. The fundamental question of this work is: can we perform a similar procedure to realize a single-particle mapping for lattice dynamics? In this paper, we provide a positive answer in the form of the effective single particle picture for anharmonic lattice dynamics (ESPALD). We will start from the harmonic case.

First, let us formulate the many-body problem for lattice dynamics, i.e.\ the equivalent of the ``all-electron'' Hamiltonian for the ionic case.
The dynamics of dielectric crystals is regulated by the time-dependent Born-Oppenheimer (BO) Hamiltonian
\begin{equation} \label{H BO}
      H_\text{BO}(t) =\sum_{I\a} \frac{  P_{I\a}^2}{2M_I}+   V_\text{BO}({\bf R}) +  {H}_\text{pert}(t),
\end{equation}
where ${  { P}_{I\a}}$ is the quantum mechanical operator that describes the momentum of atom $I {=} 1,\dots,N_{\text{at}}$ of mass $M_I$ along the cartesian direction $\alpha{=}x,y,z$,  and $  V_{\text{BO}}$ is the BO potential which depends on all the atomic coordinates ${\bf R} {=} (R_{1x},\dots,R_{N_{\text{at}}z} )$. We indicate with bold lettering $3N_{\text{at}}$ vectors. 
$  H_\text{pert}(t)$ can contain any kind of perturbation acting on the ions. In Sec.\ \ref{sec: conductivities}, we discuss e.g.\ the effect of a uniform electric field $\mathcal{E}(t)$ on the lattice is captured by the Hamiltonian
$  H_\text{pert}(t) =- \sum_\a{\mathcal{E}}_\a(t) \cdot  {d}_{\mathrm{el}\gamma}$
where $ {d}_{\mathrm{el}\gamma}$ is the electric dipole of the ions. 
{ A notable distinction between the time-dependent perturbations is that while in the electronic case the external term is typically separable in their single-particle $\sum_i H_{\text{pert},i}(t)$, in the lattice case $ {H}_\text{pert}(t)$ is in general many-body, depending on products of ionic variables such as $\sum_{ij}\mathbf{R}_i\mathbf{R}_j$ .
}

Before a fixed time $t{=}0$, the system is at equilibrium at temperature $T$, described by a canonical statistical operator $  \rho_\text{BO}\unpert$. For $t{\geq}0$ its dynamics is regulated by a Liouville equation 
\begin{subequations} \label{BO liouville}
    \begin{align}
   & i\hbar\pdv{  \rho(t)}{t}  = [  H_{\text{BO}}(t), \rho(t)] \qquad &t{\geq}0\\
   &   \rho\unpert= {e^{-\frac{1}{k_BT}  H_{\text{BO}}\unpert}}/{\Tr[e^{-\frac{1}{k_BT}  H_{\text{BO}}\unpert}]} \qquad &t{<}0,
    \end{align}
\end{subequations}
where $  H_{\text{BO}}\unpert$ stands for $  H\unpert_\text{BO} =   H_\text{BO}(t{<}0)$, and $k_B$ is the Boltzmann constant. 

The complexity of the many-body BO potential $  V_{\text{BO}}$ makes the dynamical system of Eq.\ \eqref{BO liouville} practically unsolvable. Thus, any theoretical approach for lattice dynamics necessitates an approximation for the BO potential. 
The simplest possible approximation for the BO potential is the harmonic approximation (HA). Modern strategies to solve lattice dynamics use mean-field approximations of $  H_{\text{BO}}(t)$, where atomic interactions are described beyond the simple HA by means of self-consistency. To present our formalism, we start from the simplest nontrivial case, which is the HA including external fields. We discuss self-consistent mean-field approaches in Sec.\ \ref{sec: espald}.

\subsection{Harmonic crystal subject to external fields}
{
In the HA, we suppose that a set of equilibrium positions $\tens{R}\unpert$, defined by 
\begin{equation}
    \label{equilbrium HA}
    \pdv{ V_\text{BO}({\bf R})}{ R_{I\a}}\eval_{{\bf R} = \tens{R}\unpert } = 0 ,
\end{equation}
provide a minimum for the BO potential, which we expand at second order as 
\begin{equation}
      V_{\text{BO}}({\bf R}) =   V_{\text{BO}}(\tens{R}\unpert) + \frac{1}{2} \sum_{I\a,J\b} (  R_{I\a} - \tenscomp{R}\unpert_{I\a})\phi\unpert_{I\a,J\b}(  R_{J\b} - \tenscomp{R}\unpert_{J\b})
\end{equation}
where the harmonic interatomic-force constant matrix is defined as 
\begin{equation}
    \label{phi2 harmonic}
    \phi\unpert_{I\a,J\b} = \pdv{V_\text{BO}({\bf R})}{R_{I\a}}{R_{J\b}}\eval_{{\bf R} = \tens{R}\unpert }.
\end{equation}
Typically, the equilibrium condition Eq.\ \eqref{equilbrium HA} that defines $\tens{R}\unpert$ is obtained by minimizing the forces acting on the nuclei. Such forces are obtained e.g.\ from DFT calculation, where the nuclei are considered classical objects,  without thermal or quantum fluctuations.
The HA has many limitations, one for all the fact that the equilibrium condition Eq.\ \eqref{equilbrium HA} does not imply stability, namely that $\phi\unpert$ in Eq.\ \eqref{phi2 harmonic} is positive definite. For the time being, we will work in the HA, supposing that $\phi\unpert$ is non-negative, and the resulting normal mode frequencies are real. }
With these assumption, in the HA we approximate the equilibrium BO Hamiltonian as (discarding constant terms)
\begin{align}
   &  H_\text{BO}\unpert \simeq   H_\HA\unpert= \nonumber \\
  &  \sum_{I\a} \frac{  P_{I\a}^2}{2M_I} + 
 \frac{1}{2} \sum_{I\a,J\b} (  R_{I\a} - \tenscomp{R}\unpert_{I\a})\phi\unpert_{I\a,J\b}(  R_{J\b} - \tenscomp{R}\unpert_{J\b})      . 
\end{align}
Accordingly, the state of the system is approximated as
\begin{equation}
  \rho\unpert= {e^{-\frac{1}{k_BT}  H_{\HA}\unpert}}/{\Tr[e^{-\frac{1}{k_BT}  H_{\HA}\unpert}]}.
\end{equation}
Let us introduce a compact notation we will use for the remainder of the paper. We indicate with a $\leftrightarrow$ overset $3N_\text{at}{\times} 3N_\text{at}$ matrices, while and bold lettering for $3N_\text{at}$ vectors. Linear algebra operations are defined as 
\begin{subequations} 
    \begin{align}
       ( \mtrx{\bf A}{\bf v} )_{I\a} &= \sum_{J\b}A_{I\a,J\beta}v_{J\b}\\
       ( \mtrx{\bf A} \mtrx{\bf B} )_{I\a,J\beta}&=\sum_{L\gamma}A_{I\a,L\gamma}B_{L\gamma,J\beta}\\
        {\bf w}^T\cdot {\bf v} &=\sum_{I\a}w_{I\a}v_{I\a}\\
       ( {\bf w} {\bf v}^T )_{I\a,J\beta} &=  {w}_{I\a} {v}_{J\b} 
    \end{align}
\end{subequations}
In this notation, the equilibrium Hamiltonian reads
\begin{equation} \label{H0 HA}
      H\unpert_\HA = \tfrac{1}{2} {\bf P}^T \cdot \mtrx{\bf M}{}^{-1}  {\bf P} + \tfrac{1}{2}[ {\bf R}- \tens{R}\unpert]^T \cdot \mtrx{\bm{\phi}}{} \unpert[ {\bf R}- \tens{R}\unpert]
\end{equation}
where $[\mtrx{\bf M}{}^{-1}]_{I\a,J\beta} = M^{-1}_{I}\delta_{IJ}\delta_{\a\b}$.

Now we describe the dynamics in the HA. As mentioned, the external Hamiltonian $  H_\text{pert}(t)$ drives the dynamics after $t{=}0$. { In the HA}, we consider external perturbations of the form
\begin{align}
    \label{Vext harmonic}
      H_\text{pert}(t) &=  \tfrac{1}{2}{\bf P}^T \cdot \mtrx{\mathbf{M}}{}^{-1}_\text{pert}(t) {\bf P} \\
     & + \tfrac{1}{2}(  {\bf R} - \tens{R}\unpert)^T \cdot \mtrx{\bm{\phi}}_\text{pert}(t) (  {\bf R} - \tens{R}\unpert) \nonumber  \\
    & - {\bf f}_{\mathbf{R}\,\text{pert}}(t)^T\cdot (  {\bf R}   - \tens{R}\unpert)  .
\end{align}
Terms as ${\mathbf{M}}{}^{-1}_\text{pert}$ follow from perturbation coupling to the kinetic energy of the ions, as a temperature gradient (see Sec.\ \ref{sec: conductivities}) or isotope scatterings \cite{fugallo2013ab}. Other couplings in $\mathbf{P}$ originating from magnetic fields or molecular Berry phases can be easily included in this formalism, but are neglected in this work. 
Thus, in the HA, the BO time-dependent Hamiltonian is approximated as 
\begin{align}
    H_\text{BO}(t)& \simeq H_\HA(t) \nonumber \\
    &=  \tfrac{1}{2} {\bf P}^T \cdot \mtrx{\bf M}{}^{-1}_\text{tot}(t)  {\bf P}  \nonumber \\
    & + \tfrac{1}{2}[ {\bf R}- \tens{R}\unpert]^T \cdot \mtrx{\bm{\phi}}{}_\text{tot}(t) [ {\bf R}- \tens{R}\unpert] \nonumber , \\
    & - {\bf f}_{\mathbf{R}\,\text{pert}}(t)^T\cdot (  {\bf R} - \tens{R}\unpert) \label{H(t) HA}
\end{align}
 where $\bm \phi_\text{tot}(t) = \bm \phi\unpert +\bm \phi_\text{pert}(t)$ and $\mathbf{M}^{-1}_\text{tot}(t) = \mathbf{M}^{-1} +\mathbf{M}^{-1}_\text{pert}(t)$. Finally, the BO dynamics [Eq.\ \eqref{BO liouville}] is approximated in the HA as
 \begin{subequations} \label{HA liouville}
    \begin{align}
   & i\hbar\pdv{  \rho(t)}{t}  = [  H_{\HA}(t), \rho(t)] \qquad &t{\geq}0\\
   &   \rho\unpert= {e^{-\frac{1}{k_BT}  H_{\HA}\unpert}}/{\Tr[e^{-\frac{1}{k_BT}  H_{\HA}\unpert}]} \qquad &t{<}0.
    \end{align}
\end{subequations}

\subsection{Mapping the equations of lattice dynamics} \label{sec: map of dynamics}
Even in the HA, Eqs. \eqref{HA liouville} are formally many-body, as $  H_\HA(t)$ [Eq.\ \eqref{H(t) HA}] features sum of operators on the ionic degrees of freedoms, and it solution in the presence of external potential $H_\text{pert}(t)$ [Eq.\ \eqref{Vext harmonic}] is nontrivial. 
Here, we show how Eq.\ \eqref{HA liouville} admits a single-particle mapping, in which the dynamics of the many-body density $ {\rho}(t)$ is replaced by the dynamics of the vibrational analogues of the one-electron density matrix, discussed in gKS approaches. The single-particle mapping is realized via a canonical transformation into generalized normal modes operators, outlined below.
The mapping introduced here for the HA will serve us as a stepping stone for the more complicated self-consistent approximations for the BO dynamics, discussed in Sec.\ \ref{sec: espald}.

In the HA, the dynamical matrix is 
\begin{equation}
        \mtrx{ \tens{D}}{}= \mtrx{\bf M}{}^{-\tfrac{1}{2}}\mtrx{\bm \phi}{}\unpert \mtrx{\bf M}{}^{-\tfrac{1}{2}},
\end{equation}
which can be diagonalized and admits $3N_\text{at}$ normal modes obtained via the secular equation 
\begin{equation} \label{De = w e}
    \mtrx{\tens{D}} {\tens{e}}_{\mu}= \Omega^2_{\mu} {\tens{e}}_{\mu}
\end{equation}
where $\tens{e}_\mu$ are the harmonic phonon polarization vectors.
The phonon polarization vectors satisfy an orthogonality $\tens{e}_\mu^T\cdot\tens{e}_\nu = \delta_{\mu\nu}$ and a completeness condition $\mathbb{I} = \sum_\mu \tens{e}_\mu
\tens{e}_\mu^T$, where $\mathbb{I}$ is the $3N_{\text{at}}{\times}3N_{\text{at}}$ identity.
With the dynamical matrix in Eq.\ \eqref{De = w e}, we can define the canonical transformation in the cartesian boson operators, introduced in Ref.\ \cite{simoncelli2019unified}
\begin{subequations}\label{RP to aa*}
\begin{align}
       { \bf R}- \tens{R}\unpert =&\sqrt{\hbar}
     \mtrx{\bm{\lambda}}{}_{\mathbf{R}}\left[ {\tens{a}} +  {\tens{a}}^\dagger\right] , \\
   { \bf P}= &i\sqrt{\hbar}\mtrx{\bm{\lambda}}{}_{\mathbf{P}}\left[ {\tens{a}} -  {\tens{a}}^\dagger\right],
\end{align}
\end{subequations}
where the change of basis matrices are 
\begin{subequations} \label{lambdas}
    \begin{eqnarray}
        \lambda_{\mathbf{R}\,I\a,J\beta} {=}\frac{1}{\sqrt{2}}\sqrt{M}^{-1}_{I\a} \tenscomp{D}^{-1/4}_{I\a,J\beta},\\
\lambda_{\mathbf{P}\,I\a,J\beta} {=}\frac{1}{\sqrt{2}} \sqrt{M}_{I\a} \tenscomp{D}^{1/4}_{I\a,J\beta}\:.
    \end{eqnarray}
\end{subequations}
The cartesian phonon operator $ {\tenscomp{a}}^\dagger_{I\a}$ ($ {\tenscomp{a}}_{I\a}$) represents the creation (destruction) of a vibrational excitation centered around atom $I$ along the direction $\a$ \cite{simoncelli2019unified,simoncelli2022wigner}. 
{They are related to $ {\mathbf{R}},  {\mathbf{P}}$ through the relation (notice that $\mtrx{\bm{\lambda}}{}_{\mathbf{R}}^T\mtrx{\bm{\lambda}}{}_{\mathbf{P}} = {1}/{2}$ )
\begin{subequations}\label{aa* to RP}
\begin{align}
 {\tens{a}}  =&  \frac{1}{\sqrt{\hbar}}\left[ \mtrx{\bm{\lambda}}{}_{\mathbf{P}}( { \bf R}- \tens{R}\unpert )  -i \mtrx{\bm{\lambda}}{}_{\mathbf{R}} { \bf P}  \right], \\
  {\tens{a}}{}\herm  =&  \frac{1}{\sqrt{\hbar}}\left[ \mtrx{\bm{\lambda}}{}_{\mathbf{P}}( { \bf R}- \tens{R}\unpert )  +i \mtrx{\bm{\lambda}}{}_{\mathbf{R}} { \bf P}  \right] .
\end{align}
\end{subequations}
}From the commutation relation $[ {R}_{I\a}, {P}_{I\a}] =i\hbar \delta_{IJ}
\delta_{\a\b}$ it follows that $[ {\tenscomp{a}}_{I\a},  {\tenscomp{a}}^\dagger_{J\b}] = \delta_{IJ}\delta_{\a\b}$. 
The standard phonon operators are related to the cartesian boson operator as \cite{simoncelli2019unified,simoncelli2022wigner}
\begin{equation} \label{cartesian to phonons}
      a_\mu = \tens{e}^T_\mu \cdot  {\tens{a}}
\end{equation}
The advantage of the cartesian boson operators basis is that, while capturing the bosonic nature of atomic oscillations, their representation does not depend on the vectors $\{\tens{e}_\mu\}_{\mu = 1}^{3N_\text{at}}$. In the basis of the phonon polarization vectors, the roots of the dynamical matrix  in Eqs.\ \eqref{lambdas} are $ \tenscomp{D}^{\pm1/4}_{I\a,J\beta}{=}\sum_\mu \Omega^{\pm1/2}_\mu {\tenscomp{e}_{\mu,I\a}}{\tenscomp{e}_{\mu,J\beta}}$ \cite{simoncelli2022wigner}. Both the definition of the canonical transformation and the cartesian boson operators depends on the approximation used. If the approximation of the Hamiltonian results in a different dynamical matrix, both the cartesian boson operators, and the normal mode basis must be  modified accordingly. In Sec.\ \ref{sec: espald}, we use the SCHA for the dynamical matrix.

The canonical transformation in Eqs.\ \eqref{RP to aa*} defines the mapping between the many-body quadratic operators and their single-particle equivalent. Using Eq.\ \eqref{RP to aa*} in the definition of the Hamiltonian [Eq.\ \eqref{H(t) HA}], we recast $  H_\HA(t)$ as 
\begin{align}  \label{H 2x2}
      H_\HA(t) &= \frac{\hbar}{2}\mqty[ {\tens{a}}\\  {\tens{a}}^\dagger]^\dagger{\cdot} \mqty[\mtrx{\bf h}(t) & \mtrx{\bf \Delta}(t)^* \\ \mtrx{\bf\Delta}(t) & \mtrx{\bf h}(t)^*]\mqty[ {\tens{a}}\\  {\tens{a}}^\dagger] \nonumber \\
    &- \hbar\mqty[{\bf f}(t) \\ {\bf f}(t)^*]^\dagger {\cdot} \mqty[ {\tens{a}}\\  {\tens{a}}^\dagger] 
\end{align}
which defines the $6N_\text{at}{\times}6N_\text{at}$ single-particle Hamiltonian  and the $6N_\text{at}$ vector of single-particle forces  
\begin{subequations} \label{H , F single particle}
\begin{align} 
    \H(t) &= \mqty[\mtrx{\bf h}(t) & \mtrx{\bf \Delta}(t)^* \\ \mtrx{\bf\Delta}(t) & \mtrx{\bf h}(t)^*],\\ \kket{F(t)} &= \mqty[{\bf f}(t) \\ {\bf f}(t)^*]  ,
\end{align}
\end{subequations}
where we have introduced a Dirac  notation in which $6N_\text{at}$ vectors are represented as bra/ket while $6N_\text{at}{\times}6N_\text{at}$ matrices are operators, that will be denoted with calligraphic fonts characters ($\H,\V,\mathcal{O}$). As in the standard Dirac notation, the scalar product is defined as 
\begin{equation}
    \braket{E}{F} = \sum_{\lambda=1}^{6N_\text{at}} E_\lambda^*F_\lambda,
\end{equation}
while the outer product is defined as 
\begin{equation}
    \dyad{E}{F}_{\lambda,\lambda'} = E_\lambda F^*_{\lambda'}
\end{equation}
This notation will further elucidate the analogies between the gKS theory and the ESPALD.
The matrices ${\bf h}(t)$ and ${\bf \Delta}(t)$ and the single-particle forces ${\bf f}(t)$ are obtained from the coupling coefficients of $  H_\HA(t)$ as
\begin{subequations} \label{operators RP to aa*}
    \begin{align}
        \mtrx{{\bf h}}(t) &=[\mtrx{\bm{\lambda}}{}_{\mathbf{R}}]^T \mtrx{\bm \phi}{}_\text{tot}(t)\mtrx{\bm{\lambda}}{}_{\mathbf{R}}+ [\mtrx{\bm{\lambda}}{}_{\mathbf{P}}]^T \mtrx{\bf{ M}}{}^{-1}_\tot(t)\mtrx{\bm{\lambda}}{}_{\mathbf{P}}, \\
        \mtrx{\bf\Delta}(t) &=[\mtrx{\bm{\lambda}}{}_{\mathbf{R}}]^T \mtrx{\bm \phi}{}_\text{tot}(t)\mtrx{\bm{\lambda}}{}_{\mathbf{R}}- [\mtrx{\bm{\lambda}}{}_{\mathbf{P}}]^T \mtrx{\bf{ M}}{}^{-1}_{\tot}(t)\mtrx{\bm{\lambda}}{}_{\mathbf{P}}, \\
       {\bf f}(t) &= \frac{1}{\sqrt{\hbar}}[\mtrx{\bm{\lambda}}{}_{\mathbf{R}}]^T {\bf f}_{{\bf R}\,\text{}}(t).
    \end{align}
\end{subequations}
With the transformation in Eqs.\ \eqref{operators RP to aa*}, the single-particle equivalent of the harmonic Hamiltonian $  H\unpert_\HA$ [Eq.\ \eqref{H0 HA}] is diagonal, namely 
\begin{equation}\label{H0 1b}
    \H(t{<}0) = \H\unpert_\HA = \mqty[\mtrx{\tens{D} }{}^{1/2} & 0 \\ 0 & \mtrx{\tens{D}} {}^{1/2}   ].
\end{equation}

The notation $\bf h,\bf \Delta$ in Eqs.\ \eqref{operators RP to aa*} is reminiscent of the theory of superconductors. In fact, $\mathcal{H}$ is a time-dependent Bogolubov-de Gennes (BdG) Hamiltonian for lattice dynamics, and $\bf h$  and $\bf \Delta$ represent the normal and anomalous parts  of the Hamiltonian, i.e.\ respectively the phonon number conserving and nonconserving channel of lattice dynamics. However, while Cooper pairs and Bose-Einstein condensates are typically treated in second quantization, lattice dynamics can be fully described within first quantization. 
{ From Eqs.\ \eqref{operators RP to aa*}, it follows that ${\bf f}(t), {\bf\Delta}(t)$  are real for every $t$. However, for a more general Hamiltonian with perturbation terms in the ionic momenta ${\bf P}$, ${\bf f}(t)$ and $ {\bf\Delta}(t)$ acquire imaginary parts. Not to lose generality, we keep the complex conjugate sign in the components of $\H(t)$ and $\ket{F(t)}$.}

To complete the single-particle mapping, we introduce the generalized single-particle density matrix \cite{blaizot}
\begin{equation} 
      \varrho(t) = \mqty[ \mtrx{ \varrho}{}^\text{n}(t)& \mtrx{  \varrho}{}^\text{an}(t) \\
   \mtrx{  \varrho}{}^\text{an}(t)^*& \mathbb{I}+ [\mtrx{ \varrho}{}^\text{n}(t)]{}^T   ]
\end{equation}
where respectively, the $3N_\text{at}{\times}3N_\text{at}$ normal {(equal to the one used in Refs.\ \cite{simoncelli2019unified,simoncelli2022wigner}) and anomalous densities are defined as
\begin{subequations} \label{rhoC elements}
    \begin{align}
         \varrho_{Ii,J\beta}^\text{n}(t) &=  \langle  {\tenscomp{a}}^\dagger_{J\b} {\tenscomp{a}}_{I\a}\rangle_{  \rho(t)}- \langle  {\tenscomp{a}}^\dagger_{J\b}\rangle_{  \rho(t)}\langle {\tenscomp{a}}_{I\a}\rangle_{  \rho(t)} , \\
          \varrho_{Ii,J\beta}^\text{an}(t) &=   \langle {\tenscomp{a}}_{J\b} {\tenscomp{a}}_{I\a}\rangle_{  \rho(t)} -\langle {\tenscomp{a}}_{J\b}\rangle_{  \rho(t)}\langle {\tenscomp{a}}_{I\a}\rangle_{  \rho(t)} ,
    \end{align}
\end{subequations}
}and the generalized one-body propagator
\begin{equation} \label{G = gg}
    \ket{G(t)} = \mqty[\mathbf{g}(t) \\ 
    \mathbf{g}(t)^*]
\end{equation}
where 
\begin{equation} \label{g(t)}
    g_{I\a}(t) = \langle {\tenscomp{a}}_{I\a}\rangle_{  \rho(t)}.
\end{equation}
{In light of the comparison with the BdG formalism, we dub $\ket{G}$ \emph{phonon condensate}}.
In Eqs.\ \eqref{rhoC elements},\eqref{g(t)} we have introduced the short-hand notation for the trace operation
\begin{equation}
    \left\langle  O \right\rangle_{  \rho} = \Tr[  O   \rho].
\end{equation}
{ 
In Eqs.\ \eqref{rhoC elements}-\eqref{g(t)}, the  density matrix $  \rho(t)$ used to evaluate the averages depends on the approximation used. In the HA discussed in this Section, it will be $  \rho(t)$, solution of Eq.\ \eqref{HA liouville}.   Nonetheless, the definitions of $ \varrho(t)$ and $\ket{G(t)}$ are valid for any density matrix.}
{
By using the Liouville equation [Eq.\ \eqref{HA liouville}] in the definition of $ \varrho(t)$ and of $\ket{G(t)}$ and exploiting the commutation relations, the many-body dynamics of the system can be mapped in two coupled equations (details in Appendix \ref{app: calcoli 1b liouville})
\begin{subequations} \label{1b mapping}
    \begin{align}
        i\hbar \pdv{ \varrho(t)}{t}  &=\hbar[\sigma_z \H(t), \varrho(t)]_\dagger , \label{rhoC liouville}\\
        i\hbar \pdv{}{t} \ket{G(t)}&= \hbar \sigma_z \H(t)\ket{G(t)} - \hbar\sigma_z\kket{F(t)}. 
    \end{align}
\end{subequations}
where the $[,]_\dagger$ is the generalized commutator{
\begin{align}
    [\mathcal{A},\mathcal{B}]_\dagger &{=} \mathcal{A}\mathcal{B} - (\mathcal{A}\mathcal{B})^\dagger \nonumber \\
    &{=} \mathcal{A}\mathcal{B} - \mathcal{B}^\dagger\mathcal{A}^\dagger,
\end{align}
}and $\sigma_z$ is the {($6N_{\text{at}}{\times}6N_{\text{at}}$)} Pauli matrix 
\begin{equation}
    \sigma_z = \mqty[\mathbb{I} & 0 \\ 0 & -\mathbb{I}] . 
\end{equation}

Eqs.\ \eqref{1b mapping} are the single-particle equivalent of the Liouville equation \eqref{HA liouville}. We discuss its many interesting features below. 

The solution of Eqs.\ \eqref{1b mapping} are used to describe the evolution of the physical observables. In fact, the transformations we used to obtain $\H(t), \ket{F(t)}$  from $ {H}_\HA(t)$ [Eqs.\ \eqref{operators RP to aa*}] allow the single-particle mapping of any quadratic operator.
Thus, an operator $ {O}$ with linear and quadratic terms in $ {\bf R}, {\bf P}$, is expressed in terms of the cartesian boson operator as  
\begin{equation} \label{O = aa* 2x2}
       O = \frac{\hbar}{2}\mqty[ {\bf a}\\  {\bf a}^\dagger]^\dagger{\cdot} \mqty[\mtrx{\bf O}{}^{\text{n}} & \mtrx{\bf O}{}^{\mathrm{an}*} \\ \mtrx{\bf O}{}^{\mathrm{an}} & \mtrx{\bf O}{}^{\text{n}*}]\mqty[ {\bf a}\\  {\bf a}^\dagger] + \hbar\mqty[\bf o \\ \bf o^*]^\dagger {\cdot} \mqty[ {\bf a}\\  {\bf a}^\dagger],
\end{equation}
which defines a matrix and a vector in the augmented space 
\begin{equation} 
    \mathcal{O} = \mqty[\mtrx{\bf O}{}^{\text{n}} & \mtrx{ \bf O}{}^{\mathrm{an}*} \\ \mtrx{\bf O}{}^{\mathrm{an}} & \mtrx{\bf O}{}^{\text{n}*}],\quad \ket{o} = \mqty[{\bf o} \\ {\bf o}^*],
\end{equation}
where the terms ${\bf O}^{\text{n}}$, ${\bf O}^{\mathrm{an}}$, and ${\bf o}$ are obtained with the relations Eq.\ \eqref{RP to aa*} applied to the coefficients of $  O$. We give some specific examples of quadratic operators in Sec.\ \ref{sec: conductivities}. 
The expectation value $ O(t)$ of any quadratic operator $   O$ is expressed in terms of the single-particle density matrix $  \varrho(t)$ and the phonon condensate $\ket{G(t)}$ \cite{blaizot}
\begin{align}
     O(t) &= \Tr[  O  \rho(t)] \nonumber \\
     &= \frac{\hbar}{2} \Tr[\mathcal{O}  \varrho(t)] +  \frac{\hbar}{2} \mel{G(t)}{\mathcal{O}}{G(t)} + \hbar \braket{o}{G(t)}.  \label{O(t) = rho(t) + G(t)}
\end{align}

}
{
In the case of quadratic Hamiltonian and  observables only (i.e.\ $\mathbf{f}_\text{pert} {=}0$ in  Eq.\ \eqref{H(t) HA} and $\mathbf{o}{=}0$ in Eq.\ \eqref{O = aa* 2x2}) , we define the derivative of the single-particle mapping  operators. Using Eq.\ \eqref{1b mapping} and the cyclic property of the trace, we obtain 
\begin{equation}
    \pdv{O(t)}{t} = \frac{\hbar}{2} \Tr[\dot{\mathcal{O}}(t)  \varrho(t)] +  \frac{\hbar}{2} \mel{G(t)}{\dot{\mathcal{O}} (t)}{G(t)} 
\end{equation}
where 
\begin{equation} \label{dot O}
    \dot{\mathcal{O}}(t)  = \frac{1}{i}[\mathcal{O},\sigma_z\H(t)]_\dagger. 
\end{equation}
}

\subsubsection{Discussion of the effective single-particle Liouville equation}
In Eqs.\ \eqref{1b mapping}, we mapped the dynamics of the many-body density matrix $  \rho_\BO(t)$ regulated by the Hamiltonian $  H_\BO(t)$ which has a $3N_{\text{at}}$- fold continuous spectrum into the dynamics of a single-particle density matrix $ \varrho(t)$  and a phonon condensate $\ket{G(t)}$ regulated by $\H(t)$ and $\ket{F(t)}$, which has a discrete spectrum in the  $6N_{\text{at}}$ augmented space. 

Notice how the $\hbar$ in Eqs.\ \eqref{1b mapping} can be simplified; as thoroughly discussed in Ref.\ \cite{siciliano2023wigner}, the mean-field evolution of a lattice follows classical dynamics, and quantum mechanics is included in the initial conditions, as we show in Sec.\ \eqref{sec: equilibrium rho}.

As we show in Appendix \ref{app: eom RP}, a closed system of dynamical equations as Eqs.\ \eqref{1b mapping} can be derived directly in terms of the ionic variables $ {\bf R}, {\bf P}$. Such a system of equations is the one presented in Eqs.\ (23) of Ref.\ \cite{siciliano2023wigner} in terms of ionic position and momentum correlation functions. In Ref.\ \cite{siciliano2023wigner}, the equations for the autocorrelation functions of ionic displacement and momenta ($\langle\delta \bm R(t)\delta \bm R(t)\rangle_{ \rho(t)},\langle\delta \bm P(t)\delta \bm P(t)\rangle_{ \rho(t)},\langle\delta \bm R(t)\delta \bm P(t)\rangle_{ \rho(t)}$ in the notation of Ref.\ \cite{siciliano2023wigner}) replace the dynamical equation for  $ \varrho(t)$, while the equations for the expectation value of the ionic position and momenta ($\langle \bm R(t)\rangle_{ \rho(t)},\langle \bm P(t)\rangle_{ \rho(t)}$ ) replace the equation for $\ket{G(t)}$.
However, the standard representation for the ionic position and momenta hides the (pseudo)unitary nature of the time-evolution of the single-particle dynamical variables $\ket{G(t)},  \varrho(t)$, that is unveiled by the cartesian phonon operator basis [Eq.\ \eqref{RP to aa*}]. Moreover, as we show in Appendix \ref{app: calcoli 1b liouville}, Eqs.\ \eqref{1b mapping} corresponds to 6 equations in the original $3N_\text{at}$ space. Among those 6, 3 can be obtained from the others with simple transpose or complex conjugate operations. This leaves us with 3 independent equations for anharmonic lattice dynamics at variance with the 5 proposed in Ref.\ \cite{siciliano2023wigner}.

It is also worth mentioning that the first block of Eq.\ \eqref{rhoC liouville} i.e.\ the equation  for $ \varrho^{\text{n}}(t)$, corresponds to the equation of motion for the single-particle density matrix used in the Wigner formulation of thermal transport \cite{simoncelli2019unified,simoncelli2022wigner}, in the limit of vanishing collision integral. Collisional integral in terms of higher order reduced density matrix would appear when cubic terms in $ {\bf R}, {\bf P}$ are included in the Hamiltonian \eqref{H(t) HA}, as per the BBGKY hierarchy \cite{imre1967wigner}.

Interestingly, Eqs.\ \eqref{1b mapping} correspond exactly to the dynamics of the condensed ($\ket{G(t)}$) and the uncondensed ($ \varrho$) part of a system undergoing Bose-Einstein condensation, described within the time-dependent self-consistent Hartree-Fock-Bogolubov mean field theory \cite{proukakis2001self,proukakis2008finite,castin1998low}.
Although phonons do not condense, this analogy follows directly from the mean-field nature of our approach and the bosonic character of phonons. This parallelism gives further context to our formalism and may serve as a bridge for future developments. 

The dynamics described by Eqs.\ \eqref{1b mapping} is non-Hermitian (Krein-Hermitian \cite{flynn2020deconstructing}), and describes a pseudounitary evolution \cite{mostafazadeh2004pseudounitary} where the generalized scalar product $\mel{\Psi(t)}{\sigma_z}{\Psi(t)}$  is conserved \cite{lein2019krein}. It follows that the trace of $ \varrho$ can change in time. This is typical of the dynamics of bosons with nonconserved number; while in electronic dynamics the conservation of the trace of the single-particle density matrix is related to the charge conservation, the number of phonons is not conserved. Thus, the trace of the one-body phonon density matrix can change.
However, since the equilibrium Hamiltonian $\H\unpert$ [Eq.\ \eqref{H0 1b}] commutes with $\sigma_z$, they admit simultaneous eigenvectors. We show below how a generalized version of the phonon eigenvectors in the augmented space --- phonon spinors ---  provides a basis of vectors with conserved length in the linear response regime.

\subsection{Phonon spinors}
The equilibrium one-body Hamiltonian $\H\unpert$  [Eq.\ \eqref{H0 1b}] provides a basis of the augmented space to solve the dynamics. Using the secular equation for the dynamical matrix $\tens{D}$ [Eq.\ \eqref{De = w e}], it is straightforward to show that the Hamiltonian $\H\unpert$ admits $2{\times} 3N_\text{at}$ eigenvectors 
\begin{equation} \label{H E = Omega E}
    \H\unpert \ket{E_{\mu\sigma}\unpert} = \Omega_\mu \ket{E_{\mu\sigma}\unpert}, \qquad \sigma =\pm,
\end{equation}
where  
\begin{equation} \label{E+, E-}
    \ket{E_{\mu+}\unpert} = \mqty[{\tens{e}_\mu} \\ 0] \quad , \quad \ket{E_{\mu-}\unpert} = \mqty[ 0 \\  {\tens{e}_\mu}]  .
\end{equation}
The subsctipt $\sigma=\pm(1)$ indicates the eigenvalue of $\sigma_z$
\begin{equation}
    \sigma_z  \ket{E_{\mu\sigma}\unpert} = \sigma  \ket{E_{\mu\sigma}\unpert}. 
\end{equation}
In fact, for $t{<}0$ the Hamiltonian $\H\unpert$ and $\sigma_z$ commute, and $\left\{ \ket{E\unpert_{\mu\sigma}}\right\}_{\mu=1,\sigma = \pm}^{3N_\text{at}}$ is a simultaneous basis for $\H\unpert$ and $\sigma_z$. The eigenvalue $\sigma$ is a quantum number by which we can label the states, and behaves as a pseudospin. Therefore, we dub $ \ket{E\unpert_{\mu\sigma}}$  phonon spinors, as they are the lattice dynamical equivalent of Nambu spinors used in superconductivity \cite{nambu1960quasi}. As we  discuss in Sec.\ \ref{sec: linear response}, the physical meaning of the phonon pseudospin $\sigma$ is to distinguish the positive and negative poles of the standard phonon propagator, which is quadratic in the phonon frequency.
From Eq.\ \eqref{E+, E-} and from the orthogonality and completeness conditions of the phonon eigenvectors, it follows that the phonon spinors are a complete basis for the augmented space, namely
\begin{subequations}
    \begin{align}
         \braket{E_{\mu \sigma_1}\unpert}{E_{\nu\sigma_2}\unpert} &= \delta_{\mu\nu}\delta_{\sigma_1\sigma_2}\\
         \sum_{\mu,\sigma=\pm} \dyad{E_{\mu\sigma}\unpert}{E_{\mu\sigma}\unpert} &= \mqty[\mathbb{I}&0\\ 0 & \mathbb{I}] 
    \end{align}
\end{subequations}
where the sum over $\mu$ runs over the $3N_\text{at}$ modes.

A definition of phonon spinors similar to  Eq.\ \eqref{E+, E-} can be found elsewhere (cf.\  Ref.\ \cite{flynn2020deconstructing} and references therein), typically in terms of a vector of creation and annihilation phonon operators $  a_\mu/  a^\dagger_\mu$. To our knowledge, the characterization of the state of the lattice in terms of generalized phonon polarization vectors --- which follows from the transformation in the cartesian boson operators [Eq.\ \eqref{RP to aa*}]--- and their classification in terms of their eigenvalue with respect to $\sigma_z$ represents one of the new perspectives of this work.

{
The equilibrium harmonic Hamiltonian $H_\HA\unpert$ in Eq.\ \eqref{H0 HA} results in an auxiliary one-body Hamiltonian $\H\unpert$ that is degenerate in the Nambu pseudospin sector, while the dynamical generator $\sigma_z\H\unpert$ has eigenvalues $\pm\Omega_\mu$. Nontrivial geometrical terms such as molecular Berry curvature in the equilibrium Hamiltonian $H_\HA\unpert$ can lift this pseudospin degeneracy \cite{saparov2022lattice}. In this case, chiral phonons might be obtained as linear combinations of the phonon spinors $\ket{E\unpert_{\mu\pm}}$. This shows how the formalism presented here represents a valuable theoretical tool to investigate exotic phenomena associated with phonon angular momentum \cite{zhang2014angular,zhang2026comprehensive,ren2024adiabatic,bistoni2021intrinsic}.
}

\subsection{Equilibrium condition in the phonon spinor representation} \label{sec: equilibrium rho}
The boundary conditions for $ \varrho, \ket{G}$ needed to solve Eqs.\ \eqref{1b mapping} follow from the equilibrium condition imposed at $t<0$.

Considering that at $t{<}0$ the lattice is described by a canonical statistical operator [in the HA, $ \rho\unpert_\HA$ of Eq.\ \eqref{equilbrium HA}], phonon modes are populated according to the Bose-Einstein distribution, thus 
\begin{subequations} 
\begin{align}
     \Tr[  a^\dagger_\nu   a_\mu    \rho\unpert
    ] &= \delta_{\mu\nu} n(\Omega_\mu) \label{<a*a>0},\\
    \Tr[  a_\nu   a_\mu    \rho\unpert
    ] &= 0,\label{<aa>0}\\
    \Tr[  a_\mu    \rho\unpert
    ] &= 0, \label{<a>0}
\end{align}
\end{subequations}
where 
\begin{equation} \label{bose einstein}
    n(\Omega_\mu) = \frac{1}{e^{{\hbar\Omega_\mu}/{k_BT}}-1}. 
\end{equation}
Using the relation between phonon $  a_\mu$ and cartesian operators $ {\tens{a}}$ [Eq.\ \eqref{cartesian to phonons}] in Eqs.\ \eqref{<a*a>0}-\eqref{<aa>0}, we obtain the initial conditions for the normal and anomalous densities [Eqs.\ \eqref{rhoC elements}]
\begin{subequations}
    \begin{align}
       { \varrho}{}^\text{n}_{Ii,J\beta}(t{<}0) &= \sum_\mu \tenscomp{e}_{\mu,Ii} n(\Omega_\mu) \tenscomp{e}_{\mu,J\beta} ,\\ 
        { \varrho}{}^\text{an}_{Ii,J\beta}(t{<}0) &= 0,
    \end{align}
\end{subequations}
from which, using the definition of the phonon spinors \eqref{E+, E-}, we get the initial condition $  \varrho(t{<}0) {=}  \varrho\unpert $ for the single-particle density 
\begin{equation} \label{rhoc 0 = EE+ + EE-}
    \varrho\unpert= \sum_\mu n(\Omega_\mu) \dyad{E\unpert_{\mu +}}{E\unpert_{\mu+}} +\sum_\mu [1+n(\Omega_\mu) ]\dyad{E\unpert_{\mu-}}{E\unpert_{\mu-}},
\end{equation}
while from Eq.\ \eqref{<a>0}, we get the initial conditions $\ket{G(t{<}0)} {=} \ket{G\unpert} $ for the phonon condensate
\begin{equation} \label{G0 = 0}
  \ket{G\unpert} =  0.
\end{equation}
Exploiting the relation 
\begin{equation}
    -n(-\Omega_\mu) = 1+n(\Omega_\mu),
\end{equation}
the single-particle density $ \varrho\unpert$ is expressed in terms of a summation on the phonon pseudospin as 
\begin{equation} \label{rhoc 0 = dft}
     \varrho\unpert = \sum_{\mu,\sigma=\pm} \sigma n(\sigma \Omega_\mu) \dyad{E_{\mu\sigma}\unpert}{E_{\mu\sigma}\unpert}.
\end{equation}

Although just an algebraic manipulation of the standard phonon representation, Eq.\ \eqref{rhoc 0 = dft} provides an important physical insight. In fact, the initial condition for the one-body phonon density is now expressed exactly as the initial condition for the electronic density matrix used in gKS [Eq.\ \eqref{n(t) gks}].
The fundamental feature of Eq.\ \eqref{rhoc 0 = dft} is that exploiting the (pseudo)unitary single-particle Liouville equation [Eq.\ \eqref{1b mapping}], the single-particle density $ \varrho(t)$ for $t{>}0$ is expressed as a product of constant statistical weights $\sigma n(\sigma \Omega_\mu)$ times evolving states $\ket{E_{\mu\sigma}(t)}$. We elaborate on this below.

{
As anticipated, the Bose-Einstein distribution is the only element in the single particle picture of anharmonic lattice dynamics that explicitly features $\hbar$. Thus, the quantum-mechanical nature of anharmonic phonon dynamics resides in their equilibrium distribution, but their mean-field dynamics is classical \cite{siciliano2023wigner}.
}

Eq.\ \eqref{rhoc 0 = EE+ + EE-} is equivalent to Eq.\ (7.138) of Ref.\ \cite{blaizot} for a generalized density matrix in the Hartree-Bogolubov approximation of a system of interacting bosons.

\subsection{Time-dependent Schrodinger equations for  lattice dynamics}
Using the initial conditions given in Eq.\ \eqref{rhoc 0 = dft} in the single-particle Liouville equation [Eq.\ \eqref{rhoC liouville}], we can express the single-particle density matrix for $t{>}0$ as 
\begin{equation} \label{rhoc(t) = E(t)E(t)}
     \varrho(t) = \sum_{\mu,\sigma=\pm} \sigma n(\sigma \Omega_\mu) \dyad{E_{\mu\sigma}(t)}{E_{\mu\sigma}(t)},
\end{equation}
where the phonon spinors evolve via
\begin{subequations} \label{espald schrodinger td}
\begin{align}
     i\hbar \pdv{}{t} \ket{E_{\mu\sigma}(t)} &= \hbar \sigma_z \H(t) \ket{E_{\mu\sigma}(t)} , \\
     i\hbar \pdv{}{t} \ket{G(t)}&= \hbar \sigma_z \H(t)\ket{G(t)} - \hbar\sigma_z\kket{F(t)}.
\end{align}
\end{subequations}
Eqs.\ \eqref{espald schrodinger td} are the lattice equivalent of the time-dependent Schrodinger equation for a system of noninteracting electrons subject to an external field. 
At variance with the fermionic problem, lattice dynamics possesses a further degree of freedom: the phonon condensate $\ket{G(t)}$. Due to conservation of charge (conservation of the fermionic number), electrons can not propagate without creating a hole. The dynamics of electrons is thus described in terms of electron-hole fluctuations. Instead, since the number of phonons is not conserved, $\Tr[ {\tenscomp{a}}_{I\a}  \rho(t)]{\neq}0$ away from thermal equilibrium.

Eqs.\ \eqref{espald schrodinger td} represent our key result for the analogy between lattice  and electronic dynamics. Most importantly, formulating Eqs.\ \eqref{espald schrodinger td} beyond the HA in the self-consistent case (see Sec.\ \ref{sec: espald}), provides  a coincise one-to-one correspondence between the mean-field equations for interacting ions and interacting electrons.

{ As in Eq.\ \eqref{O(t) = rho(t) + G(t)}, the solution of the equation of motion of the phonon spinors \eqref{espald schrodinger td} can be used to express the evolution of the observables. In fact, using Eq.\ \eqref{rhoc(t) = E(t)E(t)} in \eqref{O(t) = rho(t) + G(t)} we get 
\begin{align} \label{O = <EE> <G>}
  O(t)  &=  \frac{\hbar}{2}\sum_{\mu,\sigma=\pm} \sigma n(\sigma \Omega_\mu) \mel{E_{\mu\sigma}(t)}{\mathcal{O}}{E_{\mu\sigma}(t)} \nonumber \\ & + \frac{\hbar}{2} \mel{G(t)}{\mathcal{O}}{G(t)} + \hbar \braket{o}{G(t)}.
\end{align}
Eq.\ \eqref{O = <EE> <G>} shows how in the ESPALD the evolution of the expectation value of any (quadratic) observable $  O$ is expressed in terms of $6N_\text{at}{+} 1$ vectors: the $6N_\text{at}$ spinors $\{\ket{\tenscomp E_{\mu\sigma}(t)}\}$ and the phonon condensate $\ket{G(t)}$.
}

If we further separate the atomic index $I$ into the unit cell vector ${ R}\unpert_L$ and the atomic species in the cell $b$,  the translational invariance of $  H(t)$ of Eq.\ \eqref{H 2x2} results in a block diagonal $\H(t)$, which conserves the phonon quasimomentum $\bf{q}$, akin to how the time-dependent Schrodinger equation conserves electronic quasi-momentum in gKS. We elaborate on this in Sec.\ \ref{sec: bloch thm}.
Thus, the index $\mu {=} 1,\dots,3N_\text{at}$ is further separated into the phonon quasimomentum ${\bf q}$ and an index for the phonon branch $m$, that is $\mu \to ({\bf q},m) $.

{
We note that kinetic equations for phonons virtually equivalent to Eqs.\ \eqref{espald schrodinger td} were derived in Ref.\ \cite{wang2022coupled} using second-quantization. While the focus of Ref.\ \cite{wang2022coupled} was the influence of lattice dynamics in the electron-phonon coupling, here we focus on lattice anharmonicity, i.e.\ phonon-phonon scattering. Moreover, the phonon spinors were not identified as a complete set of the augmented space where the phonon dynamics takes place.
}

\subsection{Bloch theorem for phonon spinors}
\label{sec: bloch thm}

In an ordered solid, we can further exploit the translational symmetry of the Hamiltonian to block-diagonalize the equations for the phonon spinors.
We decouple the supercell atomic index $I$ ($I = 1, \dots, N_\text{at}$) into a unit cell index $L$ ($L = 1, \dots, N_{\mathrm{c}}$) and a basis atom index $b$ ($b = 1, \dots, N_\text{base}$). Consequently, the equilibrium position of atom $I$ along the Cartesian direction $\alpha$ is decomposed as $\tenscomp{R}_{I\alpha}\unpert {\to} \tenscomp{R}_{Ib\a}\unpert = R_{L\alpha}\unpert + \tau_{b\alpha}\unpert$, where $\mathbf{R}_L\unpert$ is the direct lattice vector of the $L$-th cell and $\boldsymbol{\tau}_b\unpert$ is the position of the $b$-th atom within the basis.  Here we use the bold lettering for cartesian vectors.

{
We consider a position operator $\mathbfscr{R}\unpert$ representing the equilibrium position of the ions. It is defined in the augmented space via its action on the basis of phonon spinors as 
\begin{equation} \label{R spinors}
    \mel{E_{\mu\sigma}\unpert}{\mathbfscr{R}\unpert}{E_{\nu\sigma'}\unpert}  = \delta_{\sigma\sigma'} \sum_{Lb\a}( \mathbf R_{L}\unpert + \bm\tau_{b}\unpert) \tenscomp{e}_{\mu,Lb\a} \tenscomp{e}_{\nu,Lb\a},
\end{equation}
where $\tenscomp{e}_{\mu,Lb\a}$ are the components of the polarization vectors defined in Eq.\ \eqref{De = w e}.
Next, we define the translation operators $\mathcal{T}_L$ ($L{=}1,{\dots},N_\text{c}$) on  the augmented space as
\begin{equation}
   \mel{E_{\mu\sigma}\unpert}{\mathcal{T}_L}{E_{\nu\sigma'}\unpert}  = 
       \delta_{\sigma\sigma'} \sum_{L'b\a}\tenscomp{e}_{\mu,Lb\a} \tenscomp{e}_{\nu,[(L+L')\text{mod}N_c] b\a} 
\end{equation}
If the interatomic force constants $\phi\unpert$ of Eq.\ \eqref{phi2 harmonic} satisfy the translational invariance property 
\begin{equation} \label{phi IJ = phi I-J}
    \phi\unpert_{Ib\a,Jb\b} =  \phi\unpert_{I-Jb\a,0b\b},
\end{equation}
it follows that the single-particle Hamiltonian $\H\unpert$ is translational invariant 
\begin{equation} 
  \mathcal{T}_L^\dagger \H\unpert\mathcal{T}_L = \H\unpert.
\end{equation}
Thus, as  the usual result of the Bloch theorem (see e.g.\ Ref.\ \cite{ziman1979principles}),  we relabel the states by their quasimomentum i.e.\ $\mu \to ({\bf q,}m)$:
\begin{equation} 
  \ket{E_{{\bf q} m\sigma}\unpert} = \frac{1}{\sqrt{N_{\mathrm{c}}}}  {e}^{i{\bf q }\cdot { \mathbfscr{R}}\unpert}\ket{\tenscomp{e}_{{\bf q}m\sigma}\unpert}
\end{equation}
where $\ket{\tenscomp{e}\unpert_{{\bf q}m\sigma}}$ is the ``periodic part'' of the phonon spinor, namely such that $\mathcal T_L\ket{\tenscomp{e}\unpert_{\mathbf qm\sigma}} = \ket{\tenscomp{e}\unpert_{\mathbf  q m\sigma}}$ for every $L$.
The $N_c$ phonon quasimomenta $\mathbf{q}$ are contained in the first Brillouin zone of the crystal, while the quantum number $\mathrm{m} =1,\dots,3N_\text{base} $ now labels phonon branches. The equation that defines the phonon spinors [Eq.\ \eqref{H E = Omega E}] can thus be block-diagonalized as 
\begin{equation}
    \mathcal{\H}\unpert_{\mathbf{q}} \ket{\tenscomp{e}_{\mathbf{q}m\sigma}\unpert} = \Omega_{\mathbf q m} \ket{\tenscomp{e}_{\mathbf{q}m\sigma}\unpert }
\end{equation}
where $\H_{\bf q}\unpert$ is the periodic part of the Hamltonian 
\begin{equation}
     \H\unpert_{{\bf q}} ={e}^{-i{\mathbf q }\cdot \mathbfscr{R}\unpert} \H\unpert {e}^{i{\mathbf q \cdot \mathbfscr{R}\unpert}}.
\end{equation}
}

\section{Response theory for harmonic crystals} \label{sec: response}

To complete the outline of our formalism,  here we show how to compute observables and response functions using the ESPALD in the HA. The generalization to the to the self-consistent anharmonic case is presented in Sec. \ref{sec: SC response}.

In the HA, we separate the  time-dependent external field  in $  H_\HA(t)$  as 
\begin{equation} \label{H(t) = H + Vext}
      H_\HA(t) =   H\unpert_\HA +   H_\text{pert}(t).
\end{equation}
The single particle equivalent of $  H_\HA(t)$ [Eq.\ \eqref{H(t) = H + Vext}] are the one-body Hamiltonian $\H(t)$ and forces $\ket{F(t)}$ we derived in Sec.\ \ref{sec: II}, that we separate in the equilibrium and the external part
\begin{subequations} 
    \begin{align}
        \H(t) &= \H\unpert + \H_\text{pert}(t),\\
        \kket{F(t)}& = \ket{F\unpert} + \kket{F_\text{pert}(t)},
    \end{align}
\end{subequations}

where the equilibrium condition [Eq.\ \eqref{H0 1b}] implies that 
\begin{subequations}
    \begin{align}
        \H_\text{pert}(t{<}0) & =0,\\
        \ket{F(t{<}0)} & = 0.
    \end{align}
\end{subequations}

{
The response to the external perturbation can be evaluated considering how the expectation value $O(t)$ [Eq.\ \eqref{O(t) = rho(t) + G(t)}] changes with respect to its equilibrium value 
\begin{equation} \label{delta O(t) nonlinear}
    \delta O(t) = O(t) - O\unpert ,
\end{equation}
where  $O\unpert$ is 
\begin{equation}
    O\unpert = \frac{\hbar}{2}\Tr[\mathcal{O} \varrho\unpert] = \hbar \sum_\mu [n(\Omega_\mu) {+} \frac{1}{2}] {\bf e_\mu}\!\!^T\!\!\cdot \mtrx{\bf O}{\bf e_\mu}.
\end{equation}
Solving the dynamics of $ \varrho(t)$ and of$\ket{G(t)}$ [Eqs.\ \eqref{1b mapping}] --- or equivalently, of $\ket{E_{\mu\sigma}(t)}$ and of $\ket{G(t)}$ [Eqs.\ \eqref{espald schrodinger td}] --- yields the response function $ \delta O(t)$. Most importantly, it follows from Eq.\ \eqref{O(t) = rho(t) + G(t)} that the response function $\delta O(t)$ is completely determined when the dynamics of $ \varrho(t)$ and of $\ket{G(t)}$ are solved. In this sense, with Eq.\ \eqref{delta O(t) nonlinear}, we have formulated the ionic response as a functional of the single-particle density $ \varrho(t)$ and of the phonon condensate $\ket{G(t)}$.

The coefficients of the Taylor expansion of $\delta O(t)$ near $\H_\text{pert} {=} 0$ represent different orders of response functions. Below, we will focus on the linear response regime.

\subsection{Linear response} \label{sec: linear response}
In the linear response regime, we consider the external perturbation $  H_\text{pert}$ as the product of its operatorial part $  H_\text{pert}\pert$ and a time-dependent stimulus $s(t)$, namely
\begin{equation} \label{Vext = v1 s(t)}
      H_\text{pert}(t)=   H_\text{pert} \pert s(t),
\end{equation}
and $s(t{<}0) {=} 0$. The linear variation of the observable $O(t)$ can be expressed in terms of the linear response function $\chi_{O,H_\text{pert}\pert}$ as 
\begin{equation} 
    \delta O\pert(t) = \int\limits^{\infty}_{-\infty}\dd{t'} \chi_{O,H_\text{pert}\pert} (t') s(t-t').
\end{equation}
Considering the Fourier transform in complex frequency $z{=}\w{+}i\eta$ for retarded quantities ($\eta{\geq} 0$) functions {
\begin{equation}
    f(z) = \int\limits_{0^-}^\infty \dd{t} e^{izt} f(t)
\end{equation}}
and the fact that $\delta O(t{<}0){=}0$, the linear response function is expressed in frequency domain as
\begin{equation} \label{chi = O/s}
    \chi_{O,H_\text{pert}\pert} (z)  = \frac{\delta O\pert(z)}{s(z)}.
\end{equation}
For simplicity, we set $s(t){=}\delta(t)$ so that $s(z) {=}1$ without loss of generality, since linear response functions do not depend on the shape of the stimulus.

To obtain $\delta O\pert(z)$, we perform the expansion of the one-body dynamical quantities retaining only linear terms 
\begin{subequations} \label{expansion E and G}
\begin{align}
    \ket{E_{\mu\sigma}(t)} & = \ket{E_{\mu\sigma}\unpert(t)} + \ket{E_{\mu\sigma}\pert(t)}  \\
    \ket{G(t)}&= \ket{G\unpert} + \ket{G\pert(t)}
\end{align}
\end{subequations}
and accordingly, of the single-particle Hamiltonian and of the forces as
\begin{subequations} \label{expansion H and F}
    \begin{align}
        \H(t) &= \H\unpert + \H\pert(t),\\
        \kket{F(t)}& = \ket{F\unpert} + \kket{F\pert(t)} ,
    \end{align} 
\end{subequations}
where  $\ket{F\unpert}{=}\ket{G\unpert}{=}0$, while 
\begin{subequations} \label{H,F harm = s H,F ext}
    \begin{align}
        \H\pert(t) &= \delta(t) \H_\text{pert}\pert,\\
        \kket{F\pert(t)} &= \delta(t) \ket{F_\text{pert}\pert} .
    \end{align}
\end{subequations}

Using \eqref{expansion E and G}, it follows from Eq.\ \eqref{delta O(t) nonlinear} that $\delta O\pert(t)$ can be computed as 
\begin{align}
    \delta O\pert(t) &= \frac{\hbar}{2}\sum_{\mu,\sigma=\pm}\sigma n(\sigma\Omega_\mu) \biggl[\mel{E_{\mu\sigma}\pert(t)}{\mathcal{O}}{E_{\mu\sigma}\unpert(t)} \nonumber \\
    &+ \mel{E_{\mu\sigma}\unpert(t)}{\mathcal{O}}{E_{\mu\sigma}\pert(t)} \biggr] + \hbar \braket{o }{G\pert(t)} \label{delta O pert (t)}
\end{align}
where we have used the fact that the single-particle density matrix can be expressed as 
\begin{align} 
     \varrho\pert(t) {=}\sum_{\mu,\sigma=\pm}& \sigma n(\sigma \Omega_\mu) \biggl[\dyad{E_{\mu\sigma}\pert(t)}{E_{\mu\sigma}\unpert(t)} \nonumber  \\&+ \dyad{E_{\mu\sigma}\unpert(t)}{E_{\mu\sigma}\pert(t)} \biggr]
\end{align}

Using Eqs.\ \eqref{expansion E and G}-\eqref{expansion H and F} in the time-dependent Schrodinger equations for lattice dynamics [Eqs.\ \eqref{espald schrodinger td}] and discarding quadratic terms in $\V_\text{pert}$, we get a 0-th order equation 
\begin{equation}
     i \pdv{}{t} \ket{E_{\mu\sigma}\unpert(t)} =  \sigma_z \H\unpert\ket{E_{\mu\sigma}\unpert(t)}
\end{equation}
from which it follows that the phonon spinor basis [Eq.\ \eqref{E+, E-}] follow a plane-wave evolution
\begin{equation} \label{E0 = plane waves}
    \ket{E\unpert_{\mu\sigma}(t)} = e^{-i\sigma\Omega_\mu t}  \ket{E\unpert_{\mu\sigma}},
\end{equation}
and two 1st order equations, {
\begin{subequations} \label{espald schrodinger (1) t}
\begin{align}
     i \pdv{}{t} \ket{E_{\mu\sigma}\pert(t)} &=  \sigma_z \H\unpert\ket{E_{\mu\sigma}\pert(t)} +  \sigma_z \H\pert(t)\ket{E_{\mu\sigma}\unpert(t)}   , \label{E pert(t)}\\
     i \pdv{}{t} \ket{G\pert(t)}&=  \sigma_z \H \unpert\ket{G\pert(t)} - \sigma_z\kket{F\pert(t)}. 
\end{align}
\end{subequations}}
Since $\ket{E_{\mu\sigma}\pert(t)}$ is first order in the perturbation, its quadratic norm does not contribute to linear-response observables. The exactly conserved quantity of the full pseudounitary evolution is the Krein norm $\mel{E_{\mu\sigma}(t)}{\sigma_z}{E_{\mu\sigma}(t)}$.

As an important outcome of the reformulation in the language of electronic dynamicsAs an important outcome of the reformulation in the language of electronic dynamics, the lattice dynamical Eqs.\ \eqref{espald schrodinger (1) t} are solved in frequency space exactly as in textbook perturbation theory of quantum mechanics. 
Crucially, as discussed in Sec.\ \ref{sec: linear response SC}, this simplification still holds for the self-consistent approaches to anharmonicity [see Sec.\ \ref{sec: linear response SC} and Appendix \ref{app: linear resp SC}].

We get the induced phonon spinors and the induced  phonon condensate in frequency space as 
\begin{subequations} \label{espald schrodinger (1) z}
\begin{align}
    \ket{E_{\mu\sigma}\pert(z)}_I &=\sum_{\nu,\sigma'} \sigma'\ket{E\unpert_{\nu{\sigma'}}} \frac{\mel{E\unpert_{\nu\sigma'}}{\H\pert(z)}{E\unpert_{\mu\sigma}}}{ z - (\sigma' \Omega_\nu - \sigma \Omega_\mu) },\\
   \ket{G\pert(z)}&= -\sum_{\mu,\sigma }\sigma\ket{E_{\mu\sigma}\unpert}\frac{\braket{E\unpert_{\mu\sigma}}{F\pert 
   (z)}}{ z-\sigma\Omega_\mu}
\end{align}
\end{subequations}
where we have introduced the interaction representation for the time-evolution $\ket{E\pert_{\mu\sigma}(t)}_I {=} e^{i\sigma\Omega_\mu t}\ket{E\pert_{\mu\sigma}(t)}$. In frequency space, the induced phonon density reads 
\begin{align}
    \varrho\pert(z){=} \sum_{\mu\nu, \sigma\sigma'}&\!\! \sigma'\sigma \frac{n(\sigma'\Omega_\nu) {-} n(\sigma\Omega_\mu)}{ z -(\sigma\Omega_\mu - \sigma' \Omega_\nu)}\dyad{E_{\mu\sigma}\unpert}{E\unpert_{\nu\sigma'}} \nonumber \\ & \times \mel{E_{\mu\sigma}\unpert}{\H\pert(z)}{E\unpert_{\nu\sigma'}}\:.  \label{rho(z) pert}
\end{align}

Using Eqs.\ \eqref{espald schrodinger (1) z} in Eq.\ \eqref{delta O pert (t)} we obtain the linear variation of the observable in frequency space $\delta O\pert(z)$, which we can relate to the harmonic linear response function via Eq.\ \eqref{chi = O/s},  yielding
\begin{align} \label{chi 1(z)}
    \chi_{O,H\pert_\text{pert}}^\HA(z) =&\frac{\hbar}{2
    }\!\!\sum_{\mu\nu, \sigma\sigma' } \mel{E\unpert_{\nu\sigma'}}{\mathcal{O}}{E\unpert_{\mu\sigma}} \sigma'\sigma \frac{n(\sigma'\Omega_\nu) {-} n(\sigma\Omega_\mu)}{ z -(\sigma\Omega_\mu - \sigma' \Omega_\nu)} \nonumber \\ 
    &{\times}\mel{E\unpert_{\mu\sigma}}{\H\pert_\text{pert}}{E\unpert_{\nu\sigma'}}\nonumber \\
    &- \hbar\sum_{\mu,\sigma }\sigma\braket{o}{E_{\mu\sigma}\unpert}\frac{\braket{E\unpert_{\mu\sigma}}{F_\text{pert}\pert}}{ z-\sigma\Omega_\mu}  ,
\end{align}
The superscript $\HA$ reminds that this response function is obtained within the harmonic approximation. 
Eq.\ \eqref{chi 1(z)} represents one of the main results of this paper. The most important feature of Eq.\ \eqref{chi 1(z)} is that it features matrix elements of the observables and external potentials on the phonon spinors, exactly as in the case of noninteracting electronic wavefunctions subject to external fields. Thus, as in the electronic case, selection rules can be formulated to discern how certain phonon transitions are silent due to symmetry or conservation arguments. 

In the HA explored here, the ionic response function $\chi_{O,H_\text{pert}}^\HA(z)$  is given by ``bare'' vertices, i.e.\ the ``unscreened'' external field  $\mel{E\unpert_{\mu\sigma}}{{\H}\pert_\text{pert}}{E\unpert_{\nu\sigma'}}$ and forces $\braket{E_{\mu\sigma}\unpert}{F_\text{pert}\pert}$. In fact, as phonons are noninteracting in the HA, they have no means of screening the external perturbation acting on the lattice. Instead, we discuss in Sec.\ \ref{sec: linear response SC} how in  self-consistent approaches, anharmonicity provides a source of screening of the external perturbation. Thus, in the anharmonic case, the static  ``bare'' potential and forces in Eqs.\ \eqref{chi 1(z)} are replaced by frequency dependent anharmonically ``screened'' vertices.

{
The response \eqref{chi 1(z)} has a two-phonon and a one-phonon part, which can be written in terms of two and one-phonon bare propagators, namely 
\begin{align} \label{chi 1(z) L0 g0}
    &\chi_{O,H\pert_\text{pert}}^\HA(z) =\frac{\hbar}{2
    }\!\!\sum_{\mu\nu, \sigma\sigma' } \mel{E\unpert_{\nu\sigma'}}{\mathcal{O}}{E\unpert_{\mu\sigma}}   L^0_{\mu\sigma,\nu\sigma'}(z) \nonumber \\ & \times 
    \mel{E\unpert_{\mu\sigma}}{\H\pert_\text{pert}}{E\unpert_{\nu\sigma'}}- \hbar\sum_{\mu,\sigma }\braket{o}{E_{\mu\sigma}\unpert}g^0_{\mu\sigma}(z){\braket{E\unpert_{\mu\sigma}}{F_\text{pert}\pert}}  ,
\end{align}
From the two-phonon propagator 
\begin{equation} \label{L0 ss' (z)}
    L^0_{\mu\sigma,\nu\sigma'}(z) = \sigma\sigma'\frac{n(\sigma'\Omega_\nu) {-} n(\sigma\Omega_\mu)}{ z -(\sigma\Omega_\mu - \sigma' \Omega_\nu)} ,
\end{equation}
it appears clear how the classification of the phonon eigenvectors in terms of their pseudospin $\sigma$ allow for the distinction between the resonant $\sigma {=} \sigma'$ and the antiresonant $\sigma {\neq} \sigma'$ two-phonon response. In the zero-temperature limit, the resonant sector $\sigma{=}\sigma'$ vanishes, whereas the antiresonant sector $\sigma{\neq}\sigma'$ remains finite because of the zero-point contribution. Thus, coherence between the $\sigma{=}+$ and $\sigma{=}- $ sectors is needed to reproduce the ionic response at zero temperature.

The one-phonon response in Eq.\ \eqref{chi 1(z) L0 g0} features the one-phonon bare propagatorThe one-phonon response in Eq.\ \eqref{chi 1(z) L0 g0} features the one-phonon bare propagator 
\begin{equation} \label{g0(z)}
    g^0_{\mu\sigma}(z) = \frac{\sigma}{z-\sigma\Omega_\mu},
\end{equation}
where the pseudospin $\sigma$ distinguishes between the positive and the negative poles of the one-phonon propagator. The one-phonon part of the response function in Eq.\ \eqref{chi 1(z)} has no electronic analogue. In fact, electronic excitations only exist paired with hole excitations, and the electronic system does not posses such degree of freedom. }

Depending on the analytical form of the external field $  H_\text{pert}$ and of the observable $  O$ the one-phonon response or the two-phonon response are zero. For example, if both $  O$ and $  H_\text{pert}$ have only linear terms in $  {\bf R},   {\bf P} $, the two-phonon response is zero. 
However, we show in Sec.\ \ref{sec: linear response SC} how the self-consistency couples the one-phonon response to the two-phonon propagator and viceversa. In Sec.\ \ref{sec: conductivities}, we present some notable examples of one- and two-phonon only response.

{Here, we report briefly the ESPALD analogue of the usual result for time derivatives of response functions $\dv{t} \chi_{O,H\pert_\text{pert}}^{\HA}(t){=}\chi_{\dv{O}{t},H\pert_\text{pert}}^{\HA} (t)$(see e.g.\ \cite{drigo2023}), or, in frequency space $-iz \chi_{O,H\pert_\text{pert}}^{\HA}(z){=}\chi_{\dv{O}{t},H\pert_\text{pert}}^{\HA} (z)$. For the case $\ket{o}{=}0$ (only quadratic terms), using Eq.\ \eqref{dot O} in Eq.\ \eqref{delta O pert (t)} it is straightforward to show that  
\begin{align} 
    -iz&\chi_{O,H\pert_\text{pert}}^\HA(z) =\frac{\hbar}{2
    }\!\!\sum_{\mu\nu, \sigma\sigma' } \mel{E\unpert_{\nu\sigma'}}{ \dot{\mathcal{O}} \unpert}{E\unpert_{\mu\sigma}}  \nonumber \\ &\times\sigma'\sigma \frac{n(\sigma'\Omega_\nu) {-} n(\sigma\Omega_\mu)}{ z -(\sigma\Omega_\mu - \sigma' \Omega_\nu)} \mel{E\unpert_{\mu\sigma}}{\H\pert_\text{pert}}{E\unpert_{\nu\sigma'}} \label{-iz chi(z)}
\end{align}
where $\dot{\mathcal{O}} \unpert$ is the single-particle equivalent of $\dv{O}{t}$ in linear response
\begin{equation}
    \dot{\mathcal O} \unpert = \frac{1}{i}[\mathcal{O},\sigma_z\H\unpert]_\dagger .
\end{equation}
}

\section{Optical, Thermal conductivity, and displacement autocorrelations} \label{sec: conductivities}
{
Here, we show how our formalism is apt to describe response functions often investigated in material science. 
To elucidate how our model connects to standard theories of phonons, we first present the ionic displacement autocorrelation functions. 
Next, we present the derivation of the  optical and thermal conductivity in the ESPALD.
Lattice optical conductivity is related to the infrared absorption spectra.
Thermal conductivity is the key property for materials used in thermoelectric and thermal insulating applications.

{To obtain response functions, we specialize Eq.\ \eqref{chi 1(z)} for $ \chi_{O,H\pert_\text{pert}}^\HA(z)$  to different external perturbations $\ket{F_\text{pert}\pert}, \H\pert_\text{pert}$ and observables $\ket{o}, \mathcal{O}$.} We recap in Table \ref{tab:response_recipe} the single-particle equivalent of the external perturbation/ observables for the displacement correlation functions, optical conductivity, and thermal conductivity.}

\newcommand{\twolinecell}[2]{\shortstack[c]{#1\\[-0.25ex]#2}}

\begin{table*}[htbp]
    \centering
    \setlength{\tabcolsep}{2.2pt}
    \renewcommand{\arraystretch}{1.6}
    \begin{tabular}{cc|c|c|c|c|c|c}
        \hline
        \multicolumn{2}{c|}{\textbf{Response function}}
        &
        &
        \multicolumn{2}{c|}{\textbf{Perturbation}}
        &
        \multicolumn{2}{c|}{\textbf{Observable}}
        &
        \textbf{Eq.}
        \\
        \hline
        \multirow{3}{*}{\twolinecell{Displacement}{autocorrelation}}
        &
        \multirow{3}{*}{$\chi_{\mu\nu}(z)$}
        &
        Many-body
        &
        $H_{\text{pert}}$
        &
        $\tens{e}_\nu^T {\cdot}\mtrx{\bf M}{}^{1/2} {\cdot} \tens u$
        &
        $O$
        &
        $\tens{e}_\mu^T {\cdot}\mtrx{\bf M}{}^{1/2} {\cdot} \tens u$
        &
        \multirow{3}{*}{\eqref{chi rr 0}}
        \\
        &
        &
        \multirow{2}{*}{Single-particle}
        &
        $\mathcal{H}_{\text{pert}}$
        &
        $0$
        &
        $\mathcal{O}$
        &
        $0$
        &
        \\
        &
        &
        &
        $\ket{F_{\text{pert}}}$
        &
        $-\sum_\sigma
        \frac{1}{\sqrt{2\hbar\Omega_\nu}}
        \ket{E\unpert_{\nu\sigma}}$
        &
        $\ket{o}$
        &
        $\sum_\sigma
        \frac{1}{\sqrt{2\hbar\Omega_\mu}}
        \ket{E\unpert_{\mu\sigma}}$
        &
        \\
        \hline
        \multirow{3}{*}{\twolinecell{Variance}{autocorrelation}}
        &
        \multirow{3}{*}{$\chi_{\mu\nu,\theta\eta}(z)$}
        &
        Many-body
        &
        $H_{\text{pert}}$
        &
        $\frac{1}{2}
        \tens u^T
        \mtrx{\bf M}{}^{1/2}
        \tens{e}_\theta
        \tens{e}_\eta^T
        \mtrx{\bf M}{}^{1/2}
        \tens{u}$
        &
        $O$
        &
        $\frac{1}{2}
        \tens u^T
        \mtrx{\bf M}{}^{1/2}
        \tens{e}_\mu
        \tens{e}_\nu^T
        \mtrx{\bf M}{}^{1/2}
        \tens{u}$
        &
        \multirow{3}{*}{\eqref{chi0 2ph}}
        \\
        &
        &
        \multirow{2}{*}{Single-particle}
        &
        $\mathcal{H}_{\text{pert}}$
        &
        $\frac{1}{2\sqrt{\Omega_\theta\Omega_\eta}}
        \sum_{\sigma\sigma'}
        \dyad{
            \tenscomp E_{\eta\sigma}\unpert
        }{
            \tenscomp E_{\theta\sigma'}\unpert
        }$
        &
        $\mathcal{O}$
        &
        $\frac{1}{2\sqrt{\Omega_\mu\Omega_\nu}}
        \sum_{\sigma\sigma'}
        \dyad{
            \tenscomp E_{\nu\sigma}\unpert
        }{
            \tenscomp E_{\mu\sigma'}\unpert
        }$
        &
        \\
        &
        &
        &
        $\ket{F_{\text{pert}}}$
        &
        $0$
        &
        $\ket{o}$
        &
        $0$
        &
        \\
        \hline
        \multirow{3}{*}{\twolinecell{Optical}{conductivity}}
        &
        \multirow{3}{*}{$\sigma_{xx}(z)$}
        &
        Many-body
        &
        $H_{\text{pert}}$
        &
        $d_{\text{el}x}$
        &
        $O$
        &
        $d_{\text{el}x}$
        &
        \multirow{3}{*}{\eqref{sigma(z)}}
        \\
        &
        &
        \multirow{2}{*}{Single-particle}
        &
        $\mathcal{H}_{\text{pert}}$
        &
        $0$
        &
        $\mathcal{O}$
        &
        $0$
        &
        \\
        &
        &
        &
        $\ket{F_{\text{pert}}}$
        &
        $\sum_{\mu\sigma}
        \frac{1}{\sqrt{2\hbar \Omega_\mu}}
        \ket{E\unpert_{\mu\sigma}}
        \bbrakket{
            E\unpert_{\mu\sigma}
        }{
            \widetilde{Z}^\star_\gamma
        }$
        &
    $\ket{o}$
        &
        $\sum_{\mu\sigma}
        \frac{1}{\sqrt{2\hbar \Omega_\mu}}
        \ket{E\unpert_{\mu\sigma}}
        \bbrakket{
            E\unpert_{\mu\sigma}
        }{
            \widetilde{Z}^\star_\gamma
        }$
        &
        \\
        \hline
        \multirow{3}{*}{\twolinecell{Thermal}{conductivity}}
        &
        \multirow{3}{*}{$\kappa_{xx}(z)$}
        &
        Many-body
        &
        $H_{\text{pert}}$
        &
        $d_{\text{en}x}$
        &
        $O$
        &
        $\frac{1}{V}\pdv{}{t}d_{\text{en}x}$
        &
        \multirow{3}{*}{\eqref{k(z,q)}}
        \\
        &
        &
        \multirow{2}{*}{Single-particle}
        &
        $\mathcal{H}_{\text{pert}}$
        &
        $\mathcal{D}_x$
        &
        $\mathcal{O}$
        &
        $\frac{1}{iV}
        [\mathcal{D}_x,\sigma_z \H\unpert]_\dagger$
        &
        \\
        &
        &
        &
        $\ket{F_{\text{pert}}}$
        &
        $0$
        &
        $\ket{o}$
        &
        $0$
        &
        \\
        \hline
    \end{tabular}
    \caption{
    Recap of the single particle equivalent of observables and perturbations to be used in Eq.\ \eqref{chi 1(z)} to obtain typical response functions in the ESPALD.
    $\H_\text{pert}, \ket{F_\text{pert}}$ represent the single-particle Hamiltonian and forces resulting from the single-particle mapping [Eq.\ \eqref{H 2x2}--\eqref{H , F single particle}] of the external perturbation $H_\text{pert}$.
    $\mathcal{O},\ket{o}$ represent the single-particle mapping of, respectively, the quadratic and linear part of the operator $O$ [Eq.\ \eqref{O = aa* 2x2}].
    The displacement autocorrelation function and variance autocorrelation function are related to the harmonic one-phonon and two-phonon propagators, respectively.
    $\tens{u}{=}\mathbf{R}{-}\tens{R}\unpert$ has been used for brevity.
    For the optical conductivity, $d_{\text{el}x}$ is the electric dipole operator [Eq.\ \eqref{el dipole}], while $\kket{\widetilde{Z}^\star}$ is the Born-effective charge spinor [Eq.\ \eqref{Z spinor}].
    For the thermal conductivity, $d_{\text{en}x}$ is the energy dipole [Eq.\ \eqref{energy dipole}], while $\mathcal{D}_x$ is its single-particle equivalent [Eq.\ \eqref{D gamma = R}]. For off-diagonal mode pairs, the quadratic single-particle operators shown in the table are understood to be Hermitian symmetrized as in Eqs.\ \eqref{Vext dR2 singlepart} and \eqref{O dR2 singlepart}.
    }
    \label{tab:response_recipe}
\end{table*}
Although presented here in the HA, generalization to anharmonic cases are straightforward using the ESPALD formulation presented in Sec.\ \ref{sec: espald}, where anharmonicity is included via self-consistency. We present here the formalism that enables the calculation of optical and thermal lattice response and leave the generalization to the anharmonic case for future work, where such equations will be applied to real materials.

\subsection{$\bf R$ and ${\bf R}^2$ autocorrelation functions: harmonic phonon propagators} 
Here, we connect to standard theories of phonons by presenting the $\mathbf{R}$ an $\mathbf{R}^2$ autocorrelation functions in the ESPALD. 

{
To compute the harmonic displacement autocorrelation function, we consider as the perturbation a displacement along a given eigenmode $\tens{e}_\mu$
\begin{align} 
     H_\text{pert}^{(\nu)} & = \tens{e}_\nu^T \cdot\mtrx{\bf M}{}^{1/2} \cdot [ {\bf R} - \tens{R}\unpert] \nonumber 
   \\&= \sum_{I\a} \tenscomp{e}_{\nu,I\a}{\sqrt{M_{I}}}[   R_{I\a} -\tenscomp{R}\unpert_{I\a}]. 
\end{align}
Using Eqs.\ \eqref{operators RP to aa*}, its single-particle equivalents are
\begin{subequations} \label{1 phonon external}
    \begin{align}
        \H_\text{pert} ^{(\nu)}&= 0, \\ 
        \ket{F_\text{pert}^{(\nu)}} &= -\frac{1}{\sqrt{2\hbar\Omega_\nu}} 
        \sum_\sigma\ket{E\unpert_{\nu\sigma}}
    \end{align}
\end{subequations}
By considering $  O =   H_\text{pert}^{(\mu)}$, from  Eq.\ \eqref{operators RP to aa*} and Eq.\ \eqref{O = aa* 2x2} we obtain the single-particle equivalent of the observable
\begin{subequations}
    \begin{align}
       \mathcal O &= 0, \\ 
        \ket{o_\mu} &= \frac{1}{\sqrt{2\hbar\Omega_\mu}} \sum_\sigma\ket{E\unpert_{\mu\sigma}}.
    \end{align}
\end{subequations}
Thus, the harmonic displacement autocorrelation function $\chi_{\delta R_\mu,\delta R_\nu}^\HA(z)$ (which we report for simplicity as $\chi_{\mu\nu}^\HA(z)$) is:
\begin{align} 
    \chi_{\mu\nu}^\HA(z) = -\hbar\sum_{\theta\sigma} \braket{o_\mu}{E\unpert_{\theta\sigma}}\braket{E\unpert_{\theta\sigma}}{F_\text{pert}^{(\nu)}}\frac{\sigma}{z-\sigma\Omega_\theta},
\end{align}
from which we get  the standard displacement-displacement harmonic response
\begin{equation} \label{chi rr 0}
   \chi_{\mu\nu}^\HA(z) = \frac{\delta_{\mu\nu}}{z^2 - \Omega^2_\mu},
\end{equation}
}also referred to as the harmonic phonon Green's function \cite{siciliano2023wigner}. Notice that at variance with some of the literature \cite{siciliano2023wigner,berges2023phonon,monacelli2021stochastic}, here the one-phonon propagator would be  the ``pseudospin unpolarized'' version of $g^0_{\mu\sigma}(z)$ [Eq.\ \eqref{g0(z)}], namely 
\begin{equation} \label{G0mu (z)}
    G^0_{\mu}(z) =\sum_\sigma g^0_{\mu\sigma}(z) = \frac{2\Omega_\mu}{z^2 - \Omega^2_\mu},
\end{equation}
which is the one commonly found in many-body perturbation theory approaches to phonons [cf.\  Refs.\ \cite{maradudin1962scattering,maradudin1964lattice,semwal1972thermal}, Ref.\ \cite{mahan2013many} Eq.\ (3.2.16)]. 

{
The harmonic ionic variance autocorrelation function is obtained considering the external potential 
\begin{align} \label{Vext dR2}
    &     H_\text{pert}^{(\theta\eta)} =\frac{1}{2} [ {\bf R} - \tens{R}\unpert]^T\mtrx{\bf M}{}^{1/2}\tens{e}_\theta\tens{e}_\eta^T \mtrx{\bf M}{}^{1/2}[ {\bf R} - \tens{R}\unpert]  \nonumber\\
    & =\frac{1}{2}\sum_{I\a,J\beta}[   R_{I\a} -\tenscomp{R}\unpert_{I\a}]{\sqrt{M_{I}}} \tenscomp{e}_{\theta,I\a} \tenscomp{e}_{\eta,J\b} {\sqrt{M_{J}}}[   R_{J\b} -\tenscomp{R}\unpert_{J\b}].
\end{align}
Using Eq.\ \eqref{operators RP to aa*} and \eqref{H , F single particle}, the single-particle equivalents of Eq.\ \eqref{Vext dR2} are 
\begin{subequations} 
    \begin{align}
        \H_\text{pert}^{(\theta\eta)} &= \frac{1}{4\sqrt{\Omega_\theta\Omega_\eta}} \sum_{\sigma\sigma'}\left[\dyad{\tenscomp E_{\eta\sigma}\unpert}{\tenscomp E_{\theta\sigma'}\unpert}+\text{h.c.}\right], \label{Vext dR2 singlepart} \\ 
        \ket{F_\text{pert}^{(\theta\eta)}} &= 0.
    \end{align}
\end{subequations}
Using $  O =   H_\text{pert}^{(\mu\nu)}$, the single-particle equivalent of the observable is 
\begin{subequations} 
    \begin{align}
        \mathcal{O}_{\mu\nu} &= \frac{1}{4 \sqrt{\Omega_\mu\Omega_\nu}} \sum_{\sigma\sigma'} \left[\dyad{E_{\mu\sigma}\unpert}{E_{\nu\sigma'}\unpert}+\text{h.c.}\right] \label{O dR2 singlepart} \\ 
        \ket{o_{\mu\nu}} &= 0,
    \end{align}
\end{subequations}
and we get from Eq.\ \eqref{chi 1(z)} the harmonic ionic variance autocorrelation function  $\chi_{\delta R_\mu\delta R_\nu,\delta R_\theta\delta R_\eta }^\HA(z)$ (which we report for simplicity as $\chi_{\mu\nu,\theta\eta}^\HA(z)$)

\begin{align} 
    \chi^\HA_{\mu\nu,\theta\eta}(z)
    &= \frac{\delta_{\mu\theta}\delta_{\nu\eta}+\delta_{\mu\eta}\delta_{\nu\theta}}{2}
    \frac{\hbar}{4\Omega_\mu\Omega_\nu} \nonumber \\
    &\quad\times\biggl[
    \frac{[n(\Omega_\nu) {-} n(\Omega_\mu)][\Omega_\mu - \Omega_\nu]}
    {z^2 -(\Omega_\mu - \Omega_\nu)^2 } \nonumber \\
    &\qquad +
    \frac{[n(\Omega_\nu) {+} n(\Omega_\mu)+1][\Omega_\mu + \Omega_\nu]}
    {z^2 -(\Omega_\mu + \Omega_\nu)^2 }\biggr] . \label{chi0 2ph}
\end{align}
}
The expression in Eq.\ \eqref{chi0 2ph} is elsewhere defined as the bare two-phonon propagator (cf.\   Eq.\ (60) of Ref.\ \cite{siciliano2023wigner}). In the nomenclature used here, the bare two-phonon propagator would be the "pseudospin unpolarized" version of  Eq.\ \eqref{L0 ss' (z)}
\begin{align}
    L^0_{\mu\nu}(z) &= \sum_{\sigma\sigma'}L^0_{\mu\sigma,\nu\sigma'}(z) \nonumber\\
    &=2\frac{[n(\Omega_\nu) - n(\Omega_\mu)][\Omega_\mu - \Omega_\nu]}{z^2 - (\Omega_\mu -\Omega_\nu)^2} \nonumber \\
    &+ 2\frac{[1+ n(\Omega_\nu)+ n(\Omega_\mu)][\Omega_\mu + \Omega_\nu]}{z^2 - (\Omega_\mu +\Omega_\nu)^2} 
\end{align}
which is the one obtained from the convolution in Matsubara imaginary frequency space of two one-phonon propagators $G^0_\mu(z)$ of Eq.\ \eqref{G0mu (z)} [cf.\  Ref.\ \cite{maradudin1962scattering}, Ref.\ \cite{mahan2013many} Eq.\ (3.5.1)].

\subsection{Optical Conductivity}

In dielectrics, the optical conductivity is computed as the ratio between the displacement current generated by an oscillating electric dipole in the media and the macroscopic electric field.

In the dipole approximation, the Hamiltonian describing the coupling between a uniform linearly-polarized electric field $\mathcal{E}(t)$ and the lattice is
\begin{align} \label{Vext = E Z^*}
      H_\text{pert}(t) &= - \sum_{\gamma} \mathcal{E}_\gamma(t)   d_{\text{el}\gamma} 
\end{align}
where the electric dipole operator  
\begin{equation} \label{el dipole}
      d_{\text{el}\gamma}=\sum_{J\b} Z^\star_{J,\gamma\b}[  R_{J\b} - \tenscomp{R}^0_{J\b}]
\end{equation}
features the Born-effective charge tensor $ Z^\star_{J,\gamma\b}$, {which we express in a compact way as 
\begin{equation}
     d_{\text{el}\gamma}= [\cdot\mtrx{\mathbf Z}{}^\star\cdot{\mathbf{i}}_\gamma]^T \cdot[ \mathbf R - \tens{R}^0]
\end{equation}
in terms of the matrix $[\mtrx{\mathbf Z}{}^\star]_{I\gamma,J\b} = \delta_{IJ}Z^\star_{J,\gamma\b}$, and the auxiliary vector $[{\mathbf{i}}_\gamma]_{I\a} = \delta_{\gamma\alpha}$. }
{Considering the electric field as the time-dependent stimulus $ s_\gamma(t) {=}  \mathcal{E_\gamma}(t)$ [cf.\   Eq.\ \eqref{Vext = v1 s(t)}], the single-particle mapping of Eq.\ \eqref{Vext = E Z^*} is
\begin{subequations} \label{F ext = mu}
    \begin{align}
    \mathcal{H_\text{pert}^{(\gamma)}} &= 0,\\
   \ket{F_\text{pert}^{(\gamma)}} &=\mqty[{\bf f}^{(\gamma)}\\{\bf f}^{(\gamma)}{}^*],
\end{align}
\end{subequations}
where 
\begin{align}
    \mathrm{f}_{I\a}^{(\gamma)}  = \frac{1}{\sqrt{2\hbar}} \sum_{J\beta}  \tenscomp{D}^{-1/4}_{I\a,J\b}M_{J}^{-1/2} Z^\star_{J,\gamma\beta}
\end{align}
{or, in short 
\begin{equation}
    \mathbf{f}^{(\gamma)} = \frac{1}{\sqrt{\hbar}}[ \mtrx{\bm{\lambda}}{}^{\mathbf R}]^T [\mtrx{\mathbf Z}{}^\star]^T \cdot{\mathbf{i}}_\gamma,
\end{equation}}
is the perturbing single-particle force associated with the component $\gamma$ of the electric field.
The optical conductivity is obtained from the dipole-dipole response as  \begin{equation}
    \sigma_{xy}(z) = \frac{-iz}{V} \chi_{  d_{\text{el}x} ,   d_{\text{el}y}}(z),
\end{equation}
where $V= N_\text{c}V_\text{c}$ is the volume of the system.
To obtain $\sigma(z)$ from Eq.\ \eqref{chi 1(z)}, we consider the observable $  O =   d_{\text{el}x}$. From Eqs.\ \eqref{Vext = E Z^*}-\eqref{F ext = mu}, its single-particle mapping is 
\begin{subequations} \label{O ext = mu}
    \begin{align}
    \mathcal{O} &= 0,\\
   \ket{o_x} &=\mqty[{\bf f}^{(x)}\\{\bf f}^{(x)}{}^*].
\end{align}
\end{subequations}

Using Eq.\ \eqref{chi 1(z)} the lattice optical conductivity in the HA  reads 
\begin{align} 
    \sigma^\HA_{xy}(z) &=  \frac{iz\hbar}{V}\sum_{\mu\sigma}   \bbrakket{o_x}{E_{\mu\sigma}\unpert}\bbrakket{E_{\mu\sigma}\unpert}{F^{(y)}_\text{pert}}\frac{\sigma}{z-\sigma\Omega_\mu} 
\end{align}
For nonchiral crystals, the projections of on the phonon spinor basis are pseudospin independent
\begin{align} \label{<E|F> dipole}
    \bbrakket{E_{\mu\sigma}\unpert}{F^{(\gamma)}_\text{pert}} =   \bbrakket{E_{\mu\sigma}\unpert}{o_\gamma} = \frac{1}{\sqrt{2\hbar \Omega_\mu}} \sum_{I\a} \frac{\tenscomp{e}_{\mu,I\a}Z^\star_{I,\gamma\a}}{\sqrt{M_I}}.
\end{align}
{
Using the completeness of the phonon spinors, we  express
\begin{equation}
    \kket{F^{(\gamma)}_\text{pert}} = \ket{o_\gamma} = \sum_{\mu\sigma}  \frac{1}{\sqrt{2\hbar \Omega_\mu}} \ket{E\unpert_{\mu\sigma}}\bbrakket{E\unpert_{\mu\sigma}}{\widetilde{Z}^\star_\gamma}
\end{equation}
where the single-particle spinor of the (mass-reduced) Born effective charges is
\begin{equation} \label{Z spinor}
  \ket{\widetilde{Z}^\star_\gamma} = \mqty[ \mtrx{\mathbf{M}}{}^{-1/2}[\mtrx{\mathbf Z}{}^\star]^T \cdot{\mathbf{i}}_\gamma  \\ \mtrx{\mathbf{M}}{}^{-1/2}[\mtrx{\mathbf Z}{}^\star]^T \cdot{\mathbf{i}}_\gamma ]
\end{equation}
}

Considering Eq.\ \eqref{<E|F> dipole}, the optical conductivity is 
\begin{align} 
    \sigma^\HA_{xy}(z) &=  \frac{iz}{V}\sum_{\mu}  \frac{I^x_\mu I^y_\mu}{z^2-\Omega_\mu^2} 
\end{align}
where the modal  IR activity is 
\begin{equation}
    I_\mu^x = \sum_{I\a} \frac{\tenscomp{e}_{\mu,I\a}Z^*_{I,x\a}}{\sqrt{M_I}} .
\end{equation}

To express the conductivity in reciprocal space, we use the Bloch theorem construction (Sec.\ \ref{sec: bloch thm}).
The projections of $6N_\text{at}$ vectors on the spinor basis (as e.g.\ the single-particle forces) can be expressed as 
\begin{align}
    \bbrakket{E\unpert_{{\bf q} m\sigma}}{F}= \bbrakket{\tenscomp{e}\unpert_{{\bf q}m\sigma}}{F_{{\bf q}}} 
\end{align}
where the cell-periodic forces are
\begin{equation}
   \ket{F_{\bf q}}  = \frac{1}{\sqrt{N_c}} {e}^{-i{\mathbf q}\cdot {\bf \mathbfscr{R}} \unpert} \ket{{F}} .
\end{equation}
}
Since Born-effective charge tensor is cell-periodic, the IR activities in reciprocal space are simply 
\begin{equation}
 I_{\mathbf{q}m}^x =\sqrt{N_c} \delta_{\bm q,0} \sum_{b\a} \frac{\tenscomp{e}_{\mathbf qm,b\a}Z^\star_{b,x\a}}{\sqrt{M_b}} ,
\end{equation}
which yields for the $xx$ component of the conductivity the usual formula for optical phonons at the zone center
\begin{equation} \label{sigma(z)}
    \sigma^{xx}(z) =  \frac{iz }{V_c}\sum_{m}  \frac{ |\sum_{b\a} \frac{\tenscomp{e}_{\bm 0 m,b\a}Z^\star_{b,x\a}}{\sqrt{M_b}}|^2}{z^2-\Omega_{\mathbf 0 m}^2} . 
\end{equation}
}

\subsection{Thermal conductivity}
Following Luttinger/Allen-Feldman formulations \cite{luttinger1964theory,allen1993thermal}, we consider a lattice in a uniform temperature gradient $\grad T$ subject to the external perturbation
\begin{equation} \label{Hext thermal}
      H_\text{pert}=- \sum_\gamma\frac{1}{ T} \nabla_\gamma T   ,d_{\text{en}\gamma}
\end{equation}
where $   d_{\text{en}}$ is the energy dipole. As already discussed, the linear response function in the HA does not depend on the details of the stimulus $s(t)$, so we neglect it in Eq.\ \eqref{Hext thermal}. ~\cite{eich2017functional,note:thermal-perturbation}

Following Hardy \cite{hardy1963energy}, the energy dipole for a harmonic lattice reads 
\begin{align} \label{energy dipole}
      d_{\text{en}\gamma} &= \frac{1}{2}\sum_{I}\tenscomp{R}_{I\gamma}\unpert   {h}_{\HA I} + \text{h.c.}\nonumber\\
    & = \frac{1}{4}\sum_{I\gamma}\tenscomp{R}_{I\gamma}
    \unpert \left[\frac{
      p^2_{I\alpha}}{M_I} +  
     {\tenscomp{u}}_{I\a}\sum_{J\b} \phi\unpert_{I\a,J\beta}  {\tenscomp{u}}_{J\beta}\right] + \text{h.c.}
\end{align}
where ${h}^\HA_{I}$ is the harmonic energy density satisfying $  H\unpert_\HA = \sum_I {h}_{\HA,I} $ and we used the shorthand notation ${\tenscomp{u}} = R - \tenscomp{R}\unpert$. 

Using Eq.\ \eqref{operators RP to aa*} in Eq.\ \eqref{energy dipole}, we obtain the single-particle mapping of the external perturbation $  H_\text{pert}$  [Eq.\ \eqref{Hext thermal}]
\begin{subequations}
    \begin{align}
        \H_\text{pert}^{(\gamma)} &=\frac{1}{T}   \mathcal{D}_{\gamma} \\
      \ket{F_\text{pert}} &= 0 
    \end{align}
\end{subequations}
in terms of the single-particle equivalent of the energy dipole 
\begin{equation}
    \mathcal{D}_{\gamma} = \mqty[ \mtrx{\mathbf{d}}{}_{\text{res}\gamma } & \mtrx{\mathbf{d}}{}^{*}_{ \text{ares}\gamma } \\ \mtrx{\mathbf{d }}{}^{}_{ \text{ares}\gamma }  & \mtrx{\mathbf{d}}{}_{\text{res}\gamma } ]
\end{equation}
where the resonant and antiresonant parts of the energy dipole are
\begin{subequations}
    \begin{align}
        \mtrx{\mathbf{d}}{}_{\text{res}\gamma} & =  [\mtrx{\bm{\lambda}}{}_{\mathbf{R}}]^T \mtrx{[\tens{R}_\gamma \bm{\phi}\unpert]} \mtrx{\bm{\lambda}}{}_{\mathbf{R}} + [\mtrx{\bm{\lambda}}{}_{\mathbf{P}}]^T \mtrx{[\tens{R}_\gamma\mathbf{M}^{-1}]} [\mtrx{\bm{\lambda}}{}_{\mathbf{P}}], \\
        \mtrx{\mathbf{d}}_{\text{ares} \gamma}& =  [\mtrx{\bm{\lambda}}{}_{\mathbf{R}}]^T \mtrx{[\tens{R}_\gamma \bm{\phi}\unpert]} \mtrx{\bm{\lambda}}{}_{\mathbf{R}} - [\mtrx{\bm{\lambda}}{}_{\mathbf{P}}]^T \mtrx{[\tens{R}_\gamma\mathbf{M}^{-1}]}\mtrx{\bm{\lambda}}{}_{\mathbf{P}},
    \end{align}
\end{subequations}
and the matrices $[\tens{R}_\gamma \bm{\phi}\unpert],[\tens{R}_\gamma\mathbf{M}^{-1}]$ are 
\begin{subequations}
\begin{align}
[\tens{R}_\gamma \bm{\phi}\unpert]_{I\a,J\b}  &= \frac{1}{2}(\tenscomp{R}_{I\gamma} + \tenscomp{R}_{J\gamma} ){\phi}\unpert_{I\a,J\b}, \\
[\tens{R}_\gamma \mathbf{M}^{-1}]_{I\a,J\b}  &= \tenscomp{R}_{I\gamma} M^{-1}_I \delta_{IJ}\delta_{\a\b}.
\end{align}
\end{subequations}
The matrix elements of the dipole operator $\mathcal{D}$ on the spinor basis are 

\begin{equation} \label{D gamma = R}
    \mel{E\unpert_{\mu\sigma}}{\mathcal{D}_\gamma}{E\unpert_{\nu\sigma'}} = \frac{(\Omega_\mu + \sigma\sigma'\Omega_\nu)^2}{4\sqrt{\Omega_\mu\Omega_\nu}} \sum_{I\a}\tenscomp{e}_{\mu,I\a}\tenscomp{R}\unpert_{I\gamma }\tenscomp{e}_{\nu,I\a},
\end{equation}
which may be rewritten as
\begin{equation} 
    \mel{E\unpert_{\mu\sigma}}{\mathcal{D}_\gamma}{E\unpert_{\nu\sigma'}} = \frac{(\Omega_\mu + \sigma\sigma'\Omega_\nu)^2}{4\sqrt{\Omega_\mu\Omega_\nu}}  \mel{E\unpert_{\mu\sigma}}{\mathbfscr{R}_\gamma\unpert}{E\unpert_{\nu\sigma'}}
\end{equation}
where $\mathbfscr{R}_\gamma\unpert$ is the ($\gamma$ cartesian component) of the (equilibrium) position operator defined in Eq.\ \eqref{R spinors}.

The thermal conductivity can be obtained from the Fourier law $\bm J {=} {-} \kappa \grad T$ as the ratio (derivative) of the heat flux with respect to the perturbing temperature gradient.  The heat flux is related to the energy dipole as 
\begin{equation} \label{J = dD/dt}
    J_\a(t) = \frac{1}{V} \pdv{}{t} \langle   d_{\text{en}\a}\rangle_{  \rho(t)}
\end{equation}
A temperature gradient $\grad_\gamma T$ that couples to the energy dipole $  d_{ \text{en}\gamma}$ generates a perturbed density $\varrho^{(\gamma)}(t)$ in the linear regime.

{
Considering Eq.\ \eqref{J = dD/dt}, Eq.\ \eqref{Hext thermal}, and the Fourier law, the thermal conductivity corresponds to the time derivative of the dipole-dipole response \cite{drigo2023}
\begin{equation}
    \kappa_{xy}(z) =  \frac{iz}{T}   \chi_{{d}_{\text{en} x}, d_{\text{en} y} } (z) .
\end{equation}
Being a time-derivative of a response function, we obtain $\k$ from Eq.\ \eqref{-iz chi(z)} considering  $O {=}   \tfrac{1}{V} d_{\text{en} x}$, and the single-particle equivalent of its derivative $\dot{\mathcal{O}}_x$ defines a harmonic heat flux
\begin{subequations}
    \begin{align} \label{O = J}
       \dot{\mathcal{O}}_x &= \mathcal{J}_x= \frac{1}{iV}[\mathcal{D}_x,\sigma_z \H\unpert]_\dagger ,\\ 
       \ket{o} & = 0 
    \end{align}
\end{subequations}
}

Using Eq.\ \eqref{-iz chi(z)}, we get the finite frequency thermal conductivity in the harmonic approximation as (for the $xx$ component)
{
\begin{align} 
    \kappa^\HA_{xx}(z) =&\frac{-\hbar}{2T V
    }\!\!\sum_{\mu\nu, \sigma\sigma' } \mel{E\unpert_{\nu\sigma'}}{ \mathcal{J}_x}{E\unpert_{\mu\sigma}} \nonumber \\ &\times \sigma'\sigma \frac{n(\sigma'\Omega_\nu) {-} n(\sigma\Omega_\mu)}{ z -(\sigma\Omega_\mu - \sigma' \Omega_\nu)} 
    \mel{E\unpert_{\mu\sigma}}{ \mathcal{D}_x}{E\unpert_{\nu\sigma'}} . \label{K = J D}
\end{align}
Notice that, in analogy with the electrical dipole operator for electrons, the thermal dipole $\mathcal{D}^\gamma$ in Eq.\ \eqref{D gamma = R} features the (equilibrium) positions of the ions via $\mathbfscr{R}\unpert$, and it is ill-defined in an infinite crystal. To regularize Eq.\ \eqref{K = J D}, we replace the dipole with the heat flux vertex from Eq.\ \eqref{O = J}
\begin{equation}
    \mel{E\unpert_{\mu\sigma}}{ \mathcal{D}_x}{E\unpert_{\nu\sigma'}} = i\frac{\mel{E\unpert_{\mu\sigma}}{ \mathcal{J}_x}{E\unpert_{\nu\sigma'}}}{\sigma'\Omega_\nu-\sigma\Omega_\mu}
\end{equation}}
which is well-defined in the thermodynamic limit, since
\begin{align}
&\mel{E\unpert_{\mu\sigma}}{\mathcal{J}_x}{E\unpert_{\nu\sigma'}} =  \frac{(\Omega_\mu + \sigma\sigma'\Omega_\nu)^2}{4\sqrt{\Omega_\mu\Omega_\nu}}    \nonumber \\
    &\times\frac{1}{i}\sum_{I\a,J\b}\tenscomp{e}_{\mu,I\a}[\tenscomp{R}\unpert_{Ix} -\tenscomp{R}\unpert_{Jx}] \tenscomp{D}^{1/2}_{I\a,J\b} \tenscomp{e}_{\nu,J\b} 
\end{align}
depends only on difference of the ionic positions,  and can be rewritten as 
\begin{equation}
\mel{E\unpert_{\mu\sigma}}{\mathcal{J}_x}{E\unpert_{\nu\sigma'}} = \frac{(\Omega_\mu + \sigma\sigma'\Omega_\nu)^2}{4\sqrt{\Omega_\mu\Omega_\nu}} \mel{E\unpert_{\mu\sigma}}{ \frac{1}{i}[\mathbfscr{R}_x, \H\unpert]}{E\unpert_{\nu\sigma'}} \:.
\end{equation}
Thus, we recast Eq.\ \eqref{K = J D}  as
\begin{widetext}
\begin{equation}
    \kappa^\HA_{xx}(z) =\frac{-i\hbar}{2T V
    }\!\!\sum_{\mu\nu, \sigma\sigma' } \mel{E\unpert_{\nu\sigma'}}{ \mathcal{J}_x}{E\unpert_{\mu\sigma}} \sigma'\sigma \frac{1}{ z -(\sigma\Omega_\mu - \sigma' \Omega_\nu)} 
    \frac{n(\sigma'\Omega_\nu) {-} n(\sigma\Omega_\mu)}{\sigma'\Omega_\nu-\sigma\Omega_\mu}  \mel{E\unpert_{\mu\sigma}}{ \mathcal{J}_x}{E\unpert_{\nu\sigma'}} . \nonumber
\end{equation}
\end{widetext}
The procedure used here to pass from the ill-defined energy dipole to the well-defined heat flux is the vibrational equivalent of the procedure by which the matrix elements of the position $  r$ are replaced by band velocities $\mel{\psi_{\mathbf{k}m}}{  r}{\psi_{\mathbf{k}n}} = i\hbar \frac{{v}_{\mathbf{k}mn}}{\epsilon_{\mathbf{k}n} - \epsilon_{\mathbf{k}m}}$ in an electronic system \cite{shang2018all}. This algorithm, allowed by the formalism developed here, further tightens the one-to-one correspondence between electronic and ionic response sewed by the ESPALD. Moreover, our formalism makes clear the connections between charge transport by electrons and heat transport by phonons.

With the construction presented in Sec.\ \eqref{sec: bloch thm}, a matrix element of a $6N_\text{at}{\times}6N_\text{at}$ operator $\mathcal{V}$ on the spinor basis can be expressed as 
\begin{align}
     &\mel{E_{{\bf q}m\sigma}\unpert}{\mathcal{V}}{E_{{\bf q}'n\sigma'}\unpert}= \mel{\tenscomp{e}_{\mathbf qm\sigma}\unpert}{\mathcal{V}_{{\bf q},{\bf q}'}}{\tenscomp{e}_{\mathbf{q}'n\sigma'}\unpert} 
\end{align}
where the (generalized) periodic part of the operator $\mathcal{V}$ is 
\begin{equation} 
     \mathcal{V}_{{\bf q},{\bf q'}} = \frac{1}{N_{\mathrm{c}}}  {e}^{-i{\bf q}\cdot {\mathbfscr R} \unpert} \mathcal{V} {e}^{i{\bf q}'\cdot {\mathbfscr R} \unpert}.
\end{equation}
Considering the translational invariance property of the dynamical matrix [Eq.\ \eqref{phi IJ = phi I-J}], the heat-flux in reciprocal space is
\begin{align} 
\mel{E\unpert_{\mathbf{q}\mathrm{m}\sigma}}{\mathcal{J}_x}{E\unpert_{\mathbf{q}'\mathrm{n}\sigma'}} &= \delta_{\mathbf{q}\mathbf{q}'} \frac{(\Omega_{\mathbf{q}\mathrm{m}} + \sigma\sigma'\Omega_{\mathbf{q}\mathrm{n}} )^2}{4\sqrt{\Omega_{\mathbf{q}\mathrm{m}}\Omega_{\mathbf{q}\mathrm{n}}}}  \nonumber \\ & \times \mel{\tenscomp{e}_{\mathbf qm\sigma}\unpert}{\pdv{\H\unpert_{\mathbf{q}}}{ q_x}}{\tenscomp{e}_{\mathbf{q}n\sigma'}\unpert} \label{heatflux in q} 
\end{align}
where the matrix elements of $\pdv{\H\unpert_{\mathbf{q}}}{ q_x}$ can be rewritten in terms of the the generalized (Wigner) group velocity  \cite{simoncelli2019unified,simoncelli2022wigner,caldarelli2022many} 
\begin{equation}
    \mel{\tenscomp{e}_{\mathbf qm\sigma}\unpert}{\pdv{\H\unpert_{\mathbf{q}}}{ q_x}}{\tenscomp{e}_{\mathbf{q}n\sigma'}\unpert}  = \tenscomp{e}_{\textbf{q}m}^T \cdot \mtrx{\tenscomp{v}}{}^\gamma_\mathbf{q} \tenscomp{e}_{\textbf{q}n}
\end{equation}
where
\begin{equation} \label{wigner velocity}
   \mtrx{\tenscomp{v}}{}^\gamma_\mathbf{q} =\pdv{}{{q}_\gamma} \mtrx{\tens{D}}{}^{1/2}_\mathbf{q},
\end{equation}
which is a $3N_\text{base}{\times}3N_\text{base}$ matrix for every $\mathbf{q}$ and every direction $\gamma{=}x,y,z$. 

Using Eq.\ \eqref{heatflux in q}, we get the thermal conductivity in reciprocal space 
{(expressing $z{=}\w{+}i\eta$ )
\begin{align} \label{k(z,q)}
     &\kappa^\HA_{xx}(\w+i{\eta}) =\frac{-i\hbar}{TN_\text{c}V_\text{c}
    }\!\!\sum_{\mathbf{q}\mathrm{mn}, \sigma\sigma' }\!\! \frac{\sigma\sigma'}{2} \! \frac{[\Omega_{\mathbf{q}\mathrm{m}} + \sigma\sigma'\Omega_{\mathbf{q}\mathrm{n}} ]^4}{16{\Omega_{\mathbf{q}\mathrm{m}}{\Omega_{\mathbf{q}\mathrm{n}}}}} \nonumber  \\ 
     & \times |\tenscomp{v}_{\mathbf{q}\mathrm{mn}}^x|^2   \frac{ n(\sigma\Omega_{\mathbf{q}\mathrm{m}} ){-}n(\sigma'\Omega_{\mathbf{q}\mathrm{n}})}{\sigma'\Omega_{\mathbf{q}\mathrm{n}} {-}\sigma\Omega_{\mathbf{q}\mathrm{m}}} 
    \frac{1}{ \omega +i\eta {-}(\sigma\Omega_{\mathbf{q}\mathrm{m}} {-}\sigma'\Omega_{\mathbf{q}\mathrm{n}}) }
\end{align}
Common approximations of thermal conductivity are obtained by properly replacing the terms in the second line of Eq.\ \eqref{k(z,q)} and performing the static limit $\w{\to}0$. The $\eta{\to}0^+$ limit must be performed after thermodynamical limit, as appropriate for response theory \cite{zubarev1970boundary}.

The thermodynamic limit for ordered solids is realised performing $N_\text{c}{\to}\infty$ at fixed $V_\text{c}$. In this case, we obtain in the static limit}
\begin{align} \label{bte 0 k}
    \kappa^\HA_{xx} =&\frac{\pi\hbar}{k_BT^2 N_\text{c}V_\text{c}
    }\!\!\sum_{\mathbf{q}\mathrm{mn} } \Omega_{\mathbf{q}\mathrm{m}}^2|v_{\mathbf{q}\mathrm{mn}}^x|^2 \nonumber \\&\times [-\pdv{n(\Omega_{\mathbf{q}\mathrm{m}})}{\Omega_{\mathbf{q}\mathrm{m}}}]
 \delta(\Omega_{\mathbf{q}\mathrm{m}} {-}\Omega_{\mathbf{q}\mathrm{n}} ),
\end{align} 
The Dirac delta restricts the response to degenerate modes, including the intraband contribution $\mathrm{m}{=}\mathrm{n}$ and possible off-diagonal contributions within degenerate subspaces, and signals the divergence characteristic of harmonic crystals. Moreover, it cancels exactly the antiresonant transport channels $\sigma {=} -\sigma'$.  
Although divergent, $\kappa^\HA_{xx}$ of Eq.\ \eqref{bte 0 k} bears important physical insights. In particular, with this derivation, we demonstrate how the thermal response is highly sensitive on the phonon pseudospin [Eq.\ \eqref{heatflux in q},\eqref{k(z,q)}]. It has been shown that the antiresonant part of thermal conductivity is typically much smaller than the resonant part \cite{caldarelli2022many}, and can be gauged away with a proper choice of the energy density \cite{simoncelli2019unified,ercole2016gauge}.
Thus, although the phonon spectrum might be trivial in the phonon pseudospin, the response is not. Our formalism shows in a straightforward and compelling way how a vastly investigated property such as thermal conductivity is highly sensitive on the phonon pseudospin. As this quantum number encodes the sign of the phase of vibrational waves propagating in the crystal [see Eq.\ \eqref{E0 = plane waves}], it can rule out possible interference mechanisms between phonons. Thus, in a generalization of thermal conductivity beyond the HA, the phonon pseudospin plays a vital role in the microscopic description of coherent wave-like contribution to thermal conductivity, where heat flows via interference mechanisms between phonons \cite{simoncelli2019unified,isaeva2019modeling,simoncelli2022wigner,fiorentino2023from,caldarelli2022many}.  

{
The single-mode relaxation time approximation (SMRTA) is obtained by replacing (in the resonant $\sigma{=}\sigma'$ channel) the term $\pi\delta(\omega {-}\Omega_{\mathbf{q}\mathrm{m}} {+}\Omega_{\mathbf{q}\mathrm{n}})$ originating from the $\eta{\to}0^+$ limit of Eq.\ \eqref{k(z,q)} with a single-mode relaxation time  $\delta_{\mathrm{mn}}\tau_{\mathbf{q}\mathrm{m}}$ \cite{sun2010lattice} }
\begin{align}
    \kappa^{\text{SMRTA}}_{xx} =\frac{\hbar}{T V
}\!\!\sum_{\mathbf{q}\mathrm{m}}&\Omega_{\mathbf{q}\mathrm{m}} |v_{\mathbf{q}\mathrm{mm}}^x|^2[-\pdv{n(\Omega_{\mathbf{q}\mathrm{m}})}{\Omega_{\mathbf{q}\mathrm{m}}}]{\tau_{\mathbf{q}\mathrm{m}}}
\end{align}

Thermal conductivity from Wigner transport equation (WTE) in absence of hydrodynamic effects  is obtained replacing (in the resonant $\sigma{=}\sigma'$ channel) the term $ \pi\frac{ n(\sigma\Omega_{\mathbf{q}\mathrm{m}} ){-}n(\sigma'\Omega_{\mathbf{q}\mathrm{n}} ) }{-\w} \delta(\omega {-}\Omega_{\mathbf{q}\mathrm{m}} {+}\Omega_{\mathbf{q}\mathrm{n}})$  originating from the $\eta{\to}0^+$ limit of Eq.\ \eqref{k(z,q)} with the combination of the scattering rates of the modes $\Gamma_{\mathbf{\mathbf{q}}\mathrm{m}} = {1}/{2\tau_{\mathbf{\mathbf{q}}\mathrm{m}}}$ and of modal specific heat $C_{\bm qs}$ (for a detailed derivation of such combinations see e.g.\ \cite{dangic2021origin})
\begin{align}
     &\kappa^{\text{WTE}}_{xx}  {=} \frac{1}{N_\text{c}V_\text{c}}\!\sum_{\mathbf{q}, \mathrm{mn}} \frac{\Omega_{\mathbf{q}\mathrm{m}} + \Omega_{\mathbf{q}\mathrm{n}} }{4}|v_{\mathbf{q}\mathrm{mn}}^{\mathrm{H},x}|^2  \nonumber\\ & \times\left[\frac{C_{\mathbf{q}\mathrm{m}}}{\Omega_{\mathbf{q}\mathrm{m}}}\! +\! \frac{C_{\mathbf{q}\mathrm{n}}}{\Omega_{\mathbf{q}\mathrm{n}}}\right]\frac{\frac{1}{2}[\Gamma_{\mathbf{q}\mathrm{m}} + \Gamma_{\mathbf{q}\mathrm{n}} ]\!}{ [\Omega_{\mathbf{q}\mathrm{m}} {-}\! \Omega_{\mathbf{q}\mathrm{n}} ]^2\!+\! \tfrac{1}{4}[\Gamma_{\mathbf{q}\mathrm{m}}\! +\! \Gamma_{\mathbf{q}\mathrm{n}}\!]^2} 
\end{align}
where $C_{\mathbf{q}\mathrm{m}} {=}\frac{\hbar^2 \Omega^2_{\mathbf{q}\mathrm{m}}}{k_B T^2} n(\Omega_{\mathbf{q}\mathrm{m}}) [n(\Omega_{\mathbf{q}\mathrm{m}}) {+} 1]$. The Hardy velocities $v_{\mathbf{q}\mathrm{mn}}^{\mathrm{H},x}$ are related to the Wigner velocities $v_{\mathbf{q}\mathrm{mn}}^{\mathrm{W},x}$ [Eq.\ \eqref{wigner velocity}] by $v_{\mathbf{q}\mathrm{mn}}^{\mathrm{H},x} = \frac{\Omega_{\mathbf{q}\mathrm{m}}+ \Omega_{\mathbf{q}\mathrm{n}}}{2\sqrt{\Omega_{\mathbf{q}\mathrm{m}}\Omega_{\mathbf{q}\mathrm{n}}}} v_{\mathbf{q}\mathrm{mn}}^{\mathrm{W},x}$. They appear in the WTE thermal conductivity as we used a Hardy definition of the energy density in the energy dipole [Eq.\ \eqref{energy dipole}] \cite{caldarelli2022many}.

{
In the limit $N_c{=}1,V_c{\to}\infty$, we recover the case of a disordered harmonic solid. Performing the limit $V_c{\to}\infty$, $\eta{\to}{0}^+$, $\omega{\to}0$ in this order in Eq.\ \eqref{k(z,q)}, we obtain Allen-Feldman (AF) formula for thermal conductivity \cite{allen1993thermal,simoncelli2023thermal}
\begin{align} 
    \kappa^{\text{AF}}_{xx} =&\frac{\pi\hbar}{TV_\text{c}}\!\sum_{\mathbf{q}, \mathrm{m}} \delta_{\mathbf{q},0} [-\pdv{n(\Omega_{\mathbf{q}\mathrm{m}})}{\Omega_{\mathbf{q}\mathrm{m}}}] \nonumber\\ & \times\sum_{\mathrm{n}\neq m}\frac{(\Omega_{\mathbf{q}\mathrm{m}} + \Omega_{\mathbf{q}\mathrm{n}})^2 }{4}|v_{\mathbf{q}\mathrm{mn}}^x|^2 
 \delta(\Omega_{\mathbf{q}\mathrm{m}} {-}\Omega_{\mathbf{q}\mathrm{n}} ) ,
\end{align}
coupling (quasi)degenerate modes at the zone center.
}

\section{Effective single particle picture for anharmonic lattice dynamics} \label{sec: espald}
In this Section, we discuss our main objective: the formulation of an effective single particle picture for anharmonic lattice dynamics.

In self-consistent approaches, { the BO Hamiltonian that regulates the dynamics  $H_\text{BO}(t)$ is replaced by a mean-field Hamiltonian  that depends on the state of the system $H^{[\rho(t)]}_\text{scf}(t)$, i.e.\ in Eq.\ \eqref{BO liouville} we replace
\begin{equation}
      [H_\text{BO}(t),\rho(t)] \xrightarrow{\quad}   [H^{[\rho(t)]}_\text{scf}(t),\rho(t)].
\end{equation}
Here, for a mean-field Hamiltonian, we intend an Hamiltonian with linear and quadratic terms in $\mathbf{R},\mathbf{P}$ (no mixed $\mathbf{R}\mathbf{P}$) whose couplings depend on time either directly through the external fields or indirectly through self-consistency. 
}

The density matrix evolves via a self-consistent many-body Liouville equation
\begin{subequations} \label{liouville nbody SC}
    \begin{align}
   & i\hbar\pdv{  \rho(t)}{t}  = [  H^{[\rho(t)]}_{\text{scf}}(t), \rho(t)] \qquad &t{\geq}0,\\
   &   \rho\unpert= {e^{-\frac{1}{k_BT}  H\unpert}}/{\Tr[e^{-\frac{1}{k_BT}  H\unpert}]} \qquad &t{<}0. 
    \end{align}
\end{subequations}
where $  H\unpert {=}   H^{[\rho(t)]}_{\text{scf}}(t)\eval_{t{<}0} $.
Thus, a self-consistent approximation of the BO dynamics requires (i) an approximation of the mean-field density $  \rho(t)$ (ii) an approximation of the initial state of the system $  \rho\unpert$ and (iii) a description of self-consistency, thus how the Hamiltonian $  H^{[\rho(t)]}_{\text{scf}}(t)$ depends on $ \rho$. 

{We also suppose that} the self-consistent Hamiltonian $  H^{[\rho(t)]}_{\text{scf}}(t)$ depends on the density matrix $\rho(t)$ at the same instant of time, that amounts to neglecting memory effects in the anharmonic interaction, as per the adiabatic approximaton of TD-DFT \cite{runge1984density}.

Among the existing methods for lattice dynamics, the time-dependent self-consistent harmonic approximation (TD-SCHA) \cite{monacelli2021time,siciliano2023wigner} provides a clear functional description of the self-consistent Hamiltonian $  H^{[\rho(t)]}_{\text{scf}}(t)$ and of the mean-field density matrix $  \rho(t)$ that approximates $  \rho_\text{BO}(t)$, together with a precise description of the equilibrium $  \rho\unpert$. Thus, we adopt the TD-SCHA as our fundamental framework.
{In particular, in the TD-SCHA, the state of the lattice is completely determined from the single-particle variables $\varrho(t), \ket{G(t)}$ (details in Appendix \ref{app: tdscha detail}), hence 
\begin{equation} 
    H^{[\rho(t)]}_\text{scf TD-SCHA}(t) = H^{[\varrho(t),G(t)]}_{\text{scf}}(t).
\end{equation}
}
The TD-SCHA is used to study structural and spectroscopic responses in the presence of strong anharmonicity \cite{monacelli2021stochastic,lihm2021gaussian,siciliano2023wigner}. Relying on a quantum mechanical least action principle, anharmonic effects are considered from first principles beyond perturbation theory. As such, it is the state-of-the-art theoretical and numerical tool to calculate the response of crystals subject to ultrahigh anharmonicity. 
This is the case e.g.\ for solids undergoing a structural phase transition, where the standard perturbative approach to lattice anharmonicity breaks down.

Here, we show how a mean-field self-consistent theory of lattice dynamics such as the TD-SCHA admits a single particle mapping akin to the one presented in Sec.\ \ref{sec: II} for the simple harmonic case.
In fact, once the TD-SCHA initial conditions and self-consistent relations are given, we can use the formalism developed in Sec.\ \ref{sec: II}-\ref{sec: response} to express self-consistent dynamics in one-to-one correspondence to gKS equations for interacting electrons.

{ An important difference between the lattice and the electronic self-consistent approaches is in the nature of the operators. In the electronic case, where the Coulomb interaction has a universal expression $\sum_{ij}\frac{1}{|r_i - r_j|}$, the self-consistent approximation coincides with the single-particle mapping. Namely, when the gKS Hamiltonian is formulated to approximate the many-body all-electron Hamiltonian, the resulting self-consistent Hamiltonian is single particle, as it describe a unitary dynamics of the single-particle objects $\ket{\psi^{\text{KS}}_i(t)}$. Instead, since the  BO functional $V_\text{BO}(\mathbf{R})$ is implicit, a self-consistent approximation is first used to give an explicit form to the ionic interaction. The resulting $H^{[\rho(t)]}_{\text{scf}}(t)$ retain many-body terms like $\sum_{ij}\mathbf{R}_i\mathbf{R}_j$ and describes the dynamics of the many-body density matrix $\rho(t)$.  Thus, a further step is needed to obtain from $H^{[\rho(t)]}_{\text{scf}}(t)$ a true single-particle description of the anharmonic lattice dynamics. Schematically; 
\begin{subequations} 
    \begin{align}
         H_{\text{el}} (t) &\xrightarrow{\text{self-consistency, single-particle}} H^\text{KS}_\scf(t),  \\
         H_{\text{BO}} (t)&\xrightarrow{\text{self-consistency}} H_{\text{scf}}(t) \xrightarrow{\text{single-particle}} \mathcal{H}_{\text{scf}}(t), \ket{F_\text{scf}(t)} .\label{sp phonon arrows}
    \end{align}
\end{subequations}
In a nutshell, the major theoretical achievement of this work consists in a clear recipe in how to realize the last step in  \eqref{sp phonon arrows}:  unveiling how self-consistent approaches to anharmonicity allow a mapping where lattice dynamics is encoded in the evolution of single-particle degrees of freedom $\varrho(t),\ket{G(t)}$ that evolve with effective single-particle Hamiltonian $\mathcal{H}_\text{scf}(t)$ and forces $\ket{F_\text{scf}(t)}$.
}

We start by formulating the equilibrium condition for the mean-field approximation. The equilibrium Hamiltonian of the TD-SCHA corresponds to the SCHA equilibrium Hamiltonian \cite{monacelli2021stochastic,siciliano2023wigner}
\begin{equation} \label{H0 scha}
      H\unpert = \tfrac{1}{2} {\bf P}^T \cdot \mtrx{\bf M}{}^{-1}  {\bf P} + \tfrac{1}{2}[ {\bf R}- \bm \R\unpert]^T \cdot \mtrx{\bf{\Phi}}{} \unpert[ {\bf R}- \bm \R\unpert]
\end{equation}
where the equilibrium positions $\R\unpert$ are obtained from the SCHA equilibrium density matrix as
\begin{equation}
    \R\unpert_{I\a} = \left\langle  R_{I\a} \right\rangle_{  \rho\unpert},
\end{equation}
and $  \rho\unpert$ is 
\begin{equation}
      \rho \unpert = e^{-\frac{1}{k_BT}  H\unpert}/\Tr[ e^{-\frac{1}{k_BT}  H\unpert}]
\end{equation}
{The terms $\mathcal{\R}\unpert,\bm\Phi\unpert$ in Eq.\ \eqref{H0 scha} are determined by} the SCHA equilibrium conditions \cite{errea2014anharmonic,bianco2017second,monacelli2021stochastic}
\begin{subequations} \label{scha eq}
\begin{align}
    \Phi_{I\a,J\beta}\unpert &= \left\langle \pdv{ V_\text{BO} ({\bf R})}{R_{I\a}}{R_{J\b}}\right \rangle_{ \rho\unpert}, \label{phi2 scha} \\
    0 &=  \left\langle \pdv{ V_\text{BO} ({\bf R})}{R_{I\a}}\right\rangle_{  \rho\unpert} .\label{f = 0 scha}
\end{align}    
\end{subequations}
Eqs.\ \eqref{scha eq} are derived using quantum variational principle to minimize the free-energy of the crystal \cite{errea2014anharmonic}. The variational nature of the approach ensures that the matrix in Eq.\ \eqref{phi2 scha} is non-negative and can be defined even far from the regime of applicability of the HA. In fact, the SCHA interatomic force-constant matrix ${\Phi}$ defines the auxiliary SCHA phonon frequencies that are partially renormalized by anharmonic couplings with respect to the ones obtained e.g.\ from standard density-functional perturbation theory calculations \cite{bianco2017second,siciliano2023wigner,baroni2001phonons}. Eq.\ \eqref{f = 0 scha} provides the equilibrium positions $\bm \R\unpert $ of the ions renormalized by anharmonicity, thus allowing for structural configurations of the unit cell stabilized by temperature, anharmonic, or zero-point motion effects \cite{errea2014anharmonic,errea2015high,errea2016quantum}.

Having formulated the equilibrium condition, we move to the dynamics in the TD-SCHA.
{We assume that the external many-body exact Hamiltonian $H_\text{pert}(t)$ in Eq.\ \eqref{H BO} is a sum of a contribution that depends only on $\mathbf{P}$ and a potential contribution that depends only on the atom positions $\mathbf{R}$, namely
$H_\text{pert}(t)=F(\mathbf{P},t)+V(\mathbf{R},t)$. Under this hypothesis there are no mixing $\mathbf{RP}$ terms and the TD-SCHA Hamiltonian is: }
\begin{align} 
    &  H^{[\varrho(t),G(t)]}_\text{scf}(t) = \tfrac{1}{2} {\bf P}^T\cdot\mtrx{\bf M}{}^{-1[\varrho(t),G(t)]}_\scf(t){ {\bf P}} 
    \nonumber \\
    &+ \tfrac{1}{2}[ {\bf R}{-} \bm \R\unpert]^T\cdot\mtrx{\bf \Phi}{}_\scf^{[\varrho(t),G(t)]}(t)[ {\bf R} {-} \bm \R\unpert] \nonumber \\
    &- {\bf f}_{\bf R\, \scf}^{[\varrho(t),G(t)]}(t)^T\cdot [ {\bf R} {-} \bm \R\unpert]
    \label{H = dRP compact SC}
\end{align}
where
\begin{subequations} \label{tdscha couplings}
\begin{align}
     \Phi_{\scf\,I\a,J\beta}^{[\varrho(t),G(t)]}(t) &= \left\langle  \pdv{ {H}_{\text{BO}}(t)}{R_{I\a}}{R_{J\b}}\right\rangle_{ \rho(t)}  ,\\
    M^{-1{[\varrho(t),G(t)]}}_{\scf\,I\a,J\beta}(t) &= \left \langle  \pdv{  {H}_{\text{BO}}(t)}{ P_{I\a}}{P_{J\b}}\right\rangle_{ \rho(t)} ,\\
      f_{\text{\textbf{R}scf}\,I\a}^{[\varrho(t),G(t)]}(t) &= -\left\langle  \pdv{  H_\text{BO}(t)}{ R_{I\a}}\right\rangle_{ \rho(t)}\nonumber\\+\sum_{J\b}&\left\langle\pdv{  H_\text{BO}(t)}{ R_{I\a}}{ R_{J\b}}\right\rangle_{ \rho(t)} \!  [\langle { R}_{J\b}\rangle_{ \rho(t)} -  \R\unpert_{J\b}] .
\end{align}    
\end{subequations}
{In what follows, we drop the superscript $[\rho(t), G(t)]$ on top of $\scf$ quantities for brevity.}

{$\bf{{\Phi}}_\scf$ and ${\bf M}^{-1}_\scf$ are $3N_\text{at}{\times}3N_\text{at}$ real symmetric matrices, while $\bf{ f}_{\textbf{R}}$ is a real vector. 
Couplings as $\bm \Phi_\scf$ and ${\bf f}_{\textbf{R}}$ may arise from anharmonic interactions  or external fields directly coupling to the ionic positions \cite{siciliano2023wigner}. Couplings as ${\bf {M}}_\scf^{-1}$ may come from anharmonic interactions {(e.g.\ isotope scatterings \cite{fugallo2013ab})} or external fields coupling directly to the kinetic energy of the ions, as in the case of an energy dipole \cite{drigo2023} (see Sec.\ \ref{sec: conductivities}). }
{The averages in Eqs.\ \eqref{tdscha couplings} are computed with the density matrix $ \rho(t)$ that is the solution of the mean-field Liouville equation [Eq.\ \eqref{liouville nbody SC}].}
{Notice that external potential $H_\text{pert}(t)$ is contained in $H_\text{BO}(t)$ [Eq.\ \eqref{H BO}] is not restricted to a quadratic form. The TD-SCHA couplings in Eqs.\ \eqref{tdscha couplings} probe the derivatives of any potential, and the resulting time-dependent Hamiltonian in Eq.\ \eqref{H = dRP compact SC} is quadratic.} 

Although formulated here for the TD-SCHA, we note on passing that the ESPALD is valid for every Hamiltonian of the form Eq.\ \eqref{H = dRP compact SC}, once the self-consistent equations between the Hamiltonian and the density matrix are formulated (here, Eqs.\ \eqref{tdscha couplings}). This mirrors how gKS equations can be formulated without an explicit description of the exchange-correlation functional, but an explicit expression is needed for numerical implementations.

{In the TD-SCHA framework, many-body operators $O$ are considered through the average of their derivatives on the time-dependent Gaussian $\rho(t)$. Thus, every operator acquires a time dependence through self-consistency 
\begin{equation}
     O \xrightarrow{\text{self-consistency}}   O^{[\varrho(t),G(t)]}.
\end{equation}
Assuming separate momentum and position dependence (i.e. $O = T(\mathbf{P}) + W(\mathbf{R}) )$, an operator $O$ is expressed in TD-SCHA as  
\begin{align}  \label{O tdscha}
    &  O^{[\varrho(t),G(t)]}= \tfrac{1}{2} {\bf P}^T\cdot   \mtrx{O}{}_{{\mathbf{P}}{\mathbf{P}}}^{[\varrho(t),G(t)]}{\bf P} 
    \nonumber \\
    &+ \tfrac{1}{2}[ {\bf R}{-} \bm \R\unpert]^T\cdot\mtrx{O}{}_{\mathbf{RR}}^{[\varrho(t),G(t)]}[ {\bf R} {-} \bm \R\unpert] \nonumber \\
    &- {\bf o}_{\bf R}^{[\varrho(t),G(t)]T}\cdot [ {\bf R} {-} \bm \R\unpert]
\end{align}
where
\begin{subequations} \label{O SC couplings}
\begin{align}
     O_{{\mathbf{RR}}\,I\a,J\beta}^{[\varrho(t),G(t)]}(t) &= \left\langle  \pdv{ {O}}{R_{I\a}}{R_{J\b}}\right\rangle_{ \rho(t)}  ,\\
    O^{{[\varrho(t),G(t)]}}_{\mathbf{PP}\,I\a,J\beta}(t) &= \left \langle  \pdv{  {O}}{ P_{I\a}}{P_{J\b}}\right\rangle_{ \rho(t)} ,\\
      o_{\textbf{R}\,I\a}^{[\varrho(t),G(t)]}(t) &= -\left\langle  \pdv{  O}{ R_{I\a}}\right\rangle_{ \rho(t)}\nonumber\\+\sum_{J\b}&\left\langle\pdv{O}{ R_{I\a}}{ R_{J\b}}\right\rangle_{ \rho(t)} \!  [\langle { R}_{J\b}\rangle_{ \rho(t)} -  \R\unpert_{J\b}] .
\end{align}    
\end{subequations}
The operators in the TD-SCHA were assumed to have the form Eq.\ \eqref{O tdscha} in Ref.\ \cite{siciliano2023wigner}. A formal proof of the structure in Eq.\ \eqref{O tdscha} can be obtained by performing an appropriate Legendre transform from the SCHA free energy to a function that imposes the expectation value of $O$ on the many-body state via a Lagrange multiplier. 

The important aspect of Eq.\ \eqref{O tdscha} is that at variance with the HA, the TD-SCHA can calculate operators that are not quadratic. The TD-SCHA framework ``harmonizes'' every many-body operator $O$ through its derivatives.
}

\subsection{Mapping anharmonic lattice dynamics}
Here, we rework the single-particle mapping outlined in Sec.\ \eqref{sec: map of dynamics} for the case where anharmonicity is included via self-consistency.

The starting point of the mapping is the definition of the dynamical matrix. In the TD-SCHA, the dynamical matrix is 
\begin{equation} \label{D scha}
    \mtrx{\bf D} = \mtrx{\bf M}{}^{-\tfrac{1}{2}}\mtrx{\bf \Phi}{}\unpert \mtrx{\bf M}{}^{-\tfrac{1}{2}}
\end{equation}
which defines the SCHA eigenvectors and normal modes frequencies
\begin{equation} \label{De = w e SC} 
    \mtrx{\bf D} {\bf e}_\mu = \w^2_\mu {\bf e}_\mu\,.
\end{equation}
At variance with the phonons $\Omega_\mu$ obtained from the HA, the normal modes $\omega_\mu$ of the SCHA are nonnegative by construction and already contain a class of anharmonic renormalization.
For a detailed description of the difference between SCHA and standard harmonic approximation of phonons, we refer the reader to e.g.\ Refs.\ \cite{bianco2017second,siciliano2023wigner}.

{With the SCHA dynamical matrix of Eq.\ \eqref{D scha}, the cartesian boson operators are redefined by the inverse canonical transformation, in the same form as Eq.\ \eqref{aa* to RP},
\begin{subequations}\label{aa* to RP SC}
\begin{align}
 {\bf a}  =&  \frac{1}{\sqrt{\hbar}}\left[ \mtrx{\bm{\Lambda}}{}_{\mathbf{P}}( { \bf R}- {\bm \R}\unpert )  -i \mtrx{\bm{\Lambda}}{}_{\mathbf{R}} { \bf P}  \right], \\
  {\bf a}{}\herm  =&  \frac{1}{\sqrt{\hbar}}\left[ \mtrx{\bm{\Lambda}}{}_{\mathbf{P}}( { \bf R}- {\bm \R}\unpert )  +i \mtrx{\bm{\Lambda}}{}_{\mathbf{R}} { \bf P}  \right] .
\end{align}
\end{subequations}
where now
\begin{subequations} \label{lambdas SC}
    \begin{eqnarray}
        \Lambda_{\mathbf{R}\,I\a,J\beta} {=}
\frac{1}{\sqrt{2}
}\sqrt{M}^{-1}_{I\a} D^{-1/4}_{I\a,J\beta},\\
\Lambda_{\mathbf{P}\,I\a,J\beta} {=}\frac{1}{\sqrt{2}
} \sqrt{M}_{I\a} D^{1/4}_{I\a,J\beta}\:.
    \end{eqnarray}
\end{subequations}
This is the only change in the definition of the bosonic variables with respect to Sec.\ \ref{sec: map of dynamics}: the harmonic dynamical matrix is replaced by the equilibrium SCHA one. The rest of the single-particle mapping is unchanged. This is the reason for formulating the harmonic mapping first.

In particular, the TD-SCHA Hamiltonian is mapped exactly as its harmonic counterpart in Eqs.\ \eqref{H 2x2}--\eqref{H , F single particle}, using the TD-SCHA coefficients of Eq.\ \eqref{tdscha couplings}. Since the TD-SCHA density matrix is Gaussian, and is completely specified by the connected single-particle density matrix $\varrho(t)$ and by the condensate $\ket{G(t)}$, the self-consistent mapping can be summarized as
\begin{equation}
    H_\text{scf}(t)
    \xrightarrow{\text{SP mapping}}
    \H_\scf(t),
    \kket{F_\scf(t)}.
\end{equation}
Here $\H_\scf(t)$ and $\kket{F_\scf(t)}$ are built from the TD-SCHA coefficients $\mtrx{{\bf \Phi} }{}_\scf(t)$, $\mtrx{\bf M}{}^{-1}_\scf(t)$, and $\mathbf f_{\mathbf{R}\scf}(t)$ by the same formulas used in the harmonic mapping, Eqs.\ \eqref{operators RP to aa*} with the matrices $\Lambda_{\mathbf R}, \Lambda_{\bf P}$ of Eq.\ \eqref{lambdas SC}. For illustration, the single-particle equilibrium Hamiltonian in the self-consistent case is
\begin{equation}
    \H(t{<}0) = \H\unpert = \mqty[\mtrx{\bf D }{}^{1/2} & 0 \\ 0 & \mtrx{\bf D} {}^{1/2}   ] ,
\end{equation}
{which is the single-particle mapping of $H\unpert$ in Eq.\ \eqref{H0 scha}.}

Now, we retrace the procedure leading to Eq.\ \eqref{espald schrodinger td} to obtain a time-dependent Schrodinger equation for the self-consistent case.
Following such procedure, we obtain  the self-consistent single-particle Liouville equation for the single-particle density matrix $\varrho(t)$
\begin{subequations} \label{1b mapping SC}
    \begin{align}
        i\hbar \pdv{\varrho(t)}{t}  &=\hbar[\sigma_z \H_\scf(t),\varrho(t)]_\dagger , \label{rhoC liouville SC}\\
        i\hbar \pdv{}{t} \ket{{G}(t)}&= \hbar \sigma_z \H_\scf(t)\ket{{G}(t)} - \hbar\sigma_z\kket{F_\scf(t)}. 
    \end{align}
\end{subequations}
Notice how, in the SC case, Eqs.\ \eqref{1b mapping SC} are coupled by self-consistency, i.e.\ through $\H_\scf(t),\ket{F_\scf(t)}$.
The SCHA phonon spinors are defined from the SCHA dynamical matrix  eigenvectors [Eq.\ \eqref{De = w e SC}] as 
\begin{equation} 
    \ket{E_{\mu+}\unpert} = \mqty[{\bf e_\mu} \\ 0], \quad \ket{E_{\mu-}\unpert} = \mqty[ 0 \\  {\bf e_\mu}]  ,
\end{equation}
satisfying
\begin{align}
    \H\unpert \ket{E_{\mu\sigma}\unpert} &= \omega_\mu \ket{E_{\mu\sigma}\unpert},\\
     \sigma_z \ket{E_{\mu\sigma}\unpert} &= \sigma \ket{E_{\mu\sigma}\unpert}
\end{align}
Exactly as in Sec.\ \ref{sec: equilibrium rho}, we can express the boundary conditions for $\varrho$ in the SCHA case as 
\begin{equation} \label{rhoc 0 = dft SC}
    \varrho\unpert = \sum_{\mu,\sigma} \sigma n(\sigma \omega_\mu) \dyad{E_{\mu\sigma}\unpert}{E_{\mu\sigma}\unpert}.
\end{equation}
where $n(\sigma\omega_\mu)$ is the Bose-Einstein distribution [Eq.\ \eqref{bose einstein}] evaluated at the SCHA frequencies $\w_\mu$. {As in \ref{sec: equilibrium rho}, the boundary conditions for the phonon condensate are $\ket{G\unpert}{=}0$.}

Using the initial conditions given in Eq.\ \eqref{rhoc 0 = dft SC} in the single-particle Liouville equation [Eq.\ \eqref{rhoC liouville SC}], we can express the single-particle density matrix for $t{>}0$ as 
\begin{equation} \label{rhoc(t) = E(t)E(t) sc}
    \varrho(t) = \sum_{\mu,\sigma} \sigma n(\sigma \omega_\mu) \dyad{E_{\mu\sigma}(t)}{E_{\mu\sigma}(t)},
\end{equation}
where the phonon spinors evolve according to the coupled equations
\begin{subequations} \label{espald schrodinger td SC}
\begin{align}
     i\hbar \pdv{}{t} \ket{E_{\mu\sigma}(t)} &= \hbar \sigma_z \H_\scf(t) \ket{E_{\mu\sigma}(t)} , \\
     i\hbar \pdv{}{t} \ket{G(t)}&= \hbar \sigma_z \H_\scf(t)\ket{G(t)} - \hbar\sigma_z\kket{F_\scf(t)}.
\end{align}
\end{subequations}
Eq.\ \eqref{espald schrodinger td SC} is one of the main results of this paper. It shows how, in a mean-field approximation, the equations regulating anharmonic lattice dynamics  are equivalent to the time-dependent Schrodinger equations used to describe the response of a system of interacting electrons.
In particular, the phonon spinors $\ket{E_{\mu\sigma}(t)}$, which indicate the evolution of the normal mode eigenvectors, are the phonon equivalent of the KS states $\ket{\psi_{i}^\text{KS}(t)}$ featuring in the time-dependent gKS theory  \cite{baer2018time}.
As we will show in Sec.\ \ref{sec: sternheimer}, Eqs.\ \eqref{espald schrodinger td} are coupled by the self-consistency of the one-body Hamiltonian $\H^{[\varrho(t),G(t)]}(t)$ and force $\ket{F^{[\varrho(t),G(t)]}(t)}$.

{
As in the HA, the self-consistent operators \eqref{O tdscha} are mapped in their single-particle equivalents 
\begin{equation}
      O^{[\varrho(t),G(t)]}
    \xrightarrow{\text{SP mapping}}
    \mathcal{O}^{[\varrho(t),G(t)]},
    \kket{o^{[\varrho(t),G(t)]}}.
\end{equation}
where $\mathcal{O}^{[\varrho(t),G(t)]},\kket{o^{[\varrho(t),G(t)]}}$  are obtained from the coefficients of $ O^{[\varrho(t),G(t)]}$ in Eqs.\ \eqref{O SC couplings} via Eqs.\ \eqref{operators RP to aa*} with the matrices $\Lambda_{\mathbf R}, \Lambda_{\bf P}$ of Eq.\ \eqref{lambdas SC}.

As done in Eq.\ \eqref{O(t) = rho(t) + G(t)}, the expectation value $ O(t)$ of any quadratic operator $   O$ is expressed in terms of the SC single-particle density matrix  $ \varrho(t)$ and the SC condensate $\ket{G(t)}$ \cite{blaizot} {
\begin{align} 
     O(t) &= \Tr[  O  \rho(t)] \nonumber \\
     &= \frac{\hbar}{2} \Tr[\mathcal{O}^{[\varrho(t),G(t)]} \varrho(t)] \nonumber \\
     &+  \frac{\hbar}{2} \mel{ G(t)}{\mathcal{O}^{[\varrho(t),G(t)]}}{ G(t)} \nonumber \\ &+ \hbar \braket{o^{[\varrho(t),G(t)]}}{{G}(t)}.  
\end{align}
}
The coefficients of the Taylor expansion of $\delta O(t)$ {near zero external perturbations} represent different orders of response functions. Below, we will focus on the linear response regime. 
In linear response that we will discuss below, the operator $O^{[\varrho(t),G(t)]}$ must be evaluated at equilibrium $\varrho\unpert,\ket{G\unpert}$, hence it single-particle equivalents will be
\begin{subequations}
\begin{align}
  \mathcal{O}^{[\varrho(t),G(t)]} &\simeq \mathcal{O}\unpert, \\
    \ket{o^{[\varrho(t),G(t)]}} &\simeq \ket{o\unpert} ,
\end{align}
\end{subequations}
and their time-dependence is eliminated.
}

A recap of the analogy between the time-dependent gKS theory of electronic dynamics and the ESPALD developed here is presented in Table \ref{tab:recap}
\renewcommand{\arraystretch}{2}

\begin{table*}[t]
\setlength{\tabcolsep}{5pt}
\centering
\begin{tabular}{lcc}
\toprule
&
\textbf{Phonons} [this work]
&
\textbf{Electrons}
\\
\midrule
\shortstack[l]{\small Equilibrium\\ \small single-particle density}
&
$\displaystyle
\varrho\unpert
=
\sum_{\mu,\sigma=\pm}
\sigma n(\sigma\omega_\mu)
\dyad{E_{\mu\sigma}\unpert}{E\unpert_{\mu\sigma}}
$
&
$\displaystyle
n\unpert
=
\sum_i f(\epsilon_i)
\dyad{\psi^{\text{KS}}_i}{\psi^{\text{KS}}_i}
$
\\
Time evolution
&
$\displaystyle
\begin{cases}
i \pdv{}{t} \ket{E_{\mu\sigma}(t)}
=
\sigma_z \H_\scf(t) \ket{E_{\mu\sigma}(t)}
\\
i \pdv{}{t} \ket{G(t)}
=
\sigma_z \H_\scf(t) \ket{G(t)}
-
\sigma_z \ket{F_\scf(t)}
\end{cases}
$
&
$\displaystyle
i\hbar \pdv{}{t}\ket{\psi^\text{KS}_i(t)}
=
H^\text{KS}_\scf(t)\ket{\psi^\text{KS}_i(t)}
$
\\
\shortstack[l]{\small Time-dependent\\ \small single-particle density}
&
$\displaystyle
\varrho(t)
=
\sum_{\mu,\sigma=\pm}
\sigma n(\sigma\omega_\mu)
\dyad{E_{\mu\sigma}(t)}{E_{\mu\sigma}(t)}
$
&
$\displaystyle
n(t)
=
\sum_i f(\epsilon_i)
\dyad{\psi^{\text{KS}}_i(t)}{\psi^{\text{KS}}_i(t)}
$
\\[2ex]
\hline
&
TD-SCHA
&
\\
\hline 
{\shortstack[l]{\small Time-dependent\\ \small observables}}
&
{
$\displaystyle
\begin{aligned}
O(t)
&=
\tfrac{\hbar}{2}\sum_{\mu,\sigma=\pm}
\sigma n(\sigma\omega_\mu)
\mel{E_{\mu\sigma}(t)}{\mathcal{O}^{[\varrho(t),G(t)]}}{E_{\mu\sigma}(t)}
\\
&
+
\tfrac{\hbar}{2}
\mel{G(t)}{\mathcal{O}^{[\varrho(t),G(t)]}}{G(t)}
+
\hbar\braket{o^{[\varrho(t),G(t)]}}{G(t)}
\end{aligned}
$}
&
{
$\displaystyle
O(t)
=
\sum_i f(\epsilon_i)
\mel{\psi^{\text{KS}}_i(t)}{O}{\psi^{\text{KS}}_i(t)}
$}
\\
\bottomrule
\end{tabular}
\caption{
One-to-one correspondence between the effective single particle picture for anharmonic lattice dynamics (ESPALD) for phonons and generalized Kohn-Sham theory (gKS) for electrons.
In the ESPALD, the phonon equilibrium is described by a single-particle density matrix $\varrho\unpert$ [Eq.\ \eqref{rhoc 0 = dft}], obtained by combining Bose-Einstein distribution factors $n(\omega_\mu)$ and projectors of the phonon spinor basis $\left\{ \ket{E_{\mu\pm}} \right\}$ [Eq.\ \eqref{E+, E-}], just as the electronic single-particle density is obtained by combining the Fermi-Dirac distribution $f(\epsilon_i)$ with the projectors on the Kohn-Sham states $\left\{\ket{\psi^\text{KS}_i}\right\}$.
The equation for the evolution of the phonon spinors is equivalent to the time-dependent Schrodinger equation for the gKS theory, where the single-particle Hamiltonian $\H(t)$ is the lattice equivalent of the time-dependent Kohn-Sham Hamiltonian $H^\text{KS}_\scf(t)$ that drives the electronic dynamics.
Lattice dynamics possesses an extra degree of freedom in the form of the generalized one-phonon propagator $\ket{G(t)}$, dubbed the phonon condensate, whose dynamics is driven by the self-consistent forces $\ket{F(t)}$.
{
In the TD-SCHA, evolutions of the expectation values of observables $O(t)$ are expressed in terms of $\ket{E_{\mu\sigma}(t)}$ and $\ket{G(t)}$, as for gKS observables are expressed in terms of the evolved KS states. The TD-SCHA introduces a time dependence of the observables through self-consistency Eq.\ \eqref{O tdscha}. Such time dependence is lost in linear response.
}
}
\label{tab:recap}
\end{table*}

\section{Response theory for anharmonic crystals} \label{sec: SC response}
Finally, we show how to compute the response of anharmonic crystals using the effective single particle picture of anharmonic lattice dynamics. We rework the procedure outlined in Sec.\ \ref{sec: response} allowing for external fields on top on a self-consistent Hamiltonian. As in gKS approaches, the external field will induce vibrational densities, that will change the self-consistent Hamiltonian. This feedback mechanism physically manifests as an anharmonic screening of the external perturbations.

\subsection{Linear response} \label{sec: linear response SC}
As in Sec.\ \ref{sec: linear response}, we separate the external perturbation {from the BO Hamiltonian Eq.\ \eqref{H BO}} in its operatorial and time-dependent part
\begin{equation} 
      H_\text{pert}(t)=   H_\text{pert} \pert s(t),
\end{equation}
and $s(t{<}0) {=} 0$. { $H_\text{pert}(t)$ enters the TD-SCHA Hamiltonian Eq.\ \eqref{H = dRP compact SC} through its derivatives as prescribed by Eq.\ \eqref{tdscha couplings}, i.e. 
\begin{subequations} \label{tdscha pert couplings}
\begin{align}
     \Phi_{\text{pert}\,I\a,J\beta}\pert(t) &= s(t)\left\langle  \pdv{ {H}_{\text{pert}}\pert }{R_{I\a}}{R_{J\b}}\right\rangle_{ \rho\unpert}  ,\\
    M^{-1}{}\pert_{\!\!\!\!\!\!\!\text{pert}\,I\a,J\beta}(t) &=s(t)\left \langle  \pdv{  {H}_{\text{pert}}\pert}{ P_{I\a}}{P_{J\b}}\right\rangle_{ \rho\unpert} ,\\
      f_{\text{\textbf{R}pert}\,I\a}\pert(t) &= -s(t)\left\langle  \pdv{  H_\text{pert}\pert}{ R_{I\a}}\right\rangle_{ \rho\unpert}.    
\end{align}    
\end{subequations}
With the coefficients in Eq.\ \eqref{tdscha pert couplings} we build the single-particle equivalents of $ H_\text{pert}\pert$
\begin{equation}
    H\pert_\text{pert}(t)
    \xrightarrow{\text{SP mapping}}
    \H\pert_\text{pert}(t),
    \kket{F\pert_\text{pert}(t)}.
\end{equation}
via Eq.\ \eqref{operators RP to aa*} through the SCHA matrices $\Lambda_{\mathbf{R}}, \Lambda_{\mathbf{P}}$ of Eq.\ \eqref{lambdas SC}. We next expand $\mathcal{H}_\scf(t),\ket{F\pert_\scf(t)}$ at linear order in $\H\pert_\text{pert},\kket{F\pert}$ obtaining 
\begin{subequations} \label{hscf(t) = hpert + hanh(t)}
    \begin{align}
    \H_\scf\pert(t) &=  \H_\text{pert}\pert(t)+ \H_\text{anh}^{(1)[\varrho(t),G(t)]} \\
    \kket{F_\scf\pert(t)} & =  \ket{F_\text{pert}\pert(t)} + \ket{F_\text{anh}^{(1)[\varrho(t),G(t)]}}
\end{align}
\end{subequations}
where $\H_{\text{anh} }\ket{F_\text{anh}}$ are obtained from the single-particle mapping of the coefficients 
\begin{subequations} \label{anharmonic tdscha couplings}
\begin{align}
     \Phi_{\text{anh}I\a,J\beta}\pert{}
     (t) &= \left\langle  \pdv{{V}_{\text{BO}}({\bf R})}{R_{I\a}}{R_{J\b}}\right\rangle_{  \rho\pert(t)} , \\
      f_{\text{anh},\textbf{R}\, I\a}\pert(t) &= -\left\langle  \pdv{ V_\text{BO}({\bf R})}{ R_{I\a}}\right\rangle_{  \rho\pert(t)}\nonumber\\+\sum_{J\b}&\left\langle\pdv{V_\text{BO}({\bf R})}{ R_{I\a}}{ R_{J\b}}\right\rangle_{ \rho\unpert}   \langle { R}_{J\b}\rangle_{  \rho\pert(t)} .
\end{align}    
\end{subequations}
from Eqs.\ \eqref{tdscha couplings}.
$\H_{\text{anh} }\ket{F_\text{anh}}$ depend on time through the single-particle variables $\varrho(t),G(t)$. They are thus related to self-consistency and how phonons interact in our  approximation, i.e.\ with anharmonicity. They are the lattice equivalent of the (linearized) Hartree-exchange correlation potential of electrons (cf.\  Table \ref{tab:recap LR}  below).
}

The response function in frequency space is evaluated as 
\begin{equation} \label{chi = O/s SC}
    \chi_{O,H_\text{pert}\pert} (z)  = \frac{\delta O\pert(z)}{s(z)}.
\end{equation}
{As in Eq.\ \eqref{chi = O/s} we set for simplicity $s(t){=}\delta(t)$ and $s(z) {=}1$.}
To obtain $\delta O\pert(z)$, we perform the expansion of the single-particle dynamical quantities retaining only linear terms in $\mathcal{H}_{\text{pert}}\pert, \ket{F_\text{pert}\pert}$
\begin{subequations} 
\begin{align}
    \ket{E_{\mu\sigma}(t)} & = \ket{E_{\mu\sigma}\unpert(t)} + \ket{E_{\mu\sigma}\pert(t)}  \\
    \ket{G(t)}&=  \ket{G\pert(t)}
\end{align}
\end{subequations}
and accordingly, of the self-consistent Hamiltonian and of the forces as
\begin{subequations} 
    \begin{align}
        \H_\scf(t) &= \H\unpert + \H_\scf\pert(t),\\
        \kket{F_\scf(t)}& =  \kket{F_\scf\pert(t)} ,
    \end{align} 
\end{subequations}
{since $\ket{F\unpert} =0$.}
Precisely as per Eq.\ \eqref{delta O pert (t)}, we express $\delta O(t)$ as
\begin{align}
    \delta O\pert(t) &= \frac{\hbar}{2}\sum_{\mu,\sigma}\sigma n(\sigma\omega_\mu) \biggl[\mel{E_{\mu\sigma}\pert(t)}{\mathcal{O}\unpert}{E_{\mu\sigma}\unpert(t)} \nonumber \\
    &+ \mel{E_{\mu\sigma}\unpert(t)}{\mathcal{O}\unpert}{E_{\mu\sigma}\pert(t)} \biggr] + \hbar \braket{o\unpert }{G\pert(t)} \label{delta O pert (t) SC}
\end{align}
In the self-consistent case, the 0-th order expansion of the time-dependent Schrodinger equation [Eq.\ \eqref{espald schrodinger td SC}] provides a plane-wave evolution for the SCHA spinors
\begin{equation} \label{E0 = plane waves sc}
    \ket{E\unpert_{\mu\sigma}(t)} = e^{-i\sigma\omega_\mu t}  \ket{E\unpert_{\mu\sigma}},
\end{equation}
and two 1st order equations in terms of the self-consistent fields
\begin{subequations} \label{espald schrodinger (1) t SC}
\begin{align}
     i \pdv{}{t} \ket{E_{\mu\sigma}\pert(t)} &=  \sigma_z \H\unpert\ket{E_{\mu\sigma}\pert(t)} +  \sigma_z \H_\scf\pert(t)\ket{E_{\mu\sigma}\unpert(t)}   , \label{E pert(t) sc}\\
     i \pdv{}{t} \ket{G\pert(t)}&=  \sigma_z \H \unpert\ket{G\pert(t)} - \sigma_z\kket{F_\scf\pert(t)}. 
\end{align}
\end{subequations}
By using perturbation theory calculations (slightly different from the derivation of Sec.\ \eqref{sec: linear response} due to the time  dependence of $\mathcal{H}_\scf(t), \ket{F_\scf(t)}$, see Appendix \ref{app: linear resp SC}}), we get the induced phonon spinors and the induced  phonon condensate in frequency space as 
\begin{subequations} \label{espald schrodinger (1) z SC}
\begin{align}
    \ket{E_{\mu\sigma}\pert(z)}_I &=\sum_{\nu,\sigma'} \sigma'\ket{E\unpert_{\nu{\sigma'}}} \frac{\mel{E\unpert_{\nu\sigma'}}{\H_\scf\pert(z)}{E\unpert_{\mu\sigma}}}{ z - (\sigma' \omega_\nu - \sigma \omega_\mu) }, \label{E(z) pert}\\
   \ket{G\pert(z)}&= -\sum_{\mu,\sigma }\sigma\ket{E_{\mu\sigma}\unpert}\frac{\braket{E\unpert_{\mu\sigma}}{F_\scf\pert 
   (z)}}{ z-\sigma\omega_\mu} \label{G(z) pert}
\end{align}
\end{subequations}

Using Eqs.\ \eqref{espald schrodinger (1) z SC} in Eq.\ \eqref{delta O pert (t) SC} we obtain the linear variation of the observable in frequency space $\delta O\pert(z)$, which we can relate to the linear response function via Eq.\ \eqref{chi = O/s SC},  yielding
\begin{align}
    \chi_{O,H_\text{pert}\pert}(z) &=\frac{\hbar}{2
    }\!\!\sum_{\mu\nu, \sigma\sigma' } \mel{E\unpert_{\nu\sigma'}}{\mathcal{O}\unpert}{E\unpert_{\mu\sigma}} \sigma'\sigma \frac{n(\sigma'\w_\nu) {-} n(\sigma\w_\mu)}{ z -(\sigma\omega_\mu - \sigma' \omega_\nu)} \nonumber \\ 
    &{\times}\mel{E\unpert_{\mu\sigma}}{\H_\scf\pert(z)}{E\unpert_{\nu\sigma'}}\nonumber \\
    &- \hbar\sum_{\mu,\sigma }\sigma\braket{o\unpert}{E_{\mu\sigma}\unpert}\frac{\braket{E\unpert_{\mu\sigma}}{F_\scf\pert(z)}}{ z-\sigma\omega_\mu}   \label{chi 1(z) SC}.
\end{align}
Eq.\ \eqref{chi 1(z) SC} represents one of the main results of this paper. It shows how the response of an anharmonic lattice  is in one-to-one correspondence with the electronic response function. 
As in gKS results, the ionic response function $\chi_{O,H_\text{pert}}\pert(z)$ that describes the variation of the observable $ O$ due to the presence of the external field $ H_\text{pert}$ is given in terms of a bare vertex and a ``screened vertex''. In the two-phonon part, the bare vertex is $\mel{E\unpert_{\nu\sigma'}}{\mathcal{O}\unpert}{E\unpert_{\mu\sigma}}$ while the screened one is $\mel{E\unpert_{\mu\sigma}}{\H_\scf\pert(z)}{E\unpert_{\nu\sigma'}}$. In the one-phonon part,  the bare one-phonon vertex is $\braket{o}{E_{\mu\sigma}\unpert}$ while the self-consistent vertex is $\braket{E\unpert_{\mu\sigma}}{F_\scf\pert(z)}$. 
Below, we discuss how phonon interactions, namely anharmonicity, play the exact same role in the screening of the response as the Hartree-exchange correlation interaction for the electronic response. A recap of the results for linear response for the ionic and electronic case is presented in Table \ref{tab:recap LR}.

The bare and the screened vertices are connected by the SCHA bare two-phonon and one-phonon propagators, { namely 
\begin{align} 
    \chi_{O,H\pert_\text{pert}}(z) &=\frac{\hbar}{2
    }\!\!\sum_{\mu\nu, \sigma\sigma' } \mel{{E}\unpert_{\nu\sigma'}}{\mathcal{O}\unpert}{{E}\unpert_{\mu\sigma}}   {L}^0_{\mu\sigma,\nu\sigma'}(z) \nonumber \\ & \times 
    \mel{{E}\unpert_{\mu\sigma}}{\H_\scf\pert(z)}{{E}\unpert_{\nu\sigma'}} \nonumber\\& - \hbar\sum_{\mu,\sigma }\braket{o\unpert}{{E}_{\mu\sigma}\unpert}{g}^0_{\mu\sigma}(z){\braket{{E}\unpert_{\mu\sigma}}{F_\scf\pert(z)}}  ,
\end{align} 
where}
\begin{subequations}
    \begin{align} 
    L^0_{\mu\sigma,\nu\sigma'}(z) &= \sigma\sigma'\frac{n(\sigma'\w_\nu) {-} n(\sigma\w_\mu)}{ z -(\sigma\omega_\mu - \sigma' \omega_\nu)},\\
    g^0_{\mu\sigma}(z) &= \frac{\sigma}{z-\sigma\omega_\mu},  
\end{align}
\end{subequations}
which are the SCHA equivalent of $L^0_{\mu\sigma,\nu\sigma'}(z),  g^0_{\mu\sigma}(z)$ [Eqs.\ \eqref{L0 ss' (z)},\eqref{g0(z)}]. This bears similarities with how in changing the description of the electronic bands (i.e.\ from DFT to DFT+U or screened Hartree-Fock), the bare electron-hole propagator is modified \cite{guandalini2025high}.

\subsection{Interaction kernels: anharmonicity as phonon-phonon screening}
To obtain a self-consistent cycle we have to express the self-consistent Hamiltonian $\H_\scf\pert(z)$ and forces $\ket{F_\scf\pert(z)}$ in terms of $\ket{E\pert_{\mu\sigma}(z)}$ and $\ket{G\pert(z)}$.
In frequency space, Eq.\ \eqref{hscf(t) = hpert + hanh(t)} reads

\begin{subequations} 
    \begin{align}
    \H_\scf\pert(z) &=  \H_\text{pert}\pert +  \H_\text{anh}^{(1)[\varrho(z),G(z)]} \\
    \kket{F_\scf\pert(z)} & = \ket{F_\text{pert}\pert} + \ket{F_\text{anh}^{(1)[\varrho(z),G(z)]}}
\end{align}
\end{subequations}

As in the gKS theory, the linear variation of the self-consistent potential with respect to the dynamical variable (in gKS, the electronic density matrix, here, $\varrho$ and $\ket{G}$) defines an interaction kernel. We thus express 
\begin{subequations}\label{Hanh = kernels}
    \begin{align}
        &{\H_\text{anh}^{(1)[\varrho(z),G(z)]}} =  \overset{(3)}{\mathcal{D}}\ket{G\pert(z)} + \overset{(4)}{\mathcal{D}}:{\varrho\pert(z)}\\
        &\ket{F_\text{anh}^{(1)[\varrho(z),G(z)]}} = \overset{(2)}{\mathcal{F}}\ket{G\pert(z)}+ \overset{(3)}{\mathcal F} :{\varrho\pert(z)}.
    \end{align} 
\end{subequations}
The number above the anharmonic terms represents the number of phonons involved in the scattering. 
{The 4 phonon kernel $\overset{(4)}{\mathcal{D}}$ is a rank 4 square tensor in the augmented space, as it connects the rank 2 tensor $\varrho\pert$ to another rank 2 tensor $\H_\anh\pert$. The 3 phonon kernel $\overset{(3)}{\mathcal{D}}$ ($\overset{(3)}{\mathcal{F}}$) is a rank 3 rectangular tensor in a $2{\times} 1$ ($1{\times}2$) shape in the augmented space as it connects the rank 1 (rank 2) tensor $\ket{G\pert}$ ($\varrho\pert$) to the rank 2 (rank 1) tensor $\H\pert_\anh$  ($\ket{F\pert_\anh}$). The colon symbolizes a rank 2 scalar product, i.e.\ contraction with two pairs of indices in the augmented space, the rightmost two indices of the right for the first term and the leftmost indices of the second operand, i.e.\ 
\begin{equation}
    [\overset{(4)}{\mathcal{A}}: \overset{(2)}{\mathcal{B}}]_{\a\b} = \sum_{\lambda\nu}^{6N_{at}} \overset{(4)}{\mathcal{A}}_{\alpha\beta\lambda\nu} \overset{(2)}{\mathcal{B}}_{\lambda \nu} .
\end{equation}

Eqs.\ \eqref{Hanh = kernels} are general and follow from the linear expansion.  To compute the expression of the anharmonic kernel in TD-SCHA, we have to evaluate the derivatives of Gaussian averages in Eqs.\ \eqref{anharmonic tdscha couplings}. }
We prove in {Appendix \ref{app: tdscha detail}} that the TD-SCHA anharmonic kernels as defined from Eqs.\ \eqref{Hanh = kernels} are { (using compact $3N_\text{at}$ indices $(I,\a){\to}I$ for brevity as in Appendix \ref{app: calcoli 1b liouville} for the indices $I, J, K, L, A, B, C, D$ which now label both atoms and a carterian directions)
\begin{subequations} \label{tdscha D kernels}
    \begin{align}
        \overset{(3)}{{\mathcal{D}}}_{\!\!\!\substack{IJK\\ \sigma_1\sigma_2\sigma_3}} & {=} \sqrt{\hbar} \sum_{ABC}^{3N_\text{at}}{\Lambda}^T_{\mathbf{R}\,IA}   {\Lambda}^T_{\mathbf{R}\,JB}  \overset{(3)}{ D}_{ABC}
        {\Lambda}{}_{\mathbf{R}\, CK} \\
         \overset{(4)}{\mathcal{D}}_{\!\!\!\substack{IJKL\\\sigma_1\sigma_2\sigma_3\sigma_4}} & {=}  \frac{\hbar}{2} \!\!\sum_{ABCD}^{3N_\text{at}} \!\!{\Lambda}^T_{\mathbf{R}\,IA}   {\Lambda}^T_{\mathbf{R}\,JB}  \overset{(4)}{ D}_{ABCD}
        {\Lambda}{}_{\mathbf{R}\, CK} {\Lambda}_{\mathbf{R}\, DL} ,
    \end{align}
\end{subequations}
while the anharmonic forces are 
\begin{subequations} \label{tdscha F kernels}
    \begin{align}
         {\overset{(2)}{\mathcal F}_{I\sigma}} & = 0,\\
         \overset{(3)}{\mathcal F}_{\!\!\!\substack{IJK\\\sigma_1\sigma_2\sigma_3}} &=-\frac{\sqrt{\hbar}}{2} \sum_{ABC}^{3N_\text{at}}{\Lambda}^T_{\mathbf{R}\,IA}\overset{(3)}{ D}_{ABC}{\Lambda}_{\mathbf{R}\,BJ}
        {\Lambda}_{\mathbf{R}\, CK} 
    \end{align}
\end{subequations}
where 3-phonon and 4-phonon SCHA scattering vertices are defined as averages of the higher-order derivatives of the BO potential  \cite{siciliano2023wigner}

\begin{subequations} \label{phi3 phi4}
    \begin{align}
   \overset{(3)}{{D}}_{IJK} &=  \left\langle \frac{\partial^3 V_\text{BO}({\bf R})}{\partial R_{I} \partial R_{J} \partial R_{K}}\right\rangle_{\!\! \rho\unpert}\!\!\!\!,\\
  \overset{(4)}{{D}}_{IJKL} &= \left\langle \frac{\partial^4 V_\text{BO}({\bf R})}{\partial R_{I} \partial R_{J} \partial R_{K}\partial R_{L}}\right\rangle_{\!\! \rho\unpert} ,
    \end{align}
\end{subequations}
and the matrices $\Lambda_{\mathbf{R}}$ are defined in \eqref{lambdas SC}.}
Using Eqs.\ \eqref{Hanh = kernels}, we can express the self-consistent Hamiltonian and forces in terms of the induced phonon density and one-phonon propagator {
\begin{subequations} \label{scf = bare + screening}
    \begin{align}
       \H_\scf\pert(z) &= {\H_\text{pert}\pert} + \overset{(3)}{{\mathcal{D}}}\ket{G\pert(z)} 
        {+}\overset{(4)}{\mathcal{D}} :{\varrho\pert(z)}\\ 
        \ket{F_\scf\pert(z)} &=\ket{F_\text{pert}\pert} +\overset{(3)}{\mathcal F} : {\varrho\pert(z)}
    \end{align}
\end{subequations}}
Eqs.\ \eqref{scf = bare + screening} are the lattice dynamical equivalent of the first order expansion of the self-consistent 
electronic potential in the electronic density matrix. In particular, the anharmonic couplings are the lattice equivalent of the Hartree-exchange correlation (Hxc) kernel. In this sense, anharmonicity acts on the phonon response by dressing the vertices; a noninteracting system of phonons would feel the entire external potential $  H\pert_\text{pert}$ in the form of the effective potential and forces $\H_\text{pert}\pert, \ket{F_\text{pert}\pert}$. Instead, the phonon-phonon interaction change the response providing an effective screening of the external potential and forces. In this spirit, Eqs.\ \eqref{scf = bare + screening} show how anharmonicity screens the ionic response exactly as the Coulomb interaction screens the dynamical electronic response.

We also note that since the matrix elements of $\varrho\pert$ in Eqs.\ \eqref{scf = bare + screening} feature the Bose-Einstein distribution factors 
\begin{align} 
   \mel{E_{\a\sigma_1}\unpert}{\varrho\pert(z)}{E_{\b\sigma_2}\unpert} = \sigma_2 n(\sigma_2 \w_\beta) \braket{E\unpert_{\alpha\sigma_1}}{E_{\beta\sigma_2}\pert(z)} \nonumber\\ + \sigma_1 n(\sigma_1 \w_\alpha)\braket{E_{\alpha\sigma_1}\pert(z)}{E_{\beta\sigma_2}\unpert},\label{rho1(z) mel}
\end{align}
anharmonicity is considered at a quantum level. That is, the phonons that are allowed to participate to a scattering event constrained by the conservation of the energy and momentum are populated according to the Bose-Einstein distribution.

{We note that in the TD-SCHA, the anharmonic scattering vertices are pseudospin insensitive. A model with polarized phonon-phonon scattering could be realized in exotic chiral or magnetic systems and represents an interesting possible future development for the theory outlined here.}

{
\begin{table*}[t]
\setlength{\tabcolsep}{3pt}
    \centering
    \begin{tabular}{lcc}
    \toprule
    & \textbf{Phonons }[this work] & \textbf{Electrons}   \\
\midrule
Linear response function
&   $\chi\pert_{O,V_\text{pert}}(z) {=}\frac{\hbar}{2}\Tr[\varrho\pert(z)\mathcal{O}\unpert] {+} \hbar\braket{o}{G\pert(z)}$
& $\chi\pert_{O,V_\text{pert}}(z) {=} \Tr[  O   n\pert(z)]$
\\[1ex]
Self-consistent fields
& $\begin{cases}
     \H_\scf\pert(z) = \H_\text{pert}\pert + \H_\text{anh}^{(1)[\varrho(z),G(z)]}\\
    {F_\scf\pert(z)}  =F_\text{pert}\pert+F_\text{anh}^{(1)[\varrho(z),G(z)]}
\end{cases}$ 
&   
$
  H_{\text{scf}}\pert(z) =   V_\text{pert}\pert +  {H}_{\text{Hxc}}^{[  n\pert(z)]}
$
\\[2ex]
\hline
&
TD-SCHA
&
LDA/GGA + Screened Exchange
\\ 
\hline 
Mean-field interaction &
$\begin{cases}
    \H_\text{anh}^{(1)[\varrho(z),G(z)]} =\overset{(3)}{\mathcal{D}}G\pert(z) +\overset{(4)}{\mathcal{D}}{\varrho\pert}(z)\\
    {F_\text{anh}^{(1)[\varrho(z),G(z)]}}  = \overset{(3)}{\mathcal{F}}\varrho\pert(z)
\end{cases}$ 
&  
$ {H}_{\text{Hxc}}^{(1)[  n(z)]} = 2f_{\text{Hxc}}   n\pert(z) - W   n\pert(z)$
\\[2ex]
\multirow{3}{*}{Interaction kernels} &
$
\overset{(3)}{\mathcal{D}} = 
\sqrt{\hbar}\Lambda_{\mathbf{R}}^T\Lambda_{\mathbf{R}}^T\left\langle\frac{\partial^3V_{BO}}{\partial R^3} \right\rangle_{ \rho\unpert}\!\Lambda_{\mathbf{R}}
$ 
&  
$f_\text{Hxc} = \frac{e^2}{|  r-  r'|} + f_{\text{xc}}^{\text{\tiny LDA/GGA}},\,$
\\ 
 &
$ \overset{(4)}{\mathcal{D}} = \tfrac{\hbar}{2}\Lambda_{\mathbf{R}}^T\Lambda_{\mathbf{R}}^T\left\langle\frac{\partial^4V_{BO}}{\partial R^4} \right\rangle_{ \rho\unpert}\Lambda_{\mathbf{R}}\Lambda_{\mathbf{R}}
$ 
&  
$ W = \frac{e^2\epsilon^{-1}(r,r'') }{|  r''-  r'|}$
\\ 
 &
$ 
\overset{(3)}{\mathcal{F}} = 
-\tfrac{\sqrt{\hbar}}{2}\Lambda_{\mathbf{R}}^T\left\langle\frac{\partial^3V_{BO}}{\partial R^3} \right\rangle_{ \rho\unpert}\!\Lambda_{\mathbf{R}}\Lambda_{\mathbf{R}}
$ 
&  
\\ 
\bottomrule
    \end{tabular}
\caption{ One-to-one correspondence between linear response equations for the effective single particle picture for anharmonic lattice dynamics (ESPALD) for phonons  and generalized Kohn-Sham (gKS) theory for electrons.
As in the electronic case, the linear response function for a lattice of atoms is obtained by tracing an operator $  O$ on the induced single-particle density matrix $\varrho\pert(z)$ [Eq.\ \eqref{rho(z) pert}]. In the case of phonons, another degree of freedom responds to external fields: $\ket{G\pert(z)}$. $\varrho\pert(z)$ and $\ket{G\pert(z)}$ carry information respectively of the dynamics of atomic bonds and atomic positions. The dynamics of $\varrho\pert(z)$ and $\ket{G\pert(z)}$ is regulated by a self-consistent Hamiltonian $\H_\scf\pert(z)$ and a self-consistent force ${F_\text{anh}^{(1)[\varrho(z),G(z)]}} $, as the electrons in gKS theory move in a self-consistent field potential $  H_{\text{scf}}\pert(z)$.
For an explicit expression of the mean-field interaction, we have to specify the approximation used. 
In the TD-SCHA, the anharmonic interaction {--- encoded in the kernels $\overset{(3)}{\mathcal{D}},\overset{(4)}{\mathcal{D}}, \overset{(3)}{\mathcal{F}} $ defined in Eqs.\ \eqref{tdscha D kernels}-\eqref{tdscha F kernels} ---} couples $\varrho\pert(z)$ and $\ket{G\pert(z)}$ via 3 and 4 phonon scattering vertices $\overset{(3)}{{D}},\overset{(4)}{D}$, obtained from higher order derivatives of the BO potential [Eq.\ \eqref{phi3 phi4}]. This mirrors the linear response in gKS, where the kernel of interaction is obtained via functional derivatives of the induced potentials, and depends on the specifics of the approximation used. For the sake of comparison, we report the kernel of interaction in LDA/GGA + screened exchange.
}
    \label{tab:recap LR}
\end{table*}
}

\subsection{Sternheimer formulation of lattice response} \label{sec: sternheimer}
By collecting the results we got so far, we express the equations for the linear response of an anharmonic lattice as a self-consistent set of equations:{
\begin{widetext}
\begin{subequations} \label{sternheimer tdscha}
    \begin{align}
        \ket{E_{\mu\sigma}\pert(z)}_I &=\sum_{\nu,\sigma'} \sigma'\ket{E\unpert_{\nu{\sigma'}}} \frac{\mel{E\unpert_{\nu\sigma'}}{\H_\scf\pert(z)}{E\unpert_{\mu\sigma}}}{ z - (\sigma' \omega_\nu - \sigma \omega_\mu) }, \label{E1 pert tdscha}\\
\ket{G\pert(z)}&= -\sum_{\mu,\sigma }\sigma\ket{E_{\mu\sigma}\unpert}\frac{\braket{E\unpert_{\mu\sigma}}{F_\scf\pert(z)}}{ z-\sigma\omega_\mu} ; \label{G1 pert tdscha}\\
{\H_\scf\pert(z)} &=   {\H_\text{pert}\pert} +  {\H_\text{anh}^{(1)[\varrho(z),G(z)]}},  \label{Hscf 1 stern tdscha}\\
  \ket{F_\scf\pert(z)} &= \ket{F\pert_\text{pert}} + \ket{F_\text{anh}^{(1)[\varrho(z),G(z)]}}; \label{Fscf 1 stern tdscha}\\
{\H_\text{anh}^{(1)[\varrho(z),G(z)]}} &= \overset{(3)}{\mathcal{D}}\ket{G\pert(z)} +\overset{(4)}{\mathcal{D}} :\sum_{\theta\sigma_1} \sigma_1 n(\sigma_1 \w_\theta) \biggl[\ddyad{E\unpert_{\theta\sigma_1}}{E_{\theta\sigma_1}\pert(z)}_I  + \kket{E_{\theta\sigma_1}\pert(z)}_I\bbra{E_{\theta\sigma_1}\unpert}\biggr] \label{Kanh tdscha},\\
    \ket{F_\text{anh}^{(1)[\varrho(z),G(z)]}} &=\overset{(3)}{\mathcal{F}} :\sum_{\theta\sigma_1} \sigma_1 n(\sigma_1 \w_\theta) \biggl[\ddyad{E\unpert_{\theta\sigma_1}}{E_{\theta\sigma_1}\pert(z)}_I  + \kket{E_{\theta\sigma_1}\pert(z)}_I\bbra{E_{\theta\sigma_1}\unpert}\biggr]. \label{Fanh tdscha}
    \end{align}
\end{subequations}
\end{widetext}}

Eqs.\ \eqref{sternheimer tdscha} represent a reformulation of the linear response equations for the TD-SCHA formally equivalent to a Sternheimer self-consistent cycle used for the calculation of the electronic response in gKS, where the three/four-phonon scattering vertices provide a source of screening for the ionic response to an external perturbation.
A recap of the correspondence between electronic gKS theory and TD-SCHA linear response is in Table \ref{tab:recap LR}.

This analogy allows for the application of theoretical and numerical methods developed in the past decades for the electronic response to the solution of Eqs.\ \eqref{sternheimer tdscha} for the ionic response. At variance with the electronic case where the response is completely determined once the induced one-electron density is known, the phonon response needs another ingredient besides the single-particle density $\varrho(t)$, namely the phonon condensate $\ket{G(t)}$. For this reason, the number of equations of the ESPALD is doubled with respect to the ones of gKS.

As in common algorithms of gKS approaches, Eqs.\ \eqref{sternheimer tdscha} can be solved with an iterative scheme. A standard algorithm is 
\begin{enumerate}
    \item At the first step, initialize the induced phonon spinors $\ket{E\pert_{\mu\sigma}}$ and phonon condensate $\ket{G\pert}$ to zero, or use the bare response obtained by setting the anharmonic kernels to zero. 
    \item Build the self-consistent potential and forces in Eqs.\ \eqref{Hscf 1 stern tdscha}-\eqref{Fscf 1 stern tdscha}  by combining the anharmonic couplings $\overset{(3)}{\mathcal{D}},\overset{(4)}{\mathcal{D}} $ to $\ket{E\pert_{\mu\sigma}}$ and $\ket{G\pert}$ via Eqs.\ \eqref{Kanh tdscha}-\eqref{Fanh tdscha},
    \item Update the induced phonon spinors $\ket{E\pert_{\mu\sigma}}$ and one-phonon propagator $\ket{G\pert}$ via Eqs.\ \eqref{E1 pert tdscha}-\eqref{G1 pert tdscha},
    \item Check for self-consistency and iterate.
\end{enumerate}
When self-consistency is met, the response function is obtained via Eq.\ \eqref{chi 1(z) SC}. 

Since the self-consistent cycle Eq.\ \eqref{sternheimer tdscha} is closed for $\ket{E\pert_{\mu\sigma}}$ and $\ket{G\pert}$, the linear response function in the ESPALD is completely determined by $\ket{E\pert_{\mu\sigma}}$ and $\ket{G\pert}$, i.e.\ it is a (linear) functional of the induced phonon spinors and phonon condensate 
\begin{equation}
    \chi\pert_{O,V_\text{pert}}(z) = \chi\pert_{O,V_\text{pert}}\left[\left\{\ket{E\pert_{\mu\sigma}(z)}\right\},\ket{G\pert(z)}\right].
\end{equation}
{As per the electronic case, a variational formulation of the response functional can be realized for the lattice dynamical case, expressing the response functional as a quadratic functional of the induced phonon quantities \cite{calandra2010adiabatic,caldarelli2025variational,berges2023phonon,stefanucci2025exact,gonze1992dielectric}. If certain conditions on the anharmonic kernels are met, a generalized Fermi-Golden rule can further be formulated to express the imaginary part of the ionic response function as a combination of a Dirac delta and a square modulus of a single-particle potential.\cite{caldarelli2025variational}}

Note that in Eqs.\ \eqref{sternheimer tdscha}, we have eliminated the dependence on the one-body phonon density $\varrho\pert(z)$  using Eq.\ \eqref{rho1(z) mel}. An alternative formulation of anharmonic lattice dynamical response equations can be expressed in terms of the single-particle density $\varrho\pert(z)$ by leveraging the one-to-one correspondence between $\varrho\pert(z)$  and the phonon spinors $\ket{E\pert_{\mu\sigma}(z)}$ (Eq.\ \eqref{rho1(z) mel}).

In Appendix \ref{app: tdscha detail}, we show how the system of self-consistent equations Eqs.\ \eqref{sternheimer tdscha} recovers the anharmonic phonon propagators of previous formulation of the TD-SCHA.

\section{Conclusions} \label{sec: conclusions}
In this work, we presented an effective single-particle picture for anharmonic lattice dynamics (ESPALD) in one-to-one correspondence with time-dependent generalized Kohn-Sham (gKS) theories developed for the dynamics of electronic systems. 

In this formalism, we mapped the self-consistent mean-field Hamiltonian $  H_\scf(t)$ used to approximate the Born-Oppenheimer Hamiltonian $  H_{\text{BO}}(t)$ into a single-particle Hamiltonian $\H_\scf(t)$ and forces $\ket{F_\scf(t)}$. A self-consistent relation between the mean-field $  H_\scf(t)$ and the approximated density matrix $  \rho(t)$ exists in the TD-SCHA, that we adopt as our fundamental framework.
In this single-particle mapping, the many-body Liouville equation between $   \rho(t) $ and $H_\scf(t)$ is replaced by two (pseudo)unitary wave equations regulated by $\H_\scf(t)$ and $\ket{F_\scf(t)}$, which are the vibrational analog of the time-dependent KS Hamiltonian $  H_\text{KS}(t)$ for electron dynamics. 

The mathematical structure of the mapping allowed us to develop an algebraic structure based on Dirac bra-ket notation. This construction directly connects the theory developed here to gKS electronic approaches, and greatly simplifies the analytical calculations.
The objects playing the role of evolving KS states are the phonon spinors $\ket{E_{\mu\sigma}(t)}$ and the one-phonon propagator $\ket{G(t)}$, respectively encoding the evolution of the atomic bonds and average positions. The vectorial space of the ESPALD is doubled with respect to the vectorial space of the phonon eigenvectors. This mimics the procedure of the BdG theory of superconductivity \cite{nambu1960quasi,degennes1966superconductivity}, where the space is doubled to distinguish electron and hole dynamics. The vibrational equivalent of electrons/holes of BdG theory are the phonon polarization vectors with positive and negative frequencies, distinct by their phonon pseudospin $\sigma =\pm$. Anomalous terms $\bf{\Delta}$ in the Hamiltonian coupling $\sigma {=}\pm$ phonon spinors naturally stem from anharmonicity.

While the equilibrium single-particle Hamiltonian $\H\unpert$ is doubly degenerate in the phonon pseudospin, such degeneracy can be broken by e.g.\ the introduction of ${\bf R P }$ couplings in the equilibrium Hamiltonian. Such terms are typically provided by molecular Berry phases \cite{zhang2022lattice,zhang2026comprehensive,ren2024adiabatic} or Lorentz forces from magnetic fields.  In fact, the quadratic mean-field Hamiltonian can be straightforwardly generalized  to include momentum-dependent external perturbations. Thus, this work paves the way to a comprehensive treatment of phonon-phonon interaction and effects that break time-reversal symmetry, 
possibly encompassing both phonon anharmonicity and phonon chirality. Our approach opens possibilities for first-principles approaches to phenomena associated with phonon angular momentum, like the phonon Einstein de Haas effect \cite{zhang2014angular}, the phonon Hall effect \cite{strohm2005phenomenological,sun2021phonon,chen2022large,li2020phonon} and vibrational circular dichroism \cite{vcd1987,vcd2016,vcd2004}.
Even in the case of pseudospin degeneracy of the phonon spectrum, we showed how a response function such as  thermal conductivity is highly sensitive to the phonon pseudospin. The coherence between equal(opposite) pseudospin states give rise to the resonant(antiresonant) thermal conductivity. Our preliminary result obtained in the HA unveils how careful consideration of phonon pseudospin is needed when tackling the phenomenon of coherent thermal transport, of pivotal important for reproducing  transport coefficients of strongly anharmonic crystals and glasses.

We have shown that the dynamical response of the lattice can be formulated as a functional of the single-particle density $\varrho(t)$ and the one-phonon propagator $\ket{G(t)}$. This exact dependence provides insights for the evaluation of higher-order phonon susceptibilities. Borrowing procedures already developed for the calculation of nonlinear electronic susceptibilities \cite{aversa1995nonlinear, dal1996density}, future development of this work can be made towards nonlinear phononics \cite{forst2011nonlinear, mankowsky2016non}, offering a robust theoretical framework for the calculation of second harmonic generation  or sum-frequency generation \cite{leisegang2018giant, shen1989surface, juraschek2018sum}, without relying on phenomenological models for anharmonicity.

In the linear response regime, we obtain the anharmonic response in terms of matrix elements of the external perturbations on the phonon spinors basis. This representation allows for the stipulation of phonon selection rules akin to those governing electronic optical transitions, allowing for the a priori identification of inactive vibrational states not participating in the response. 
Furthermore, the introduction of self-consistent single-particle Hamiltonian and forces allows us to cast the fully interacting response into a formally noninteracting susceptibility,  featuring effective potential and forces. Anharmonicity physically manifests as an electronic-like screening effect. We showed how in the linear response regime, phonon interactions are encoded in anharmonic kernels that plays the exact same role for vibrational excitations as the Hartree-exchange-correlation kernel does for interacting electrons.  
Developing on this equivalence, we formulated a self-consistent cycle for induced dynamical objects $\ket{E\pert_{\mu\sigma}}$ and $\ket{G\pert}$, akin to the Sternheimer cycle used e.g.\ in DFPT to compute the induced electronic density.
Future numerical implementation could benefit from this analogy in readapting routines developed in electronic response computational methods to solve the response of an anharmonic lattice.

In this regard, we hope that our framework may serve as a Rosetta stone connecting the extensive research efforts in electronic response theory with those in lattice dynamics.
For researchers studying the anharmonic motion of ions, it provides a direct pathway to harvest decades of advanced methodological know-how, ranging from $\mathbf{k}$-point parallelization or interpolation techniques and the application of symmetry operations to optimize numerical calculation, to the theoretical advancement done to refine the electronic mean-field response with many-body perturbation theory. Conversely, it allows electronic structure experts to apply their familiar language of density matrices, effective Hamiltonians, and screening kernels directly to the problem of anharmonic lattice dynamics.

\section*{Acknowledgements}
G.C.\ acknowledges fruitful discussions with Dr.\ Jacopo Fiore,  Dr.\ Francesco Macheda, and Dr.\ Paolo Fachin.  We acknowledge the MORE-TEM ERC-SYN project, grant agreement no.\ 951215.

\appendix
\onecolumngrid

\section{Single-particle particle mapping of the many-body Liouville equation} \label{app: calcoli 1b liouville}
In this Appendix, we show how to derive the system of dynamical equations in Eqs.\ \eqref{1b mapping}, from the Liouville equation for the many-body density matrix Eq.\ \eqref{HA liouville} with a quadratic Hamiltonian as in Eq.\ \eqref{H(t) HA}.

Here, for brevity, we combine the atomic and the cartesian indices in a single one, that is $(I,\a){\to} (I)$.
{Let us define the ``unconnected'' density matrix 
\begin{equation} \label{rho (t) 2x2 matrix app}
     \widetilde \varrho(t) = \mqty[ \mtrx{\widetilde \varrho}{}^\text{n}(t)& \mtrx{ \widetilde \varrho}{}^\text{an}(t) \\
   \mtrx{ \widetilde \varrho}{}^\text{an}(t)^*& \mathbb{I}+ [\mtrx{\widetilde \varrho}{}^\text{n}(t)]{}^T   ]
\end{equation}
where respectively, the normal and anomalous densities are defined as
\begin{subequations} \label{rho(t) mels app}
    \begin{align}
        \widetilde \varrho_{I,J}^\text{n}(t) &= \langle {\tenscomp {a}}^\dagger_{J} {\tenscomp{a}}_{I}\rangle_{ \rho(t)}, \\
         \widetilde \varrho_{I,J}^\text{an}(t) &= \langle {\tenscomp{a}}_{J} {\tenscomp{a}}_{I}\rangle_{ \rho(t)} ,
    \end{align}
\end{subequations}
}
Eqs.\ \eqref{1b mapping} are obtained using the Liouville equation Eq.\ \eqref{HA liouville} on the time-derivative of the elements of the single-particle density matrix $\widetilde \varrho(t)$ in Eqs.\ \eqref{rho (t) 2x2 matrix app}-\eqref{rho(t) mels app}, that is 
\begin{subequations} \label{drho/dt explicit}
\begin{align}
     i\hbar \pdv{}{t}  \widetilde \varrho_{AB}^\text{n}(t)& = i\hbar\Tr[  {\tenscomp{a}}_B\herm  {\tenscomp{a}}_A  \pdv{}{t}   \rho(t) ]  = \langle [\tenscomp{a}_J\herm \tenscomp{a}_I ,   H(t)] \rangle_{ \rho(t)}, \\ 
      i\hbar \pdv{}{t}  \widetilde \varrho_{AB}^\text{an}(t)& = i\hbar\Tr[  {\tenscomp{a}}_B  {\tenscomp{a}}_A  \pdv{}{t}   \rho(t) ]  = \langle [ {\tenscomp{a}}_B  {\tenscomp{a}}_A ,   H(t)] \rangle_{ \rho(t)} ,\\
        i\hbar \pdv{}{t}  \widetilde \varrho_{AB}^\text{an}(t)^*& = i\hbar\Tr[  {\tenscomp{a}}_B\herm  {\tenscomp{a}}_A\herm  \pdv{}{t}   \rho(t) ]  = \langle [ {\tenscomp{a}}_B\herm {\tenscomp{a}}_A \herm,   H(t)] \rangle_{ \rho(t)} ,\\
        i\hbar \pdv{}{t} [\widetilde \varrho_{AB}^\text{n}(t)^T+ \delta_{AB} ]& = i\hbar\Tr[  {\tenscomp{a}}_B  {\tenscomp{a}}_A\herm  \pdv{}{t}   \rho(t) ]  = \langle [ {\tenscomp{a}}_B  {\tenscomp{a}}_A \herm,   H(t)] \rangle_{ \rho(t)} ,
\end{align}
\end{subequations}
where we have used the cyclic property of the trace. For brevity, we have dropped self-consistent dependence of the Hamiltonian from the density matrix.
With the same procedure, the elements of the phonon condensate $\ket{G(t)}$ from Eq.\ \eqref{G = gg}-\eqref{g(t)} are
\begin{subequations} \label{dg/dt explicit}
    \begin{align}
         i\hbar \pdv{}{t}  g_{A}(t)&= i\hbar \Tr[ {\tenscomp{a}}_A \pdv{}{t}  \rho(t)] = \langle [ {\tenscomp{a}}_A,   H(t)] \rangle_{ \rho(t)},\\
           i\hbar \pdv{}{t}  g_{I}(t)^*&= i\hbar \Tr[ {\tenscomp{a}}_A \herm\pdv{}{t}  \rho(t)] = \langle [ {\tenscomp{a}}_A\herm,   H(t)] \rangle_{ \rho(t)}.
    \end{align}
\end{subequations}
Using expression of the Hamiltonian in the cartesian boson operators [Eq.\ \eqref{H 2x2}-\eqref{H , F single particle}], we get the following commutation relations for  products of cartesian boson operators
\begin{subequations} \label{commutators}
    \begin{align}
        [ {\tenscomp{a}}_B\herm  {\tenscomp{a}}_A ,   H(t)] &=\text{h}_{AJ}(t) {\tenscomp{a}}\herm_J  {\tenscomp{a}}_B- {\tenscomp{a}}\herm_J {\tenscomp{a}}_A\text{h}_{JB}(t)+\Delta_{AJ}(t)^* {\tenscomp{a}}_B\herm {\tenscomp{a}}_J\herm - {\tenscomp{a}}_A {\tenscomp{a}}_{J}\Delta_{JB}(t) - \text{f}_A(t)  {\tenscomp{a}}_B\herm +  {\tenscomp{a}}_A\text{f}_B^*(t),\\
        [ {\tenscomp{a}}_B  {\tenscomp{a}}_A ,   H(t)] & = {\tenscomp{a}}_A {\tenscomp{a}}_J \text{h}_{JB}(t) +  \text{h}_{AJ}(t) {\tenscomp{a}}_J {\tenscomp{a}}_B +  {\tenscomp{a}}\herm_{J} {\tenscomp{a}}_{A}\Delta_{JB}(t)^* + \Delta_{AJ}(t)^* {\tenscomp{a}}\herm_{J} {\tenscomp{a}}_{B} + \Delta_{AB}(t)^* -\text{f}_{A}(t)  {\tenscomp{a}}_B  - {\tenscomp{a}}_A\text{f}_{B}(t),\\
         [ {\tenscomp{a}}_B\herm  {\tenscomp{a}}_A\herm ,   H(t)] & =- {\tenscomp{a}}_A\herm {\tenscomp{a}}_J\herm \text{h}_{JB}(t) - \text{h}_{AJ}(t) {\tenscomp{a}}_J\herm {\tenscomp{a}}_B\herm  - {\tenscomp{a}}\herm_{J} {\tenscomp{a}}_{A}\Delta_{JB}(t) - \Delta_{AJ}(t) {\tenscomp{a}}\herm_{J} {\tenscomp{a}}_{B} - \Delta_{AB}(t) +\text{f}^*_{A}(t)  {\tenscomp{a}}_B \herm +  {\tenscomp{a}}_A\herm\text{f}^*_{B}(t),\\
         [ {\tenscomp{a}}_B  {\tenscomp{a}}_A\herm ,   H(t)] & =  {\tenscomp{a}}_J {\tenscomp{a}}_A\herm \text{h}_{JB}(t)- \text{h}_{AJ}(t)  {\tenscomp{a}}_B {\tenscomp{a}}_J\herm +  {\tenscomp{a}}_A\herm {\tenscomp{a}}_J\herm \Delta_{BJ}(t)^*-\Delta_{AJ}(t) {\tenscomp{a}}_B {\tenscomp{a}}_{J}  -   {\tenscomp{a}}_A\herm\text{f}_B(t) + \text{f}_A^*(t) {\tenscomp{a}}_B , 
    \end{align}
\end{subequations}
where we are using an Einstein convention for summation on repeated indices for brevity. Similarly, for the commutators of the cartesian boson operators, we get
\begin{subequations} \label{1b commutators}
    \begin{align}
        [ {\tenscomp{a}}_A,   H(t) ]& = \text{h}_{AJ}(t) {\tenscomp{a}}_J(t) + \Delta_{AJ}(t)^* {\tenscomp{a}}_J \herm - \text{f}_A(t) ,\\
         [ {\tenscomp{a}}_A\herm,   H(t) ]& = -\text{h}_{AJ}(t) {\tenscomp{a}}_J \herm- \Delta_{AJ}(t) {\tenscomp{a}}_J  +\text{f}_A^*(t).
    \end{align}
\end{subequations}
Now using Eqs.\ \eqref{commutators} in Eqs.\ \eqref{drho/dt explicit} we get in matrix notation 
\begin{subequations} \label{eq motion drho/dt}
    \begin{align}
        i\hbar \frac{\partial}{\partial t} \mtrx{\widetilde \varrho}{}^\text{n} (t) &= \mtrx{\text{h}}(t)  \mtrx{\widetilde \varrho}{}^\text{n} (t) -  \mtrx{\widetilde \varrho}{}^\text{n} (t)\mtrx{\text{h}}(t) + \mtrx{\Delta}(t)^* \mtrx{\widetilde \varrho}{}^{\text{an}}(t)^* - \mtrx{\widetilde \varrho}{}^{\text{an}}(t)\mtrx{\Delta}(t) - {\bf f}(t) {\bf g}(t)\herm + {\bf g}(t) {\bf f}(t)\herm, \\
        i\hbar \frac{\partial}{\partial t} \mtrx{\widetilde \varrho}{}^\text{an} (t) &= \mtrx{\text{h}}(t)  \mtrx{\widetilde \varrho}{}^\text{an} (t) +  \mtrx{\widetilde \varrho}{}^\text{an} (t)\mtrx{\text{h}}(t) +  \mtrx{\widetilde \varrho}{}^{\text{n}}(t)\mtrx{\Delta}(t)^* + \mtrx{\Delta}(t) ^* [\mtrx{\widetilde \varrho}{}^{\text{n}}(t)^T + \mathbb{I}]- {\bf f}(t) {\bf g}(t)^T - {\bf g}(t) {\bf f}(t)^T, \\
        i\hbar \frac{\partial}{\partial t} \mtrx{\widetilde \varrho}{}^\text{an} (t)^* &= -\mtrx{\text{h}}(t)  \mtrx{\widetilde \varrho}{}^\text{an} (t)^* -  \mtrx{\widetilde \varrho}{}^\text{an} (t)^*\mtrx{\text{h}}(t) - \mtrx{\Delta}{}(t)\mtrx{\widetilde \varrho}{}^{\text{n}}(t)- \mtrx{\Delta}(t)  [\mtrx{\widetilde \varrho}{}^{\text{n}}(t)^T + \mathbb{I}]+ {\bf f}(t)^* {\bf g}(t)\herm + {\bf g}(t)^* {\bf f}(t)\herm, \\
        i\hbar \frac{\partial}{\partial t}[ \mtrx{\widetilde \varrho}{}^\text{n} (t)^T +\mathbb{I}] &= -\mtrx{\text{h}}(t)  \mtrx{\widetilde \varrho}{}^\text{n} (t)^T +\mtrx{\widetilde \varrho}{}^\text{n} (t)\mtrx{\text{h}}(t) + \mtrx{\Delta}(t)^*\mtrx{\widetilde \varrho}{}^{\text{an}}(t)^*  - \mtrx{\widetilde \varrho}{}^{\text{an}}(t)\mtrx{\Delta}(t)  - {\bf g}(t)^*{\bf f}(t)^T  +  {\bf f}(t)^*{\bf g}(t)^T ,
    \end{align}
\end{subequations}
and similarly, the equations for the one-phonon propagator are obtained using \eqref{1b commutators} in \eqref{dg/dt explicit} 
\begin{subequations} \label{eq motion dg/dt}
    \begin{align}
          i\hbar \frac{\partial}{\partial t}  {\bf g}(t)&= \mtrx{\text{h}}(t) {\bf g}(t) + \mtrx{\Delta}(t)^*{\bf g}(t) ^*- {\bf f}(t),\\
           i\hbar \frac{\partial}{\partial t}  {\bf g}(t)^*&= -\mtrx{\text{h}}(t) {\bf g}(t)^* - \mtrx{\Delta}(t){\bf g}(t) + {\bf f}^*(t).
    \end{align}
\end{subequations}
Before going further, it is worth mentioning that among Eqs.\ \eqref{eq motion drho/dt}-\eqref{eq motion dg/dt}, the first two equations in Eqs. \eqref{eq motion drho/dt} and the first equation in \eqref{eq motion dg/dt} are independent and are sufficient to solve lattice dynamics. However, using half of these equation hides the (pseudo)unitary nature of the dynamics, and forbids a complete single-particle mapping in one to one correspondence with electronic dynamics.

Eqs.\ \eqref{eq motion drho/dt} can be rearranged in $2{\times}2$ shape as 
\begin{align} \label{d rho/dt =  2x2}
    i\hbar \frac{\partial}{\partial t}  \mqty[ \mtrx{\widetilde \varrho}{}^{\text{n}}(t) & \mtrx{\widetilde \varrho}{}^{\text{an}}(t) \\ \mtrx{\widetilde \varrho}{}^{\text{an}}(t)^* & \mtrx{\widetilde \varrho}{}^{\text{n}}(t)^T + \mathbb{I}] =  &\mqty[\mtrx{\text{h}}(t) & \mtrx{\Delta}(t)^* \\ -\mtrx{\Delta}(t)  & - \mtrx{\text{h}}(t)] \mqty[ \mtrx{\widetilde \varrho}{}^{\text{n}}(t) & \mtrx{\widetilde \varrho}{}^{\text{an}}(t) \\ \mtrx{\widetilde \varrho}{}^{\text{an}}(t)^* & \mtrx{\widetilde \varrho}{}^{\text{n}}(t)^T + \mathbb{I}] -  \mqty[ \mtrx{\widetilde \varrho}{}^{\text{n}}(t) & \mtrx{\widetilde \varrho}{}^{\text{an}}(t) \\ \mtrx{\widetilde \varrho}{}^{\text{an}}(t)^* & \mtrx{\widetilde \varrho}{}^{\text{n}}(t)^T + \mathbb{I}
    ]\mqty[\mtrx{\text{h}}(t) & -\mtrx{\Delta}(t)^* \\ \mtrx{\Delta}(t)  & - \mtrx{\text{h}}(t)]\nonumber \\ 
    & \nonumber- \mqty[ {\bf f}(t) {\bf g}(t)\herm & {\bf f}(t) {\bf g}(t)^T  \\  -{\bf f}(t)^* {\bf g}(t)\herm  & - {\bf f}(t)^* {\bf g}(t)^T] + \mqty[{\bf g}(t) {\bf f}(t)\herm & - {\bf g}(t) {\bf f}(t)^T  \\ {\bf g}(t) ^*{\bf f}(t)\herm &  -{\bf g}(t)^* {\bf f}(t)^T ] 
\end{align}
{which can be recast as  ~\cite{note:sigma-z-action}
\begin{subequations} 
    \begin{align}
        i\hbar \pdv{\widetilde \varrho(t)}{t}  &= \hbar[\sigma_z \H(t),\widetilde \varrho(t)]_\dagger -\hbar\sigma_z  \kket{ F(t) }\bra{G(t)} +   \hbar\ket{G(t) } \bbra{{F(t)}}  \sigma_z , \\
        i\hbar \pdv{}{t} \ket{G(t)}&= \hbar\sigma_z \H(t)\ket{G(t)} - \hbar\sigma_z\kket{F(t)} .
    \end{align}
\end{subequations}
To obtain the equation for the single-particle density matrix, we express the relation between the unconnected density matrix \eqref{rho (t) 2x2 matrix app} and the density matrix $ \varrho(t)$ defined in Eq.\ \eqref{rhoC elements} as 
\begin{equation} \label{rhoc = rho - GG}
\widetilde \varrho(t) =  \varrho(t) + \dyad{G(t)}{G(t)}
\end{equation}
we apply the equation of motion for $\widetilde \varrho(t)$ and $\ket{G(t)}$ directly on the definition [Eq.\ \eqref{rhoc = rho - GG}] of $ \varrho_C(t)$
\begin{align}
   i\hbar \pdv{}{t} \varrho(t) &= i\hbar \pdv{}{t}\widetilde \varrho(t) - i\hbar \pdv{\ket{G(t)}}{t}\bra{G(t)} - i\hbar \ket{G(t)}\pdv{\bra{G(t)}}{t} \nonumber \\ 
   &= \hbar[\sigma_z\H(t),\widetilde \varrho(t)]_\dagger - \hbar\sigma_z \H(t)\ket{G(t)}\bra{G(t)} +  \hbar\ket{G(t)}\bra{G(t)}  \H(t)\sigma_z  \nonumber \\
   &= \hbar[\sigma_z\H(t),\widetilde \varrho(t) -\dyad{G(t)}{{G}(t)}]_\dagger 
\end{align}
which is Eq.\ \eqref{rhoC liouville}.
}

\section{Equations of motion in the position-momentum basis} \label{app: eom RP}
In this Appendix, we show how to derive the system of dynamical equations for the atomic position and momentum moments. We employ the same methodology used for the Cartesian boson operators, starting from the Liouville equation for the many-body density matrix $ {\rho}(t)$. This procedure yields the equations of motion for the Gaussian parameters obtained in Ref.\ \cite{siciliano2023wigner} with a Gaussian approximation for the many-body density matrix. 

We consider a generic quadratic Hamiltonian in the standard cartesian basis 
\begin{equation} 
     {H}(t) = \frac{1}{2} \sum_{A,B}  {P}_A M^{-1}_{AB}  {P}_B + \frac{1}{2} \sum_{A,B}  {R}_A \Phi_{AB}(t)  {R}_B - \sum_A F_A(t)  {R}_A.
\end{equation}
The time dependence of the inverse-mass matrix is dropped since it was neglected in Ref.\ \cite{siciliano2023wigner}.
Applying the Liouville equation to the expectation value of a generic observable $ {O}$, we have $i\hbar \pdv{}{t} \langle  {O} \rangle_{ \rho(t)} = \langle [ {O},  {H}(t)] \rangle_{ \rho(t)}$  by using cyclic property of the trace.
For the first moments $ {R}_A$ and $ {P}_A$, evaluating the commutators yields
\begin{subequations} \label{eq:eom_1st_moments_explicit}
    \begin{align}
        \pdv{}{t} \langle  {R}_A \rangle_{ \rho(t)} &= \frac{1}{i\hbar} \langle [ {R}_A,  {H}(t)] \rangle_{ \rho(t)} = \sum_B M^{-1}_{AB} \langle  {P}_B \rangle_{ \rho(t)}, \\
        \pdv{}{t} \langle  {P}_A \rangle_{ \rho(t)} &= \frac{1}{i\hbar} \langle [ {P}_A,  {H}(t)] \rangle_{ \rho(t)} = -\sum_B \Phi_{AB}(t) \langle  {R}_B \rangle_{ \rho(t)} + F_A(t).
    \end{align}
\end{subequations}
where we used the canonical commutation relations $[ {R}_A,  {P}_B] = i\hbar \delta_{AB}$, $[ {R}_A,  {R}_B] = 0$, and $[ {P}_A,  {P}_B] = 0$.
Similarly, for the second-order correlators we get
\begin{subequations} \label{eq:eom_2nd_moments_uncentered}
    \begin{align}
        \pdv{}{t} \langle  {R}_A  {R}_B \rangle_{ \rho(t)} =& \sum_C \Big( \langle  {R}_A  {P}_C \rangle_{ \rho(t)} M^{-1}_{CB} + M^{-1}_{AC} \langle  {P}_C  {R}_B \rangle_{ \rho(t)} \Big), \\
        \pdv{}{t} \langle  {P}_A  {P}_B \rangle_{ \rho(t)} =& -\sum_C \Big( \langle  {P}_A  {R}_C \rangle_{ \rho(t)} \Phi_{CB}(t) + \Phi_{AC}(t) \langle  {R}_C  {P}_B \rangle_{ \rho(t)} \Big) \nonumber \\
        &+ \langle  {P}_A \rangle_{ \rho(t)} F_B(t) + F_A(t) \langle  {P}_B \rangle_{ \rho(t)}, \\
        \pdv{}{t} \langle  {R}_A  {P}_B \rangle_{ \rho(t)} =& -\sum_C \langle  {R}_A  {R}_C \rangle_{ \rho(t)} \Phi_{CB}(t) + \sum_C M^{-1}_{AC} \langle  {P}_C  {P}_B \rangle_{ \rho(t)} \nonumber \\
        &+ \langle  {R}_A \rangle_{ \rho(t)} F_B(t).
    \end{align}
\end{subequations}

To isolate the fluctuations from the macroscopic trajectory, we arrange the operators into column vectors $ {\bf R}$ and $ {\bf P}$. We define the centroid displacements $\overline{\bf R}(t)=\langle{\bf R}\rangle_{\rho(t)}-\tens{R}\unpert$ and $\overline{\bf P}(t)=\langle{\bf P}\rangle_{\rho(t)}$, and the centered fluctuation vectors $\delta {\bf R}(t) =  {\bf R} - \langle  {\bf R} \rangle_{ \rho(t)}$ and $\delta {\bf P}(t) =  {\bf P} - \langle  {\bf P} \rangle_{ \rho(t)}$. The connected correlators are thus represented as matrices, e.g., $\langle \delta {\bf R}(t) \delta {\bf R}(t)^T \rangle_{ \rho(t)} = \langle  {\bf R}  {\bf R}^T \rangle_{ \rho(t)} - \langle  {\bf R} \rangle_{ \rho(t)} \langle  {\bf R}^T \rangle_{ \rho(t)}$.
By differentiating these definitions with respect to time and substituting the macroscopic and uncentered equations [Eqs.\ \eqref{eq:eom_1st_moments_explicit} and \eqref{eq:eom_2nd_moments_uncentered}], the force vectors ${\bf F}(t)$ perfectly cancel out. Collecting the results, we obtain the coupled system of five equations of motion for the macroscopic moments and the centered fluctuation correlators. In matrix notation, these correspond exactly to Eq.\ 23 of Ref.\ \cite{siciliano2023wigner}:
\begin{subequations} \label{eq:eom_5_equations}
    \begin{align}
        \pdv{}{t} \overline{\bf R}(t) &= \mtrx{\mathbf{M}}{}^{-1} \overline{\bf P}(t), \\
        \pdv{}{t} \overline{\bf P}(t) &= -\mtrx{\bm{\Phi}}(t) \overline{\bf R}(t) + {\bf F}(t), \\
        \pdv{}{t} \langle \delta {\bf R} (t)\delta {\bf R}(t)^T \rangle_{ \rho(t)} &= \langle \delta {\bf R} (t) \delta {\bf P}(t)^T \rangle_{ \rho(t)} \mtrx{\mathbf{M}}{}^{-1} + \mtrx{\mathbf{M}}{}^{-1} \langle \delta {\bf P}(t) \delta {\bf R}(t)^T \rangle_{ \rho(t)}, \\
        \pdv{}{t} \langle \delta {\bf P}(t) \delta {\bf P}(t)^T \rangle_{ \rho(t)} &= -\langle \delta {\bf P}(t) \delta {\bf R}(t)^T \rangle_{ \rho(t)} \mtrx{\bm{\Phi}}(t) - \mtrx{\bm{\Phi}}(t) \langle \delta {\bf R} (t)\delta {\bf P}(t)^T \rangle_{ \rho(t)}, \\
        \pdv{}{t} \langle \delta {\bf R} (t) \delta {\bf P}(t)^T \rangle_{ \rho(t)} &= -\langle \delta {\bf R} (t) \delta {\bf R}(t)^T \rangle_{ \rho(t)} \mtrx{\bm{\Phi}}(t) + \mtrx{\mathbf{M}}{}^{-1} \langle \delta {\bf P} (t)\delta {\bf P}(t)^T \rangle_{ \rho(t)}.
    \end{align}
\end{subequations}

These equations map identically to the dynamics of the phonon condensate $ \tens{g}(t)$ and of the single-particle density matrix blocks $\mtrx{ \varrho}{}^\text{n}$ and $\mtrx{ \varrho}{}^\text{an}$ derived using the Cartesian boson operators. The formal connection is established by the substitution:
\begin{subequations}\label{app: RP to aa*}
\begin{eqnarray}
       { \bf R}- \tens{R}\unpert &=&\sqrt{\hbar}
     \mtrx{\bm{\lambda}}{}_{\mathbf{R}}\left[ {\tens{a}} +  {\tens{a}}^\dagger\right] , \\
   { \bf P}&= &i\sqrt{\hbar}\mtrx{\bm{\lambda}}{}_{\mathbf{P}}\left[ {\tens{a}} -  {\tens{a}}^\dagger\right].
\end{eqnarray}
\end{subequations}
By inserting Eq.\ \eqref{app: RP to aa*} into the centered correlators, we link the two bases explicitly:
\begin{subequations} \label{eq:covariance_to_rho_matrix}
    \begin{align}
        \overline{\mathbf{R}}(t) &=  \sqrt{\hbar}\mtrx{\bm{\lambda}}{}_{\mathbf{R}}\left[{\tens{g}(t)} + {\tens{g}(t)}^*\right],\\
        \overline{\mathbf{P}}(t) &=  i\sqrt{\hbar}\mtrx{\bm{\lambda}}{}_{\mathbf{P}}\left[{\tens{g}(t)} - {\tens{g}(t)}^*\right],\\
        \langle \delta {\bf R}(t)\delta {\bf R}(t)^T \rangle_{ \rho(t)} &= \hbar \mtrx{\bm{\lambda}}{}_{\mathbf{R}} \Big( \mtrx{ \varrho}{}^\text{n}(t) + \mtrx{ \varrho}{}^\text{n}(t)^T + \mtrx{ \varrho{}}{}^\text{an}(t) + \mtrx{ \varrho}{}^{\text{an} *}(t) + \mathbb{I} \Big) [\mtrx{\bm{\lambda}}{}_{\mathbf{R}}]^T, \\
        \langle \delta {\bf P}(t) \delta {\bf P}(t)^T \rangle_{ \rho(t)} &= \hbar \mtrx{\bm{\lambda}}{}_{\mathbf{P}} \Big( \mtrx{ \varrho}{}^\text{n}(t) + \mtrx{ \varrho}{}^\text{n}(t)^T - \mtrx{ \varrho}{}^\text{an}(t) - \mtrx{ \varrho}{}^{\text{an} *}(t) + \mathbb{I} \Big) [\mtrx{\bm{\lambda}}{}_{\mathbf{P}}]^T, \\
        \langle \delta {\bf R}(t) \delta {\bf P}(t)^T \rangle_{ \rho(t)} &= i\hbar \mtrx{\bm{\lambda}}{}_{\mathbf{R}} \Big( \mtrx{ \varrho}{}^\text{n}(t)^T - \mtrx{ \varrho}{}^\text{n}(t) + \mtrx{ \varrho}{}^\text{an}(t) - \mtrx{ \varrho}{}^{\text{an} *}(t) - \mathbb{I} \Big)[ \mtrx{\bm{\lambda}}{}_{\mathbf{P}}]^T.
    \end{align}
\end{subequations}
Using these identities within Eqs.\ \eqref{eq:eom_5_equations} reconstructs Eq.\ \eqref{d rho/dt =  2x2}.

Interestingly, Ref.\ \cite{siciliano2023wigner} we presented 5 independent equations to solve the dynamics, while the cartesian boson basis simplifies the number of independent equations to 3 [cf.\  \eqref{dg/dt explicit}]. Moreover, in Ref.\ \cite{siciliano2023wigner} we presented the equation of motion as a consequence of the Gaussian approximation on the density matrix. Here, we show how the equations of motion in the $\mathbf{R},\mathbf{P}$ basis follow from the fact that the Hamiltonian is quadratic, and are valid for any density matrix.

\section{Details of the linear response for the self-consistent case} \label{app: linear resp SC}

In this appendix, we report the details of the calculation of the linear response equations for the anharmonic self-consistent lattice dynamics.

We start by solving Eqs.\ \eqref{espald schrodinger (1) t SC}, which we recall here
\begin{subequations} 
\begin{align}
     i \pdv{}{t} \ket{E_{\mu\sigma}\pert(t)} &=  \sigma_z \H\unpert\ket{E_{\mu\sigma}\pert(t)} +  \sigma_z \H_\scf\pert(t)\ket{E_{\mu\sigma}\unpert(t)}   , \label{E pert(t) app}\\
     i \pdv{}{t} \ket{G\pert(t)}&=  \sigma_z \H \unpert\ket{G\pert(t)} - \sigma_z\kket{F_\scf \pert(t)}. \label{G pert (t) app}
\end{align}
\end{subequations}
in frequency space, where we dropped the parametric dependence on the density matrix on self-consistent terms for brevity.

First, we solve Eq.\ \eqref{E pert(t) app}. 
We use variation of constants method. We expand a spinor in the unperturbed spinor basis
\begin{equation}
    \ket{E_{\mu\sigma}(t)}  = \sum_{\lambda \sigma'}c_{\lambda\sigma',\mu\sigma}(t)  e^{-i\sigma' \omega_\lambda t}\ket{E_{\lambda\sigma'}\unpert} 
\end{equation}
where the coefficients are defined as  
\begin{equation}
c_{\lambda\sigma',\mu\sigma}(t) = \sigma'\mel{E_{\lambda\sigma'}\unpert}{\sigma_z}{E_{\mu\sigma}\unpert(t)}= \braket{E_{\lambda\sigma'}\unpert}{E_{\mu\sigma}(t)} \:.
\end{equation}
Expanding at linear order, we get
\begin{equation}
    \ket{E_{\mu\sigma}(t)}  = \sum_{\lambda \sigma'}c_{\lambda\sigma',\mu\sigma}\unpert(t)  e^{-i\sigma' \omega_\lambda t}\ket{E_{\lambda\sigma'}\unpert} + \sum_{\lambda \sigma'}c_{\lambda\sigma',\mu\sigma}\pert(t)  e^{-i\sigma' \omega_\lambda t}\ket{E_{\lambda{\sigma'}}\unpert}
\end{equation}
The first order equation of motion then becomes 
\begin{equation}
   \sum_{\lambda\sigma'}i \pdv{c_{\lambda\sigma',\mu\sigma}\pert(t)}{t}e^{-i\sigma'\omega_{\lambda}t} \ket{E_{\lambda{\sigma'}}\unpert} = \sum_{\lambda\sigma'}c_{\lambda\sigma',\mu\sigma}\unpert \sigma_z \H_\scf\pert(t)e^{-i\sigma'\omega_{\lambda}t} \ket{E_{\lambda{\sigma'}}\unpert}
\end{equation}
From the zero-th order equation it follows that
$c_{\lambda\sigma',\mu\sigma}\unpert =\delta_{\mu\lambda}\delta_{\sigma\sigma'}$, thus 
\begin{equation}
   \sum_{\lambda\sigma''}i \pdv{c_{\lambda\sigma'',\mu\sigma}\pert(t)}{t}e^{-i\sigma''\omega_{\lambda}t} \ket{E\unpert_{\lambda{\sigma''}}} = \sigma_z \H_\scf\pert(t)e^{-i\sigma\omega_{\mu}t} \ket{E_{\mu\sigma}\unpert}
\end{equation}
having changed $\sigma'{\to} \sigma ''$. Projecting on  $\ket{E_{\nu\sigma'}\unpert}$ we get 
\begin{equation}
    i\sigma' \pdv{c_{\nu\sigma',\mu\sigma}\pert(t)}{t}e^{-i\sigma'\omega_{\nu}t} =e^{-i\sigma\omega_{\mu}t} \mel{E_{\nu{\sigma'}}\unpert}{\H_\scf\pert(t)}{E_{\mu\sigma}\unpert}. 
\end{equation}
Considering that $c\pert(t=0) {=} 0$, we integrate in time
\begin{equation}
c\pert_{\nu\sigma'}(t) = \frac{1}{i\sigma' }\int_{-\infty}^t \dd{t'}e^{-i(\sigma\omega_{\mu}- \sigma'\omega_{\nu}) t'} \mel{E_{\nu{\sigma'}}\unpert}{\H\pert_\scf(t')}{E_{\mu\sigma}\unpert}
\end{equation}
where we have used the fact that $\H\pert_\scf(t{<}0){=}0$ to extend the integral.
We get the equation of the perturbed eigenvector in the interaction picture
\begin{equation}
   \ket{E_{\mu\sigma}\pert(t)}_I = \sum_{\nu\sigma'}  \ket{E_{\nu\sigma'}\unpert}\frac{1}{i\sigma' }\int_{-\infty}^t \dd{t'}e^{-i(\sigma\omega_{\mu}- \sigma'\omega_{\nu}) (t'-t)} \mel{E_{\nu{\sigma'}}\unpert}{\H\pert_\scf(t')}{E_{\mu\sigma}\unpert} .
\end{equation}
Its Fourier transform reads 
\begin{equation}
   \ket{E^\sigma_\mu(z)\pert}_I = \sum_{\nu\sigma'}  \ket{E_\nu^{\sigma'}}\frac{1}{i\sigma' }\int^\infty_{-\infty} e^{i[z -(\sigma'\omega_{\nu}- \sigma\omega_{\mu})] t}\int_{-\infty}^t \dd{t'}e^{i(\sigma\omega_{\nu}- \sigma'\omega_{\mu}) t'} \mel{E^{\sigma'}_{\nu}}{\H\pert(t')}{E^\sigma_\mu}.
\end{equation}
The integral can be solved with integration by parts (see e.g.\ Appendix A of Ref.\ \cite{caldarelli2025variational}), obtaining  (using $1/\sigma = \sigma$)
\begin{equation}
     \ket{E_{\mu\sigma}\pert(z)}_I =\sum_{\nu\sigma'} \sigma'\ket{E_{\nu\sigma'}\unpert} \frac{1}{ z - (\sigma' \omega_\nu - \sigma \omega_\mu) }\mel{E_{\nu\sigma'}\unpert}{\H_\scf\pert(z)}{E_{\mu\sigma}\unpert}.
\end{equation}
which is Eq.\ \eqref{E(z) pert}. 
To solve \eqref{G pert (t) app}, we directly pass in frequency space obtaining 
\begin{equation}
     z \ket{G\pert(z)}= \sigma_z\H\unpert\ket{G\pert(z)}- \sigma_z\ket{F_\scf\pert(z)}
\end{equation}
that is solved by 
\begin{equation}
    \ket{G\pert(z)}= -\sum_{\mu\sigma}\frac{\sigma}{ z -\sigma\omega_\mu}\braket{E\unpert_{\mu\sigma}}{F\pert(z)}
\end{equation}
which corresponds to Eq.\ \eqref{G(z) pert}.

\section{Details of the derivation of the time-dependent self-consistent harmonic approximation} \label{app: tdscha detail}
Here we collect some details of the derivation of the time-dependent self-consistent harmonic approximation employed to discuss the effective single-particle picture for anharmonic dynamics. 

{
\subsection{Dependence of the TD-SCHA Hamiltonian on single-particle quantities} 

Here, we show explicitly that the Gaussian density matrix used in the TD-SCHA is completely determined by the ESPALD variables $\varrho(t)$ and $\ket{G(t)}$. Using the Wigner representation for the time-dependent Gaussian of the TD-SCHA \cite{siciliano2023wigner}, fixing the parameters of the Wigner Gaussian is equivalent to fixing the many-body density matrix. That is, the many-body Gaussian is fully determined by its first and second moments, which we express below in terms of $\varrho(t)$ and $\ket{G(t)}$.

In the Wigner formulation of the TD-SCHA, the many-body density matrix is represented by the Gaussian phase-space distribution
\begin{equation}
\begin{split}
    \wt{\rho}({\bf R},{\bf P},t)
    =&\; \mathcal{N}(t)
    \exp\Bigg\{
    -\frac{1}{2}
    \delta \wt{\bf R}(t)^T
    \mtrx{\wt{\bm{\alpha}}}{}(t)
    \delta \wt{\bf R}(t)
    -\frac{1}{2}
    \delta \wt{\bf P}(t)^T
    \mtrx{\wt{\bm{\beta}}}{}(t)
    \delta \wt{\bf P}(t)
    +\delta \wt{\bf R}(t)^T
    \mtrx{\wt{\bm{\gamma}}}{}(t)
    \delta \wt{\bf P}(t)
    \Bigg\},
\end{split}
\end{equation}
where, following the notation of Ref.\ \cite{siciliano2023wigner},
\begin{equation}
    \wt R_A = \sqrt{M_A} R_A,
    \qquad
    \wt P_A = \frac{P_A}{\sqrt{M_A}},
    \qquad
    \delta\wt{\bf R}(t)=\wt{\bf R}-\langle \wt{\bf R}\rangle_{\rho(t)},
    \qquad
    \delta\wt{\bf P}(t)=\wt{\bf P}-\langle \wt{\bf P}\rangle_{\rho(t)} .
\end{equation}

Using the cartesian-boson transformation, Eq.\ \eqref{aa* to RP SC}, one obtains
\begin{subequations}\label{eq:first_moments_G_same_notation}
\begin{align}
    \langle {\bf R}\rangle_{\rho(t)}-\tens{R}\unpert
    &=
    \sqrt{\hbar}\LamR\left[\tens{g}(t)+\tens{g}(t)^*\right],
    \\
    \langle {\bf P}\rangle_{\rho(t)}
    &=
    i\sqrt{\hbar}\LamP\left[\tens{g}(t)-\tens{g}(t)^*\right],
\end{align}
\end{subequations}
and the centered correlators
\begin{subequations} 
    \begin{align}
        \langle \delta {\bf R}(t)\delta {\bf R}(t)^T \rangle_{ \rho(t)}
        &=
        \hbar \LamR
        \Big[
        \rhon(t)
        +
        \rhon(t)^T
        +
        \rhoan(t)
        +
        \rhoan(t)^{ *}
        +
        \mathbb{I}
        \Big]
        [\LamR]^T,
        \\
        \langle \delta {\bf P}(t) \delta {\bf P}(t)^T \rangle_{ \rho(t)}
        &=
        \hbar \LamP
        \Big[
        \rhon(t)
        +
        \rhon(t)^T
        -
        \rhoan(t)
        -
        \rhoan(t)^{ *}
        +
        \mathbb{I}
        \Big]
        [\LamP]^T,
        \\
        \langle \delta {\bf R}(t) \delta {\bf P}(t)^T \rangle_{ \rho(t)}
        &=
        i\hbar \LamR
        \Big[
        \rhon(t)^T
        -
        \rhon(t)
        +
        \rhoan(t)
        -
        \rhoan(t)^{ *}
        \Big]
        [\LamP]^T .
    \end{align}
\end{subequations}
Here $\delta {\bf R}(t)={\bf R}-\langle{\bf R}\rangle_{\rho(t)}$ and $\delta {\bf P}(t)={\bf P}-\langle{\bf P}\rangle_{\rho(t)}$, while the ESPALD single-particle variables are the elements of the single-particle density matrix
\begin{equation}
      \varrho(t) = \mqty[
      \rhon(t) & \rhoan(t) \\
      \rhoan(t)^* & \mathbb{I}+ [\rhon(t)]{}^T
      ] ,
\end{equation}
and of the phonon condensate $\ket{G(t)}=\mqty[\tens{g}(t)\\ \tens{g}(t)^*]$.

We then collect the mass-rescaled centered correlators in the covariance matrix
\begin{equation}
    {\wt{\bf C}}{}(t)
    =
    \mqty[
    \langle\delta\wt{\bf R}(t)\delta\wt{\bf R}(t)^T\rangle_{\rho(t)}
    &
    \langle\delta\wt{\bf R}(t)\delta\wt{\bf P}(t)^T\rangle_{\rho(t)}
    \\
    \langle\delta\wt{\bf P}(t)\delta\wt{\bf R}(t)^T\rangle_{\rho(t)}
    &
    \langle\delta\wt{\bf P}(t)\delta\wt{\bf P}(t)^T\rangle_{\rho(t)}
    ] .
\end{equation}
The inverse of this matrix gives the parameters appearing in the Wigner Gaussian,
\begin{equation}
    \mqty[
    \mtrx{\wt{\bm{\alpha}}}{}(t) & -\mtrx{\wt{\bm{\gamma}}}{}(t) \\
    -\mtrx{\wt{\bm{\gamma}}}{}(t)^T & \mtrx{\wt{\bm{\beta}}}{}(t)
    ]
    =
    {\wt{\bf C}}{}(t)^{-1} .
\label{eq:precision_from_covariance}
\end{equation}
Equations \eqref{eq:first_moments_G_same_notation}--\eqref{eq:precision_from_covariance} show that the Wigner Gaussian, and therefore the TD-SCHA many-body density matrix, is fixed once $\varrho(t)$ and $\ket{G(t)}$ are given.

As a consequence, the self-consistent averages entering the TD-SCHA Hamiltonian and forces can be regarded as functionals of the ESPALD variables. We can therefore write
\begin{equation}
    \H^{[\rho(t)]}(t) \equiv \H^{[\varrho(t),G(t)]}(t),
    \qquad
    \kket{F^{[\rho(t)]}(t)} \equiv \kket{F^{[\varrho(t),G(t)]}(t)} .
\end{equation}
 }
\subsection{Derivation of the anharmonic kernels in the TD-SCHA}
Here, we prove that the anharmonic kernels in the TD-SCHA are
\begin{subequations} \label{actions d3 d4 app}
    \begin{align}
        \overset{(3)}{{\mathcal{D}}}_{\!\!\!\substack{IJK\\ \sigma_1\sigma_2\sigma_3}} & {=} \sqrt{\hbar} \sum_{ABC}^{3N_\text{at}}{\Lambda}^T_{\mathbf{R}\,IA}   {\Lambda}^T_{\mathbf{R}\,JB}  \overset{(3)}{ D}_{ABC}
        {\Lambda}{}_{\mathbf{R}\, CK} \\
         \overset{(4)}{\mathcal{D}}_{\!\!\!\substack{IJKL\\\sigma_1\sigma_2\sigma_3\sigma_4}} & {=}  \frac{\hbar}{2} \!\!\sum_{ABCD}^{3N_\text{at}} \!\!{\Lambda}^T_{\mathbf{R}\,IA}   {\Lambda}^T_{\mathbf{R}\,JB}  \overset{(4)}{ D}_{ABCD}
        {\Lambda}{}_{\mathbf{R}\, CK} {\Lambda}_{\mathbf{R}\, DL} ,\\
         {\overset{(2)}{F}{}^\text{anh}_{I\sigma}} & = 0,\\
         \overset{(3)}{\mathcal F}_{\!\!\!\substack{IJK\\\sigma_1\sigma_2\sigma_3}} &=-\frac{\sqrt{\hbar}}{2} \sum_{ABC}^{3N_\text{at}}{\Lambda}^T_{\mathbf{R}\,IA}\overset{(3)}{ D}_{ABC}{\Lambda}_{\mathbf{R}\,BJ}
        {\Lambda}_{\mathbf{R}\, CK} 
    \end{align}
\end{subequations}
{
We start from the quadratic term of the TD-SCHA Hamiltonian \eqref{H = dRP compact SC}. The anharmonic kernels are obtained from the expansion of $\Phi_\scf(t)$ of Eq.\ \eqref{tdscha couplings} in absence of external perturbations in $H_\BO(t)$, which we dub}
\begin{align}
     \Phi_{\text{anh}I\a,J\beta}{}
     (t) &= \left\langle  \pdv{{V}_{\text{BO}}({\bf R})}{R_{I\a}}{R_{J\b}}\right\rangle_{ \rho(t)} .
\end{align}    
The latter is computed with the time-dependent Gaussian density matrix 
\begin{align} 
    \rho(\bm R,t) = {\frac{1}{(2\pi)^{3N_\text{at}/2}\sqrt{\det \langle\delta {\bf R}(t)\delta {\bf R}(t)^T\rangle_{ \rho(t)}}}}  \exp[-\tfrac{1}{2}\delta{\bf R}(t)^T\cdot\langle\delta {\bf R} (t)\delta {\bf R}(t)^T\rangle_{ \rho(t)}^{-1}\delta {\bf R}(t) ].
\end{align}
where $\delta  {\bf R} =   {\bf R} -\langle {\bf R} \rangle_{ \rho(t)}$. $\rho(\bm R,t)$ represents the probability of finding the ions of the system in a given position $\mathbf{R}{=}(R_{1x},\dots, R_{N_\text{at}z})$ at time $t$, and it is normalized over the configuration space $\mathbb{R}^{3N_\text{at}}$, $\int_{\mathbb{R}^{3N_\text{at}}} \dd{\mathbf R}\rho(\bm R,t) {=} 1$. Considering that the variations of a Gaussian are the variation of its variance and its mean, namely
\begin{align}\label{gaussian expansion}
    \rho\pert(\bm R,t)\simeq \pdv{\rho(\bm R,t)}{\langle\delta {\bf R} (t)\delta {\bf R}(t)^T\rangle_{ \rho(t)}}\eval_{ \rho\unpert} \!\!\langle\delta {\bf R} (t)\delta {\bf R}(t)^T\rangle_{ \rho\pert(t)}  + \pdv{\rho(\bm R,t)}{\langle {\bf R}\rangle_{ \rho(t)}}\eval_{ \rho\unpert} \!\!\langle {\bf R}\rangle_{ \rho\pert(t)}, 
\end{align}
by leveraging Price\'s theorem for Gaussian expectation values \cite{price,brown,bianco2017second}, which states 
\begin{subequations}
    \begin{align}
        \pdv{\Tr[  O(\bm R)   \rho]}{\langle\delta { R}_{I\a}\delta { R}_{J\b}\rangle_{ \rho}}  &= \frac{1}{2}\Tr[\pdv{  O(\bm R)}{  R_{I\a}}{  R_{J\b}}   \rho],\\
        \pdv{\Tr[  O(\bm R)   \rho]}{\langle  { R}_{I\a}\rangle_{ \rho}}  &= \Tr[\pdv{  O(\bm R)}{  R_{I\a}}   \rho],
    \end{align}
\end{subequations}
using derivation by parts we get the anharmonic quadratic coupling at linear order (summation on repeated indices is intended)
\begin{equation}
   {{ \Phi}}{}_{\text{anh}I\a,J\b}\pert
     (t)   =  \langle \frac{\partial^3{V}_\text{BO}(\bm R)}{\partial R_{I\a}\partial R_{J\b}\partial R_{K\delta}} \rangle_{ \rho\unpert} \langle     R_{K\delta}\rangle_{ \rho\pert(t)} +\frac{1}{2}  \langle \frac{\partial^4{V}_\text{BO}(\bm R)}{\partial R_{I\a}\partial R_{J\b}\partial R_{K\delta} \partial R_{L\gamma}} \rangle_{ \rho\unpert} \langle  \delta { R}_{K\delta} (t)\delta { R}_{L\gamma}(t)\rangle_{ \rho\pert(t)},
\end{equation}
that we express in a compact way and in frequency space as
\begin{equation}
   \mtrx{{ \Phi}}{}_{\text{anh}}\pert
     (z)   =  \overset{(3)}{\mathbf{D}}\, \langle  \mathbf R\rangle_{ \rho\pert(z)}+\frac{1}{2}  \overset{(4)}{\mathbf{D}} : \langle  \delta {\bf R} (t)\delta {\bf R} (t)^T\rangle_{ \rho\pert(z)}
\end{equation}
{where the $\overset{(3)}{D}$ and $\overset{(4)}{D}$ are defined in Eq.\ \eqref{phi3 phi4}.}
Considering the relation between the real-space moments of the Gaussian and the components of the single-particle density matrix [Eq.\ \eqref{eq:covariance_to_rho_matrix}], we can express it as 
\begin{equation}
     \mtrx{{ \Phi}}{}_{\text{anh}}\pert
     (z)   =  \sqrt{\hbar}\overset{(3)}{\mathbf{D}}  \mtrx{\bm{\Lambda}}{}_{\mathbf{R}}\sum_{\sigma_1}  \mathbf{G}_{\sigma_1}\pert(z)+\frac{\hbar}{2}  \overset{(4)}{\mathbf{D}}: [\mtrx{\bm{\Lambda}}{}_{\mathbf{R}} \sum_{\sigma_1\sigma_2}\mtrx{\varrho}{}_{\sigma_1\sigma_2}\pert(z) [\mtrx{\bm{\Lambda}}{}_{\mathbf{R}} ]^T]
\end{equation}
where $\mathbf{G}_{\sigma_1}\pert(z)$ is the $3N_{\text{at}}$ block of the phonon condensate $\ket{G\pert(z)}$ and $\mtrx{\varrho}{}_{\sigma_1\sigma_2}\pert(z)$ are $3N_{\text{at}}{\times} 3N_{\text{at}}$ blocks of the single-particle density matrix $\varrho\pert(t)$. The anharmonic Hamiltonian only depends on the ionic positions, hence using Eqs.\ \eqref{operators RP to aa*} its single particle equivalent reads 
\begin{equation}
    \H_\text{anh}\pert(z) = \mtrx{\bm{\Lambda}}{}_{\mathbf{R}}^T \left[\sqrt{\hbar}\overset{(3)}{\mathbf{D}}  \mtrx{\bm{\Lambda}}{}_{\mathbf{R}}\sum_{\sigma_1}  \mathbf{G}_{\sigma_1}\pert(z)+\frac{\hbar}{2}  \overset{(4)}{\mathbf{D}}: [\mtrx{\bm{\Lambda}}{}_{\mathbf{R}} \sum_{\sigma_1\sigma_2}\mtrx{\varrho}{}_{\sigma_1\sigma_2}\pert(z) [\mtrx{\bm{\Lambda}}{}_{\mathbf{R}} ]^T] \right] [\mtrx{\bm{\Lambda}}{}_{\mathbf{R}} ] \mqty[\mathbb{I} & \mathbb{I} \\ \mathbb{I} & \mathbb{I}  ].
\end{equation}
{
or explicitly with compact indices 
\begin{align}
     \H_{\text{anh}\, I\sigma,J\sigma'}\pert(t) &= \sqrt{\hbar} 
     \sum_{ABCD}^{3N_\text{at}}{{\Lambda}}_{\mathbf{R}\,{IA}}^T {{\Lambda}}_{\mathbf{R}\, JB} ^T\overset{(3)}{{D}}_{ABC} {{\Lambda}}_{\mathbf{R}\, CD} \sum_{\sigma_1}  {G}_{D\,\sigma_1}\pert(z) \nonumber \\& 
     + \frac{\hbar}{2} \sum_{ABCDEF}^{3N_\text{at}}{{\Lambda}}_{\mathbf{R}\,{IA}}^T {{\Lambda}}_{\mathbf{R}\, JB} ^T\overset{(4)}{{D}}_{ABCD} {{\Lambda}}_{\mathbf{R}\, CE}{{\Lambda}}_{\mathbf{R}\, DF} \sum_{\sigma_1
     \sigma_2}  {\varrho}_{EF\,\sigma_1\sigma_2}\pert(z)
\end{align}
from which the first two lines of Eqs. \eqref{actions d3 d4 app} follow.

Analogously, for the anharmonic forces we start from
\begin{equation}
     f^{R}_{\text{anh},I\a}(t) = -\left\langle  \pdv{ V_\text{BO}({\bf R})}{ R_{I\a}}\right\rangle_{ \rho(t)}+\sum_{J\b}\left\langle\pdv{V_\text{BO}({\bf R})}{ R_{I\a}}{ R_{J\b}}\right\rangle_{ \rho(t)}   [\langle { R}_{J\b}\rangle_{ \rho(t)} -  \R\unpert_{J\b}] .
\end{equation}
which we rewrite in the compact form 
\begin{equation}
     \mathbf f^{R}_{\text{anh}}(t) = -\left\langle  \pdv{ V_\text{BO}({\bf R})}{ \mathbf R}\right\rangle_{ \rho(t)}+\left\langle\pdv{V_\text{BO}({\bf R})}{\mathbf R}{\mathbf R}\right\rangle_{ \rho(t)}   [\langle {\bf R}\rangle_{ \rho(t)} -  \bm \R\unpert] .
\end{equation}
Considering the expansion  Eq.\ \eqref{gaussian expansion}, we get 
\begin{equation}
     \mathbf f^{R\, (1)}_{\text{anh}}(t) = -\frac{1}{2}\overset{(3)}{\mathbf{D}} : \langle\delta {\bf R} (t)\delta {\bf R} (t)^T\rangle_{ \rho\pert(t)} = -\frac{\hbar}{2}\overset{(3)}{\mathbf{D}} :\mtrx{\bm{\Lambda}}{}_{\mathbf{R}} \sum_{\sigma_1\sigma_2}\mtrx{\varrho}{}_{\sigma_1\sigma_2}\pert(t) [\mtrx{\bm{\Lambda}}{}_{\mathbf{R}} ]^T.
\end{equation}
Using Eqs.\ \eqref{operators RP to aa*}, we get the single-particle equivalent of the above force in frequency space as 
\begin{equation}
    \ket{F\pert_{\text{anh}}(z)} = -\frac{\sqrt{\hbar}}{2}[\mtrx{\bm{\Lambda}}{}_{\mathbf{R}} ]^T\left[\overset{(3)}{\mathbf{D}} :\mtrx{\bm{\Lambda}}{}_{\mathbf{R}} \sum_{\sigma_1\sigma_2}\mtrx{\varrho}{}_{\sigma_1\sigma_2}\pert(z) [\mtrx{\bm{\Lambda}}{}_{\mathbf{R}} ]^T \right]\mqty[\mathbb{I}\\ \mathbb{I}]
\end{equation}
or explicitly with compact indices 
\begin{equation}
    F\pert_{\text{anh}\, I\sigma}(z) = -\frac{\sqrt{\hbar}}{2}\sum_{ABCDE}^{3N_\text{at}}{{\Lambda}}{}_{\mathbf{R}\, IA}^T\overset{(3)}{{D}}_{ABC}{\Lambda}_{\mathbf{R\,} {BD}} {\Lambda}_{\mathbf{R}\, CE} \sum_{\sigma_1\sigma_2}{\varrho}_{D\sigma_1,E\sigma_2}\pert(z)  
\end{equation}
which proves the last two lines of \eqref{actions d3 d4 app}.
}

\subsection{Recovery of existing self-consistent approaches for anharmonic lattice dynamics}

Here, we show how to recover exactly the expression of the linear response function obtained in previous formulations of the TD-SCHA.

The existing formulations of the TD-SCHA follow a similar yet conceptually different approach to derive the anharmonic lattice response. In Refs.\ \cite{monacelli2021time,siciliano2023wigner,lihm2021gaussian},  anharmonicity is considered as part of the one-phonon and of the two-phonon propagator. Then, the response is obtained by combining the interacting phonon propagators with the vertices associated with the external fields and observables under study, i.e.\ bare vertices. 
However, since phonon interactions are included via a symbolic inversion of a Dyson equation of phonon propagators, existing TD-SCHA formulations lack the physical interpretation of anharmonicity as a phonon screening of the external perturbation. This physical analysis unlocked by the ESPALD will be of crucial importance for unmasking  microscopic mechanisms behind complex optical spectra of strongly anharmonic crystals or unveiling the nature of heat carriers near structural phase transitions.

{
\subsubsection{TD-SCHA one-phonon propagator}

The equivalence between ESPALD and the standard TD-SCHA one-phonon propagator is obtained by considering a mode-polarized displacement perturbation. In what follows, the superscript $(\nu)$ labels the first-order quantities induced by the external perturbation directed along the SCHA mode $\nu$,
\begin{equation}
    H_\text{pert}^{(\nu)}
    =
    \mathbf{e}_\nu^T \cdot\mtrx{\bf M}{}^{1/2}
    \cdot[{\bf R}-\tens{R}\unpert] .
\end{equation}
As in Eq.\ \eqref{1 phonon external}, this perturbation has no quadratic single-particle component,
\begin{subequations}
\begin{align}
    \H_\text{pert}^{(\nu)} &= 0,\\
    \ket{F_\text{pert}^{(\nu)}}
    &=
    -\frac{1}{\sqrt{2\hbar\omega_\nu}}
    \sum_\sigma
    \ket{E\unpert_{\nu\sigma}} .
\end{align}
\end{subequations}
Equivalently,
\begin{equation}
    \braket{E\unpert_{\mu\sigma}}{F_\text{pert}^{(\nu)}}
    =
    -\frac{\delta_{\mu\nu}}{\sqrt{2\hbar\omega_\nu}} .
\end{equation}

Since $\H_\text{pert}^{(\nu)}=0$, the external field directly drives the condensate. Combining the equations of the self-consistent cycle [Eqs.\ \eqref{sternheimer tdscha}], it is useful to introduce the pseudospin-unpolarized induced condensate
\begin{equation}
   G_\mu^{(\nu)}(z)
   =
   \sum_\sigma
   \braket{E\unpert_{\mu\sigma}}{G^{(\nu)}(z)}
\end{equation}
and the mode component of the external force
\begin{equation}
    F_{\text{pert},\mu}^{(\nu)}
    =
    -\frac{\delta_{\mu\nu}}{\sqrt{2\hbar\omega_\nu}} .
\end{equation}
The induced condensate then satisfies
\begin{equation} \label{G1 tdscha}
   G_\mu^{(\nu)}(z)
   =
   -G^0_\mu(z)
   F_{\text{pert},\mu}^{(\nu)}
   +
   \sum_{\theta}
   G^0_\mu(z)
   \Pi_{\mu\theta}(z)
   G_\theta^{(\nu)}(z) ,
\end{equation}
where the one-phonon self-energy is
\begin{equation}
{\Pi}_{\mu\nu}(z)
    =
    \sum_{\alpha\beta\gamma\delta}
    \overset{(3)}{\mathcal{D}}_{\mu\alpha\beta}
    \left[
    1-\frac{\hbar}{2}L^0(z)\overset{(4)}{\mathcal{D}}
    \right]^{-1}_{\alpha\beta\gamma\delta}
    \frac{\hbar}{2}L^0_{\gamma\delta}(z)
    \overset{(3)}{\mathcal{D}}_{\gamma\delta\nu},
\end{equation}
{ where
\begin{subequations} 
    \begin{align}
   \overset{(3)}{\mathcal{D}}_{\mu\nu\theta} &= {\sum_{\substack{IJK\\ \a\b\gamma}}}\frac{\text{e}_{\mu,I\a}\text{e}_{\nu,J\beta} \text{e}_{\theta,K\gamma}}{\sqrt{8M_IM_JM_K\w_\mu \w_\nu \w_\theta}} \left\langle \frac{\partial^3 V_\text{BO}({\bf R})}{\partial R_{I\a} \partial R_{J\b} \partial R_{K\gamma}}\right\rangle_{\!\! \rho\unpert}\!\!\!\!,\\
  \overset{(4)}{\mathcal{D}}_{\mu\nu\theta\xi} &= \sum_{\substack{IJKL\\\a\b\gamma\delta}}\frac{\text{e}_{\mu,I\a}\text{e}_{\nu,J\beta} \text{e}_{\theta,K\gamma} \text{e}_{\xi,L\delta}}{\sqrt{16M_IM_JM_KM_L \w_\mu \w_\nu\w_\theta\w_\xi}}\left\langle \frac{\partial^4 V_\text{BO}({\bf R})}{\partial R_{I\a} \partial R_{J\b} \partial R_{K\gamma}\partial R_{L\delta}}\right\rangle_{\!\! \rho\unpert} .
    \end{align}
\end{subequations}
are the anharmonic kernels of Eq.\ \eqref{tdscha D kernels} projected on the eigenmode basis.}
Eq.\ \eqref{G1 tdscha} is the Dyson equation for the displacement induced by a perturbation along mode $\nu$. Notice that we do not introduce a two-index induced condensate: the second mode index of the response function will arise from the label of the perturbation.

The displacement autocorrelation function is obtained by projecting $\ket{G^{(\nu)}(z)}$ on the single-particle observable associated with the displacement along mode $\mu$,
\begin{subequations}
\begin{align}
    \mathcal{O}_{\mu} &= 0,\\
    \ket{o_{\mu}}
    &=
    \frac{1}{\sqrt{2\hbar\omega_\mu}}
    \sum_\sigma
    \ket{E\unpert_{\mu\sigma}} .
\end{align}
\end{subequations}
Thus
\begin{equation} \label{onephonon anharm}
    \chi_{\mu\nu}(z)
    =
    \hbar
    \braket{o_{\mu}}{G^{(\nu)}(z)} .
\end{equation}
Equation \eqref{onephonon anharm} is the mode-resolved one-phonon Green's function. In the present normalization, the self-energy above is related to the one used in Refs.\ \cite{bianco2017second,lihm2021gaussian,monacelli2021time,siciliano2023wigner} by
\begin{equation}
    \Pi_{\mu\nu}(z)
    =
    \frac{1}{\sqrt{4\omega_{\mu} \omega_\nu}}
    \Pi^\text{Ref.\ \cite{lihm2021gaussian,monacelli2021time,siciliano2023wigner}}_{\mu\nu}(z),
\end{equation}
which follows from the {change of basis between the $\overset{(3/4)}{\mathcal{D}}$ and $\overset{(3/4)}{D}$ via the $\Lambda_\mathbf{R}$ matrices [Eq.\ \eqref{tdscha D kernels}]. These prefactors cancel out within the definition of $\ket{o_\mu},\ket{G(z)}$, so that  $\chi_{\mu\nu}(z)$ coincides with the anharmonic one-phonon propagator defined in Eq.\ (74) of Ref.\ \cite{siciliano2023wigner}.}

The sign difference in Eq.\ \eqref{G1 tdscha} follows from the fact that here the Hamiltonian coupling is written as
$-\mathbf{f}_{\mathbf{R}}(t)\cdot\mathbf{R}$, while in Ref.\ \cite{siciliano2023wigner} the external field is written in terms of the gradient of the external potential.

\subsubsection{TD-SCHA anharmonic two-phonon propagator}

The same logic applies to the two-phonon propagator. We now consider a mode-polarized quadratic perturbation, with the superscript $(\theta\eta)$ labeling the first-order quantities induced by
\begin{equation}
    H_\text{pert}^{(\theta\eta)}
    =
    \frac{1}{2}
    [{\bf R}-\tens{R}\unpert]^T
    \mtrx{\bf M}{}^{1/2}
    \mathbf{e}_\theta\mathbf{e}_\eta^T
    \mtrx{\bf M}{}^{1/2}
    [{\bf R}-\tens{R}\unpert] .
\end{equation}
As in Eq.\ \eqref{Vext dR2 singlepart}, this perturbation has no linear single-particle component,
\begin{subequations}
\begin{align}
    \H_\text{pert}^{(\theta\eta)}
    &=
    \frac{1}{4\sqrt{\omega_\theta\omega_\eta}}
    \sum_{\sigma\sigma'}
    \dyad{\tenscomp E_{\eta\sigma}\unpert}
          {\tenscomp E_{\theta\sigma'}\unpert}+\text{h.c.},
    \\
    \ket{F_\text{pert}^{(\theta\eta)}} &= 0 .
\end{align}
\end{subequations}

Since $\ket{F_\text{pert}^{(\theta\eta)}}=0$, the perturbation directly induces the single-particle density matrix. We denote by $\varrho^{(\theta\eta)}(z)$ the density response induced by the perturbation directed along the pair of modes $(\theta,\eta)$. Its pseudospin-unpolarized matrix elements are
\begin{equation}
     \varrho_{\mu\nu}^{(\theta\eta)}(z)
     =
     \sum_{\sigma\sigma'}
     \mel{E\unpert_{\mu\sigma}}
     {\varrho^{(\theta\eta)}(z)}
     {E\unpert_{\nu\sigma'}} .
\end{equation}
Combining the self-consistent Sternheimer equations [Eqs.\ \eqref{sternheimer tdscha}], one obtains
\begin{align} 
    \varrho_{\mu\nu}^{(\theta\eta)}(z)
    &=
    L^0_{\mu\nu}(z)
    \H_{\text{pert},\mu\nu}^{(\theta\eta)}
    \nonumber\\
    &+
    \frac{\hbar}{2}
    \sum_{\alpha\beta}
    L^0_{\mu\nu}(z)
    \Sigma_{\mu\nu\alpha\beta}(z)
    \varrho_{\alpha\beta}^{(\theta\eta)}(z),
\end{align}
where
\begin{equation}
    \Sigma_{\mu\nu\theta\xi}(z)
    =
    \overset{(4)}{D}_{\mu\nu\theta\xi}
    +
    \sum_{\lambda}
    \overset{(3)}{\mathcal{D}}_{\mu\nu\lambda}
    G^0_\lambda(z)
    \overset{(3)}{\mathcal{D}}_{\lambda\theta\xi}.
    \label{2phonon se}
\end{equation}
This is the Dyson equation for the density induced by the quadratic perturbation $(\theta\eta)$.

The ionic-variance autocorrelation function is obtained by projecting $\varrho^{(\theta\eta)}(z)$ on the observable associated with the squared displacement along the pair of modes $(\mu,\nu)$,
\begin{subequations}
\begin{align}
    \mathcal{O}{_{\mu\nu}}
    &=
    \frac{1}{4\sqrt{\omega_\mu\omega_\nu}}
    \sum_{\sigma\sigma'}
    \dyad{E_{\mu\sigma}\unpert}
          {E_{\nu\sigma'}\unpert}+\text{h.c.},
    \\
    \ket{o_{\mu\nu}} &=0 .
\end{align}
\end{subequations}
Therefore,
\begin{equation} \label{anharmonic 2ph espald}
    \chi_{\mu\nu,\theta\eta}(z)
    =
    \frac{\hbar}{2}
    \Tr\left[
    \mathcal{O}^{(\mu\nu)}
    \varrho^{(\theta\eta)}(z)
    \right] .
\end{equation}
Equation \eqref{anharmonic 2ph espald} is the mode-resolved anharmonic two-phonon propagator. The two-phonon self-energy of Eq.\ \eqref{2phonon se} is related to the one defined in Eq.\ (78) of Ref.\ \cite{siciliano2023wigner} by
\begin{equation}
     \Sigma_{\mu\nu\theta\xi}(z)
     =
     \frac{1}{\sqrt{16\omega_\mu\omega_\nu\omega_\theta\omega_\xi}}
     \Sigma^\text{Ref.\ \cite{siciliano2023wigner}}_{\mu\nu\theta\xi}(z).
\end{equation}
Thus, after projection on $\mathcal{O}^{(\mu\nu)}$, the response in Eq.\ \eqref{anharmonic 2ph espald} coincides with the anharmonic two-phonon propagator defined in Eq.\ (80) of Ref.\ \cite{siciliano2023wigner}.
}
\twocolumngrid

\bibliographystyle{unsrt}
\bibliography{espald-bib}

\begin{thebibliography}{100}
\bibitem{note:thermal-perturbation}
We adopt the sign convention associated with the preparation-and-switch-off protocol appropriate to statistical thermal perturbations. In this protocol, the system is prepared in the presence of the thermal field and the perturbation is subsequently switched off, consistently with the statistical nature of thermal transport.
\bibitem{note:sigma-z-action}
To facilitate the comparison, we recall that $\sigma_z$ changes the sign of the bottom row when acting from the left on a $2{\times}2$ matrix, while it changes the sign of the right column when acting from the right.
\end{thebibliography}

\end{document}